\documentclass[iop,appepreferably]{emulateapj}
\usepackage{natbib}
\usepackage{graphicx}
\usepackage{epstopdf}
\usepackage{lineno}
\usepackage{footnote}
\usepackage{epsfig}
\usepackage{graphics}
\usepackage{amssymb}
\usepackage{amsmath}
\usepackage{xcolor}
\usepackage{tikz}
\newcommand{\mycbox}[1]{\tikz{\path[draw=#1,fill=#1] (0,0) rectangle (0.2cm,0.2cm);}}
\usepackage{lineno}
\usepackage{rotating}
\newcommand{\noprint}[1]{}

\newcommand{\figsetend}{}
\newcommand{\figsetgrpstart}{}
\newcommand{\figsetgrpend}{}

\newcommand{\figsetgrpnum}[1]{\noprint{#1}}
\newcommand{\figsetgrptitle}[1]{\noprint{#1}}
\newcommand{\figsetplot}[1]{\noprint{#1}}
\newcommand{\figsetgrpnote}[1]{\noprint{#1}}

\shorttitle{Neutrino Expectations from short Blazar Flares}
\shortauthors{Kreter et al.}

\usepackage[utf8]{inputenc}
\usepackage[T1]{fontenc}
\usepackage{hyperref}

\begin{document}

\title{On the Detection Potential of Blazar Flares for\\ Current Neutrino Telescopes}

\author{M. Kreter$^1$, M. Kadler$^2$, F. Krau\ss{}$^3$, K. Mannheim$^2$, S. Buson$^2$, R. Ojha$^4$, J. Wilms$^5$\\ and M. B\"{o}ttcher$^1$}
\affil{$^1$Centre for Space Research, North-West University, Private Bag X6001, Potchefstroom 2520, South Africa}
\affil{$^2$Lehrstuhl f\"{u}r Astronomie, Universit\"{a}t W\"{u}rzburg, Emil-Fischer-Strasse 31, 97074 W\"{u}rzburg, Germany}
\affil{$^3$Department of Astronomy \& Astrophysics, Pennsylvania State University, University Park, PA, 16802, United States}
\affil{$^4$NASA Goddard Space Flight Center, Greenbelt, MD 20771, USA}
\affil{$^5$Dr. Remeis Sternwarte \& ECAP, Universit\"at Erlangen-N\"urnberg, Sternwartstrasse 7, 96049 Bamberg, Germany}
\email{michael@kreter.org}

\begin{abstract}
Blazar jets are extreme environments, in which relativistic proton interactions with an ultraviolet photon field could give rise to photopion production.~High-confidence associations of individual high-energy neutrinos with blazar flares could be achieved via spatially and temporally coincident detections. In 2017, the track-like, extremely high-energy neutrino event IC\,170922A was found to coincide with increased $\gamma$-ray emission from the blazar TXS\,0506+056, leading to the identification of the most promising neutrino point source candidate so far. We calculate the expected number of neutrino events that can be detected with IceCube, based on a broadband parametrization of bright short-term blazar flares that were observed in the first 6.5 years of \textit{Fermi}/LAT observations. We find that the integrated keV-to-GeV fluence of most individual blazar flares is far too small to yield a substantial Poisson probability for the detection of one or more neutrinos with IceCube. We show that the sample of potentially detectable high-energy neutrinos from individual blazar flares is rather small. We further show that the blazars 3C\,279 and PKS\,1510$-$089 dominate the all-sky neutrino prediction from bright and short-term blazar flares.
In the end, we discuss strategies to search for more significant associations in future data unblindings of IceCube and KM3NeT.
\end{abstract}

\section{Introduction}
\hspace{-14pt}
The discovery of extraterrestrial very-high-energy neutrinos by \citet{IceCube_2013} has triggered the hunt for the identification of their astrophysical sources. Jets from Active Galactic Nuclei (AGN) are promising candidates for the detected high-energy neutrino flux \citep{Mannheim_1993, Mannheim_1995}, as they are among the most extreme environments in the universe.
Especially promising are blazars, a subclass of AGN  characterized by jets pointing towards the observer that show variability on timescales from minutes to years. 
Blazars can be further subdivided into BL\,Lac objects and  Flat-Spectrum Radio Quasars (FSRQ). While typical BL Lac spectra are dominated by a featureless non-thermal continuum,
FSRQs are known for broad emission lines at optical wavelengths and a flat radio spectrum.
Due to their extreme luminosities, these sources are promising candidates for photohadronic neutrino production \citep{Mannheim_1989,Mannheim_1992, Stecker_1996}.
On 22 September 2017, the \citet{IceCube_2018a} detected a $\sim\,290$\,TeV  neutrino-induced, track-like event (called IC\,170922A) from a direction spatially and temporally consistent with enhanced $\gamma$-ray activity from the blazar TXS\,0506+056. In a separate follow-up study the \citet{IceCube_2018b} found an additional excess of $13 \pm 5$ events (within $\sim$ 6 months in 2014--2015) from the direction of TXS\,0506+056, uncorrelated to $\gamma$-ray activity observed by the \textit{Fermi} Large Area Telescope (LAT).
By using leptohadronic models \citet{Petropoulou_2019} investigated whether this excess can be explained without violating existing electromagnetic observations. They found that none of their models are consistent with the observed multimessenger emission, suggesting the presence of multiple emission regions in the jet of TXS\,0506+056.\\

By studying flaring periods in Mrk\,421, which is one of the nearest blazars to Earth, \citet{Petropoulou_2016} found that the predicted PeV neutrino flux is correlated to the photon flux throughout the electromagnetic spectrum. They further conclude that even a non-detection of  
high-energy neutrinos by IceCube does not exclude a neutrino/$\gamma$-ray correlation during bright flaring periods. Recent neutrino-blazar correlation studies indicate that the contribution from blazars to the total high-energy neutrino flux measured by IceCube is rather small \citep{Turley_2016, Aartsen_2016, Murase_2016, Padovani_2016}. \citet{Hooper_2019} concludes that the  contribution from blazars to the diffuse neutrino flux measured by IceCube is not larger than 5--15\%.
However, in our previous work we showed that the contribution from individual blazar flares is not negligible \citep{Krauss_2014, Kadler_2016}. While general neutrino point-source searches \citep{IceCube_2013, IceCube_2018a, Aartsen_2015b, ANTARES_2015} focus on a simultaneous correlation to the observed blazar $\gamma$-ray flux, we have shown in \citet{Krauss_2018} that the total high-energy fluence is a much better proxy for the expected number of observable neutrino events.\\

Besides blazars, other source classes such as starburst galaxies and $\gamma$-ray bursts (GRBs) have been suggested as potential counterparts of the observed high-energy neutrino flux 
\citep{Loeb_2006, Razzaque_2003, Waxman_1997, Waxman_2000}. So far, a coincident detection between a high-energy neutrino and any known GRB has not been found \citep{Achterberg_2008, Aartsen_2016b}. Starburst galaxies and other types of AGN such as radio-quiet quasars \citep{Stecker_1996, Alvarez_2004}, low luminosity AGN (LLAGN) \citep{Kimura_2015} or radio galaxies \citep{Becker_2014, Hooper_2016, Blanco_2017} remain potential  candidates for the observed diffuse neutrino flux.\\

Well-established jet models such as ~\citet{Bottcher_2013} describe the multiwavelength behavior of blazars in a temporary steady state. In a purely leptonic scenario, the self-consistent radiative output is calculated based on a temporary equilibrium between particle injection,
radiative cooling and electron escape from a spherical emission region. In a lepto-hadronic scenario, the radiative output is evaluated for both primary electrons and protons, taking internal $\gamma\gamma$-pair production cascades into account.
Other codes like \textit{Athena} describe relativistic accretion flows based on astrophysical magnetohydrodynamics (MHD) \citep{Stone_2008}. As \textit{Athena}  is based on directionally unsplit, higher-order Godunov methods, it is well suited to describe the rotational discontinuities typically present in blazar jets.

More recent models developed for the TXS\,0506+056 association indicate that the neutrino emission from $p-\gamma$ interactions can be best described by a lepto-hadronic model with a weak hadronic component \citep{Gao_2019, Oikonomou_2019}. Such models predict a correlation between the observed X-ray/TeV emission and the neutrino flux. 
Other studies, such as \citet{Reimer_2019}, suggest a stationary soft-X-ray photon target field in a photo-hadronic neutrino production scenario, and a decoupling of the observed $\gamma$-ray emission and measured neutrino flux. \\

In this work, we present a calorimetric approach to estimate the number of predicted high-energy neutrino events an individual blazar flare would produce.
We compute the expected  neutrino excess rates of bright blazar flares by performing a systematic search for flaring intervals on different time scales.
This paper is structured in the following way:
Section\,\ref{Method} introduces the methods. We introduce the algorithm used to identify bright short-term blazar flares and the multiwavelength analysis performed. We show how the calorimetric neutrino expectation for individual blazar flares is derived. In Sect.\,\ref{Section Results} we present the results of the calorimetric neutrino predictions for the 50 most promising flares, derived from a sample of 150 bright blazars. In Sect.\,\ref{Discussion} we calculate the total neutrino detection potential for our sample of 50 flares and compare our findings to previous neutrino point search studies. 
Limitations of the methods are discussed. We conclude in Sect.\,\ref{Conclusion} and suggest strategies to search for neutrino blazar correlation studies in the future.

\section{Method} \label{Method}

\subsection{Flare Identification} \label{Flare_identification}
\hspace{-14pt}
We used daily binned \textit{Fermi}/LAT $\gamma$-ray light curves of a sample of 150 bright blazars monitored daily by the LAT\footnote{All these light curves are publicly available at  \url{http://fermi.gsfc.nasa.gov/ssc/data/access/lat/msl_lc/}}.
Note that our sample is primarily dominated by FSRQs. This is due to the fact that the brightest $\gamma$-ray flares are always found to be associated with this class of AGN \citep[e.g.][for 3C 454.3 and 3C 279 flares]{Ackermann_2011,Ackermann_2015}.
This sample of the brightest blazars consists of all sources that have reached a daily flux level at energies above 100 MeV of $10^{-6}\,\text{cm}^{-2}\,\text{s}^{-1}$ at least once since the start of the \textit{Fermi} mission. Although these automated light curves do not have absolute calibration, we can use them just to identify times of flaring activity. ~We identify periods of enhanced activity according to:
\begin{align}
\label{flare_selection_1}
\Delta_{\text{flare}} = \left[\left(\text{Flux}  - 3 \times \text{Flux}_{\text{err}}\right) \geq 3\text{G}\right] \times \text{A}_{\text{Eff}}
 \end{align}
 The parameters $\text{Flux}$ and $\text{Flux}_{\text{err}}$ indicate the daily \textit{Fermi}/LAT photon flux and $1 \sigma$ flux uncertainty in the energy range of $100$\,MeV -- $300$\,GeV, while
$\text{G}$ corresponds to a (source-dependent) non-flaring quiescent ground level state. The quantity $\Delta_{\text{flare}}$ therefore defines a threshold selection criterion for flares of a considered blazar light curve. 
\begin{figure}[pb]
 \centering
\plotone{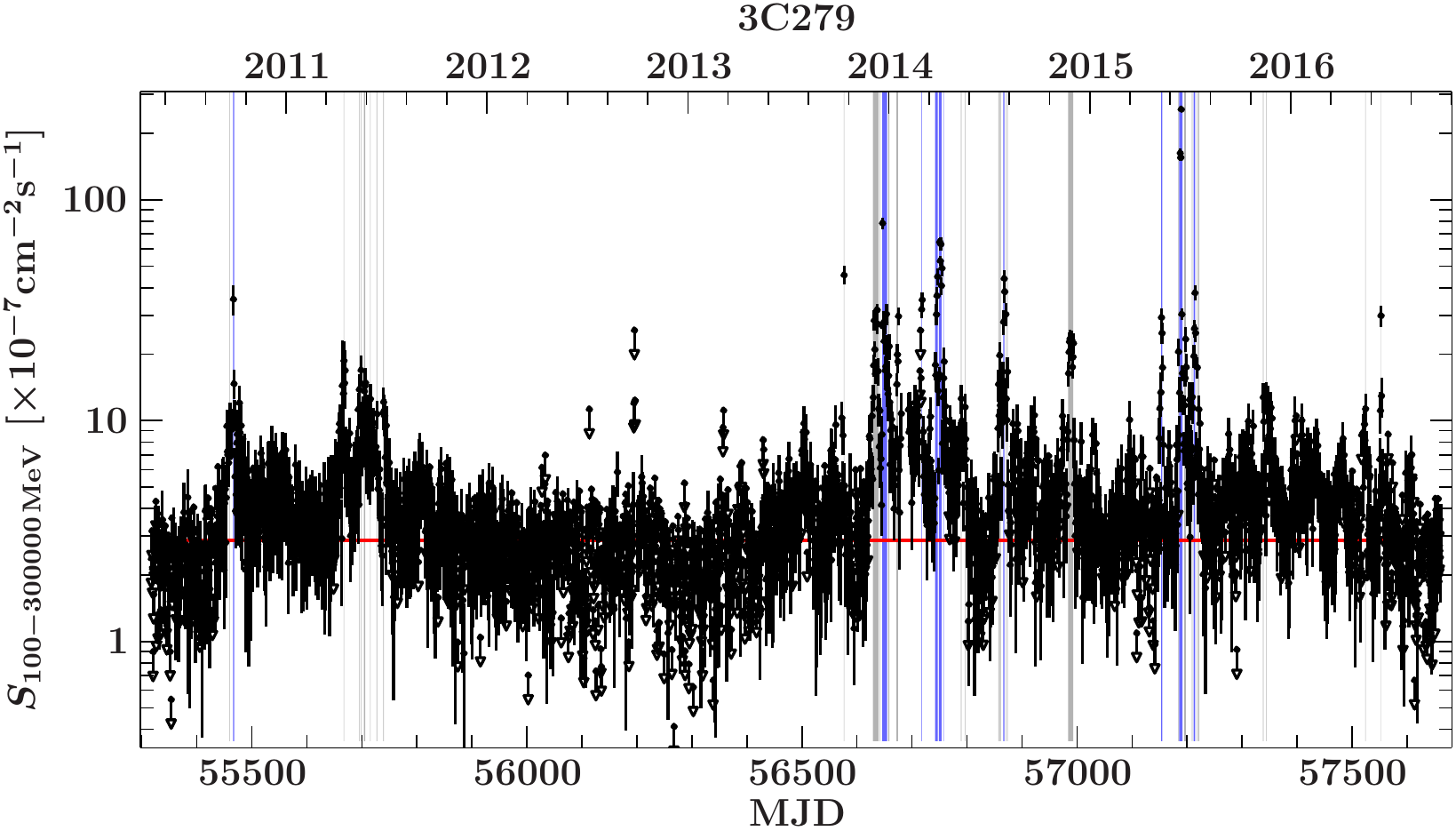}
\caption{\textit{Fermi}/LAT $\gamma$-ray light curve for 3C\,279 in the energy range  $100$\,MeV -- $300$\,GeV and a time range of $55317\,\text{MJD}$ (2010-05-01) -  $57662\,\text{MJD}$ (2016-10-01) using daily binning.
    Highlighted flares  (blue and gray) correspond to selected time intervals following the method of Sect.\,\ref{Flare_identification}, while blue intervals indicate the most promising flares, as listed in Table\,\ref{Tab_neutrino_expectation}. The red line corresponds to the identified ground level G. Zooms on flaring periods from Table\,\ref{Tab_neutrino_expectation} are provided as supplementary material.\\~\\
    (The complete figure set (20 images) is available in the online journal.)}
    \label{flare_selection_histo}
\end{figure}

\figsetgrpstart
\figsetgrpnum{1.1}
\figsetgrptitle{PKS\,0402$-$362}
\figsetplot{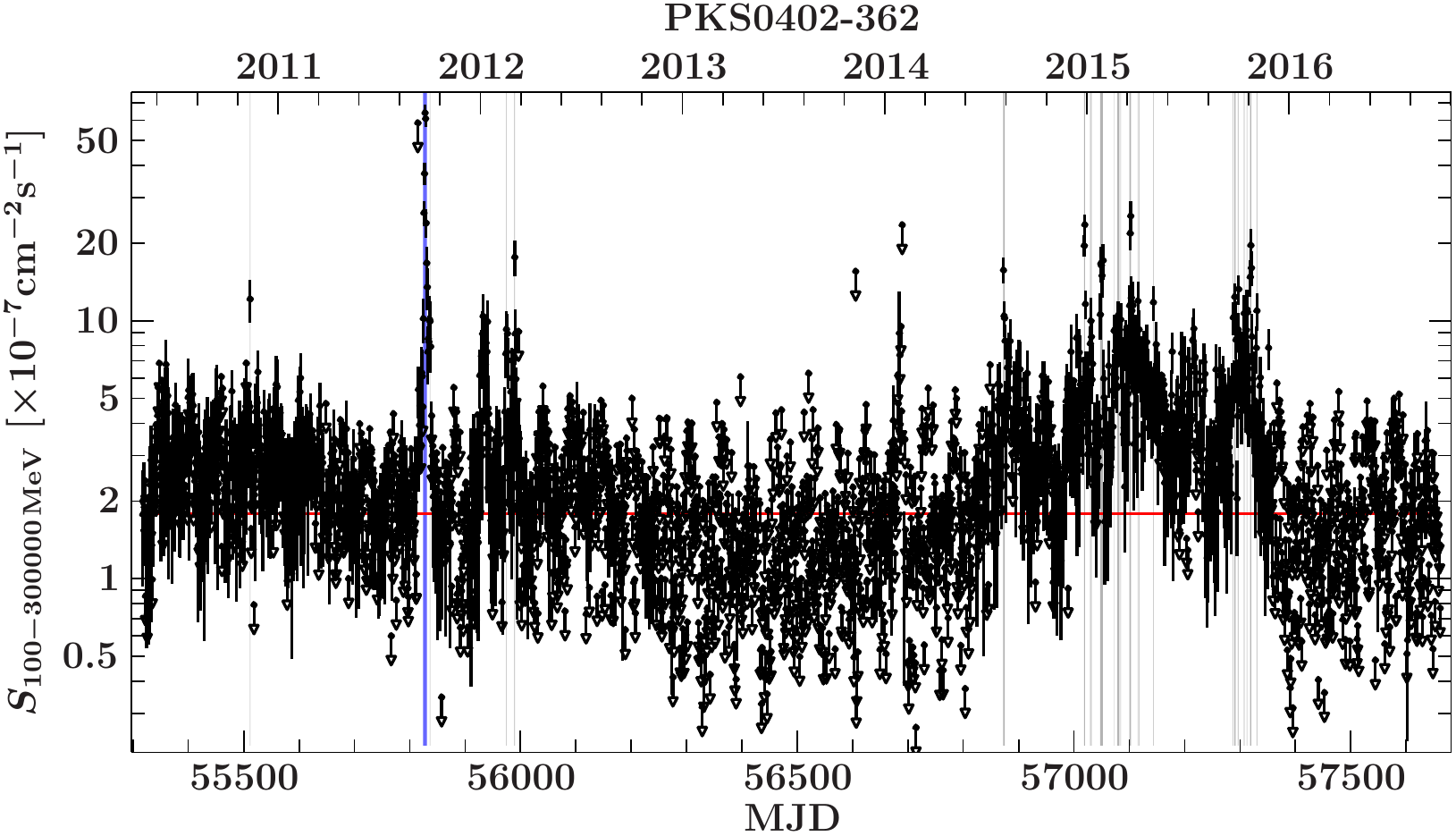}
\figsetgrpnote{Fermi/LAT γ-ray light curve for PKS\,0402$-$362  in the energy range  $100$\,MeV -- $300$\,GeV and a time range of $55317\,\text{MJD}$ (2010-05-01) -  $57662\,\text{MJD}$ (2016-10-01) using daily binning.     Highlighted flares  (blue and gray) correspond to selected time intervals following the method of Sect.\,\ref{Flare_identification}, while blue intervals indicate the most promising flares, as listed in Table\,\ref{Tab_neutrino_expectation}. The red line corresponds to the identified ground level G.}
\figsetgrpend

\figsetgrpstart
\figsetgrpnum{1.2}
\figsetgrptitle{3C\,279}
\figsetplot{./plots/Light_curves/3C279_1d_100-300000MeV.pdf}
\figsetgrpnote{Fermi/LAT γ-ray light curve for 3C\,279 in the energy range  $100$\,MeV -- $300$\,GeV and a time range of $55317\,\text{MJD}$ (2010-05-01) -  $57662\,\text{MJD}$ (2016-10-01) using daily binning.     Highlighted flares  (blue and gray) correspond to selected time intervals following the method of Sect.\,\ref{Flare_identification}, while blue intervals indicate the most promising flares, as listed in Table\,\ref{Tab_neutrino_expectation}. The red line corresponds to the identified ground level G.}
\figsetgrpend

\figsetgrpstart
\figsetgrpnum{1.3}
\figsetgrptitle{PKS\,1329$-$049}
\figsetplot{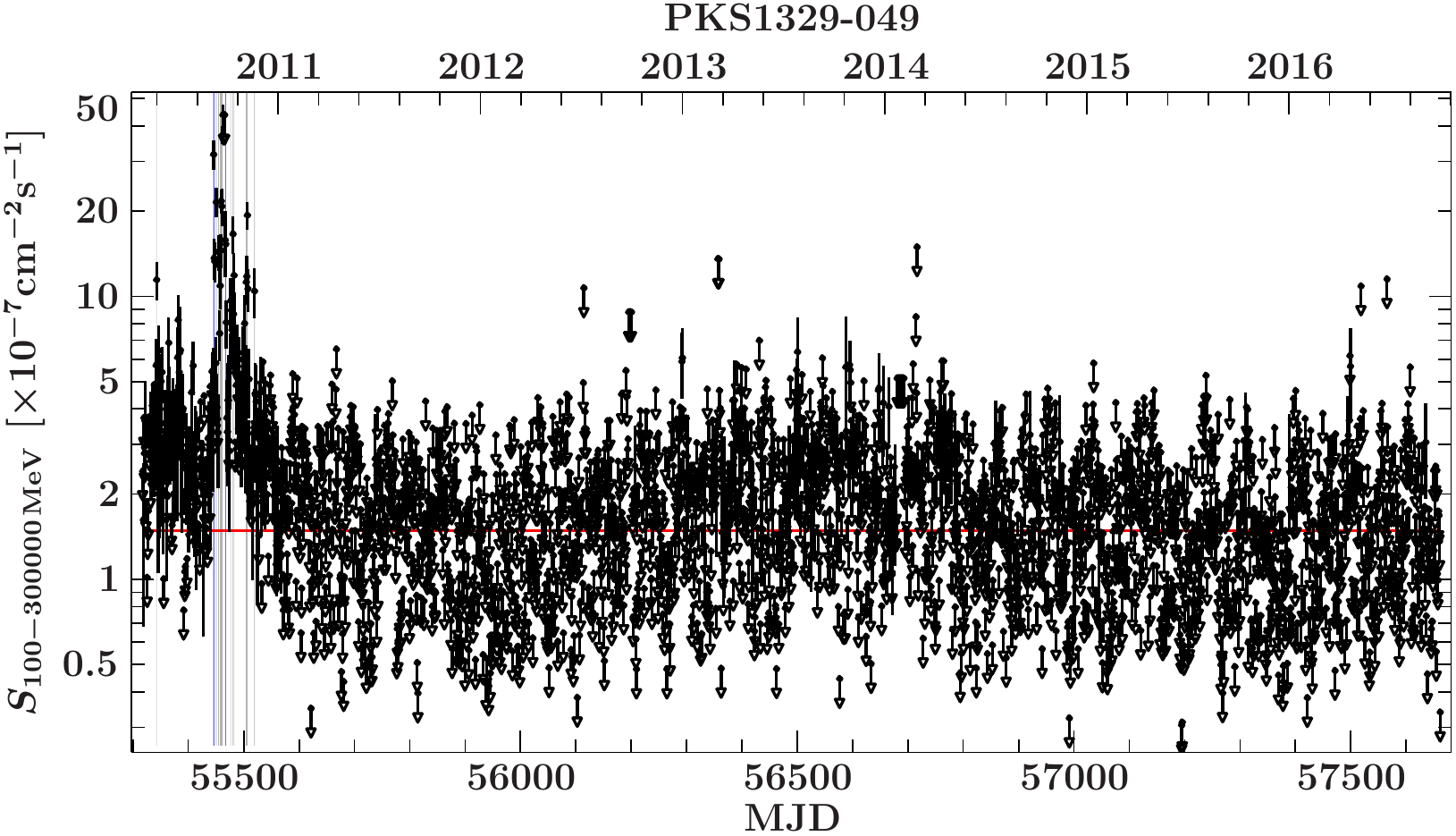}
\figsetgrpnote{Fermi/LAT γ-ray light curve for PKS\,1329$-$049 in the energy range  $100$\,MeV -- $300$\,GeV and a time range of $55317\,\text{MJD}$ (2010-05-01) -  $57662\,\text{MJD}$ (2016-10-01) using daily binning.     Highlighted flares  (blue and gray) correspond to selected time intervals following the method of Sect.\,\ref{Flare_identification}, while blue intervals indicate the most promising flares, as listed in Table\,\ref{Tab_neutrino_expectation}. The red line corresponds to the identified ground level G.}
\figsetgrpend

\figsetgrpstart
\figsetgrpnum{1.4}
\figsetgrptitle{PKS\,1424$-$418}
\figsetplot{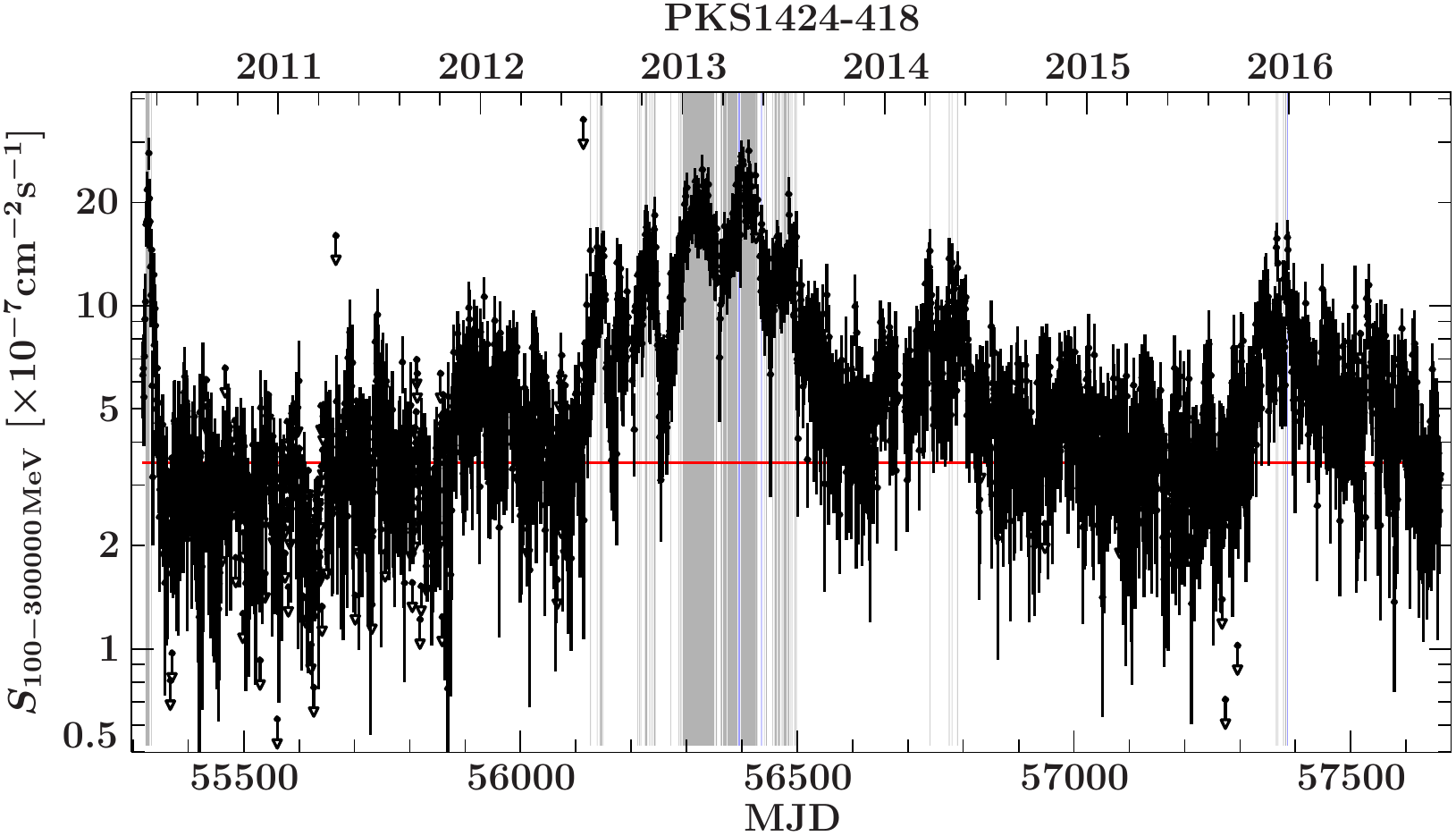}
\figsetgrpnote{Fermi/LAT γ-ray light curve for PKS\,1424$-$418 in the energy range  $100$\,MeV -- $300$\,GeV and a time range of $55317\,\text{MJD}$ (2010-05-01) -  $57662\,\text{MJD}$ (2016-10-01) using daily binning.     Highlighted flares  (blue and gray) correspond to selected time intervals following the method of Sect.\,\ref{Flare_identification}, while blue intervals indicate the most promising flares, as listed in Table\,\ref{Tab_neutrino_expectation}. The red line corresponds to the identified ground level G.}
\figsetgrpend

\figsetgrpstart
\figsetgrpnum{1.5}
\figsetgrptitle{PKS\,1510$-$089}
\figsetplot{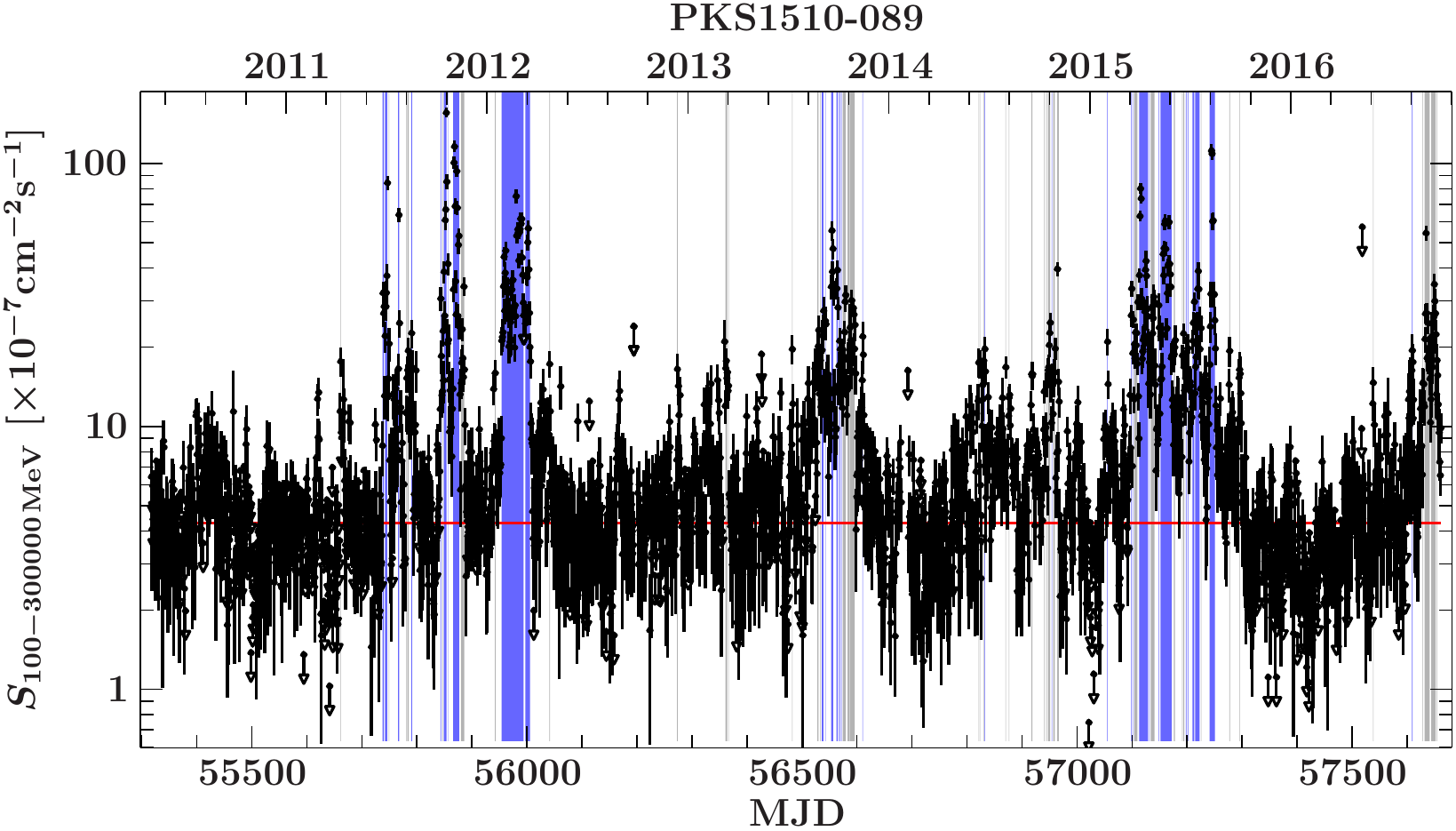}
\figsetgrpnote{Fermi/LAT γ-ray light curve for PKS\,1510$-$089 in the energy range  $100$\,MeV -- $300$\,GeV and a time range of $55317\,\text{MJD}$ (2010-05-01) -  $57662\,\text{MJD}$ (2016-10-01) using daily binning.     Highlighted flares  (blue and gray) correspond to selected time intervals following the method of Sect.\,\ref{Flare_identification}, while blue intervals indicate the most promising flares, as listed in Table\,\ref{Tab_neutrino_expectation}. The red line corresponds to the identified ground level G.}
\figsetgrpend

\figsetgrpstart
\figsetgrpnum{1.6}
\figsetgrptitle{CTA\,102}
\figsetplot{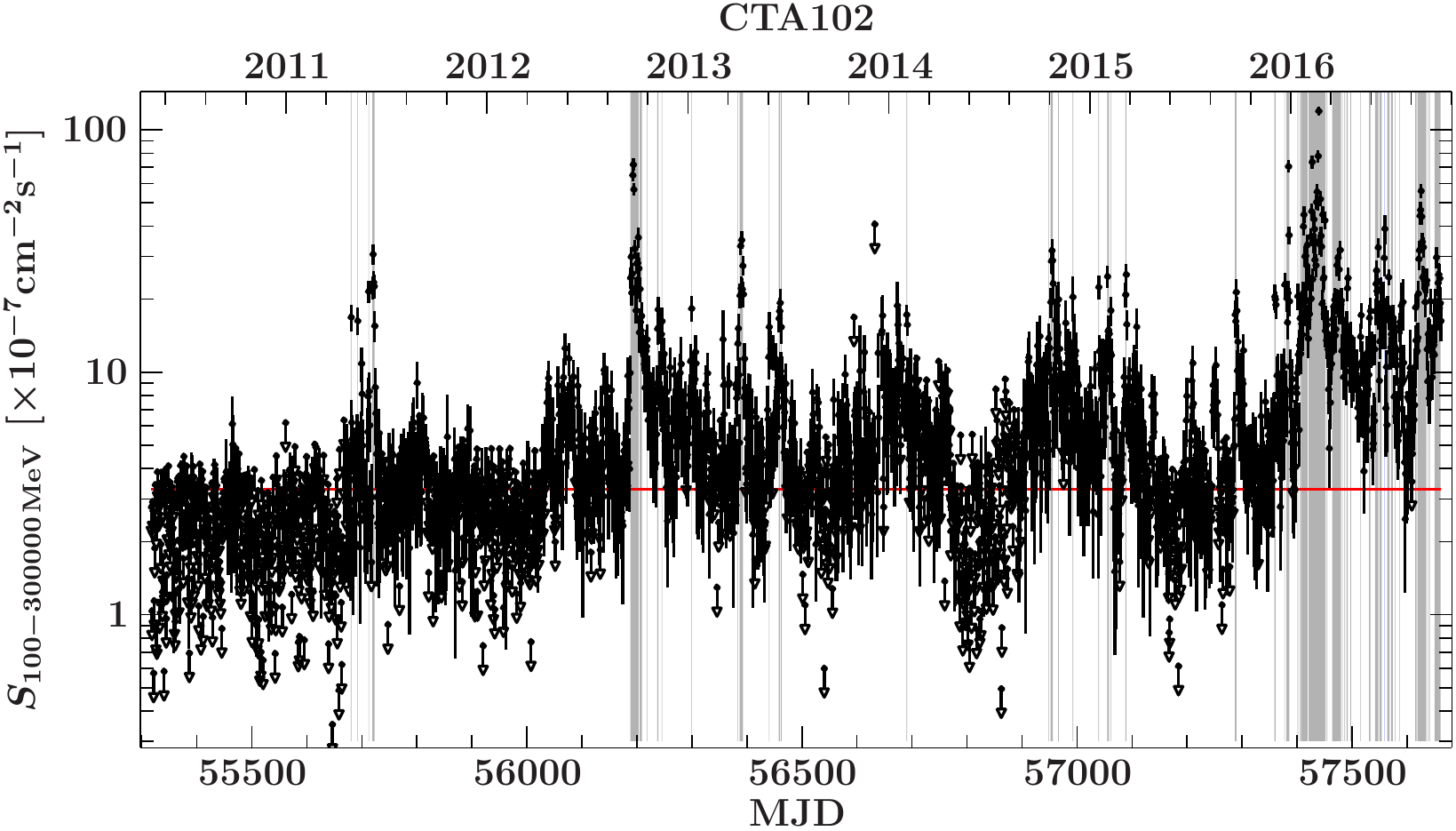}
\figsetgrpnote{Fermi/LAT γ-ray light curve for CTA\,102 in the energy range  $100$\,MeV -- $300$\,GeV and a time range of $55317\,\text{MJD}$ (2010-05-01) -  $57662\,\text{MJD}$ (2016-10-01) using daily binning.     Highlighted flares  (blue and gray) correspond to selected time intervals following the method of Sect.\,\ref{Flare_identification}, while blue intervals indicate the most promising flares, as listed in Table\,\ref{Tab_neutrino_expectation}. The red line corresponds to the identified ground level G.}
\figsetgrpend

\figsetgrpstart
\figsetgrpnum{1.7}
\figsetgrptitle{3C\,454.3}
\figsetplot{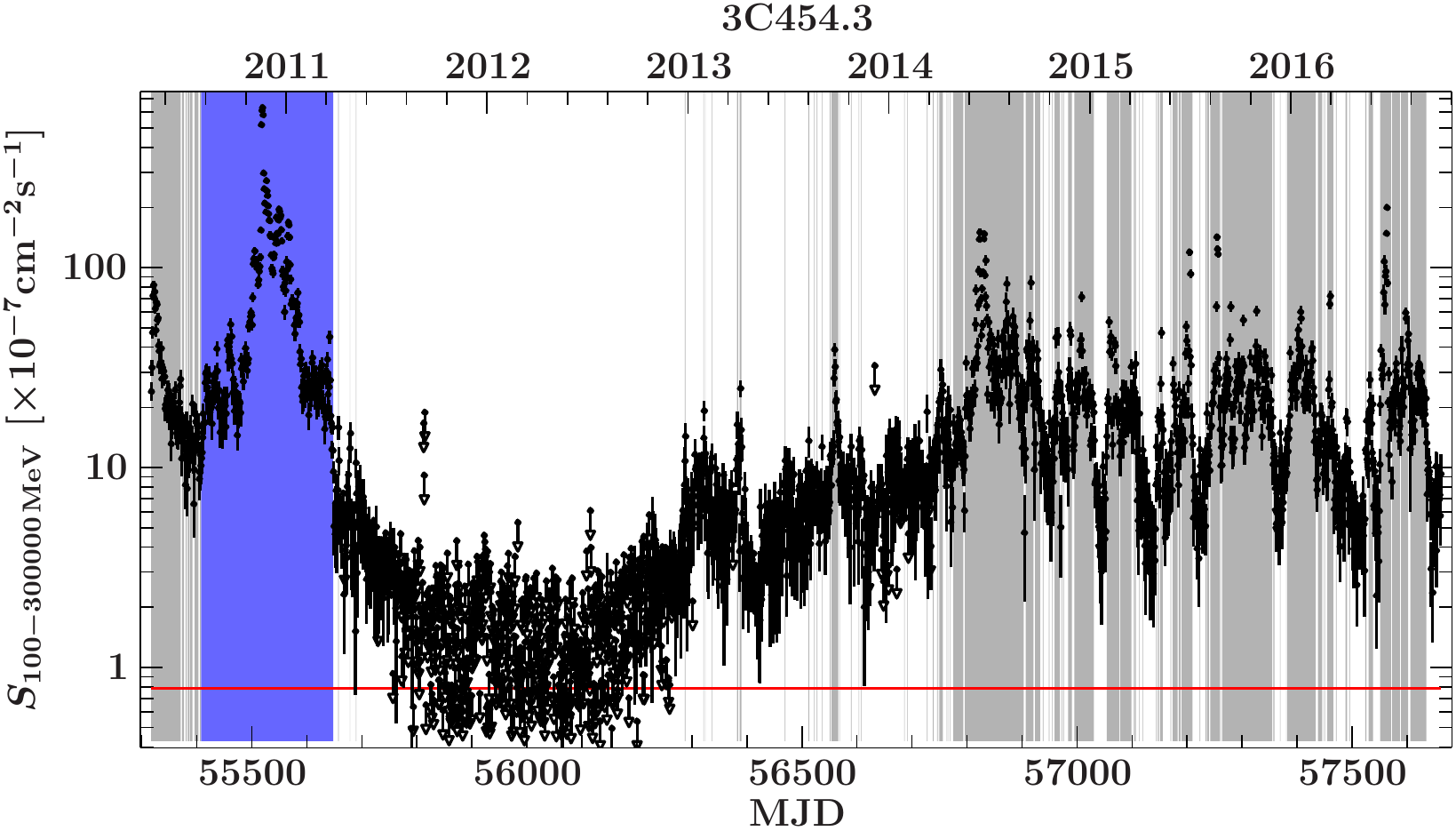}
\figsetgrpnote{Fermi/LAT γ-ray light curve for 3C\,454.3 in the energy range  $100$\,MeV -- $300$\,GeV and a time range of $55317\,\text{MJD}$ (2010-05-01) -  $57662\,\text{MJD}$ (2016-10-01) using daily binning.     Highlighted flares  (blue and gray) correspond to selected time intervals following the method of Sect.\,\ref{Flare_identification}, while blue intervals indicate the most promising flares, as listed in Table\,\ref{Tab_neutrino_expectation}. The red line corresponds to the identified ground level G.}
\figsetgrpend
\figsetend

\figsetgrpstart
\figsetgrpnum{1.8}
\figsetgrptitle{3C\,279}
\figsetplot{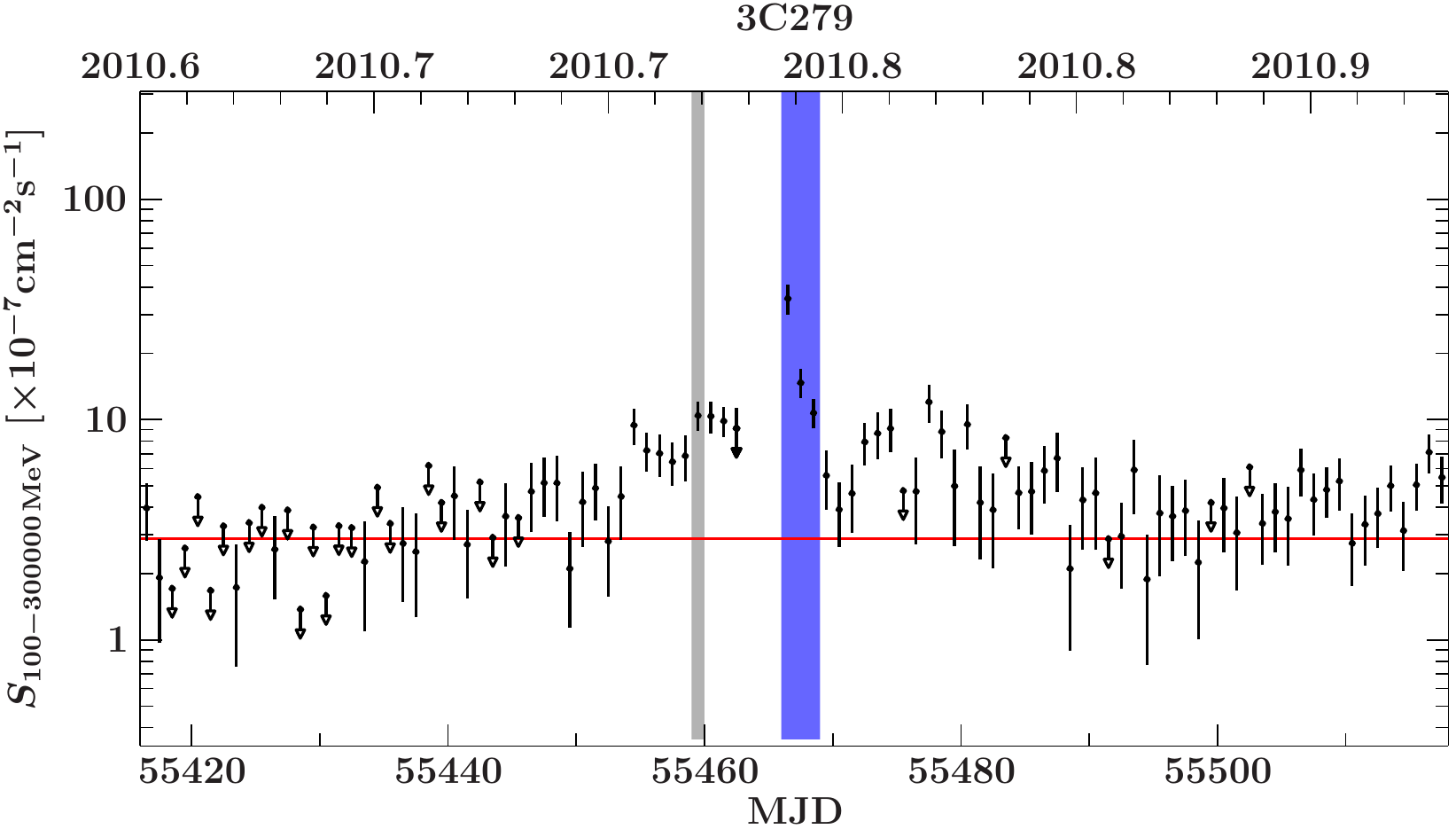}
\figsetgrpnote{Fermi/LAT  first zoomed γ-ray light curve for 3C\,279 in the energy range  $100$\,MeV -- $300$\,GeV using daily binning. Highlighted flares  (blue and gray) correspond to selected time intervals following the method of Sect.\,\ref{Flare_identification}, while blue intervals indicate the most promising flares, as listed in Table\,\ref{Tab_neutrino_expectation}. The red line corresponds to the identified ground level G.}
\figsetgrpend

\figsetgrpstart
\figsetgrpnum{1.9}
\figsetgrptitle{3C\,279}
\figsetplot{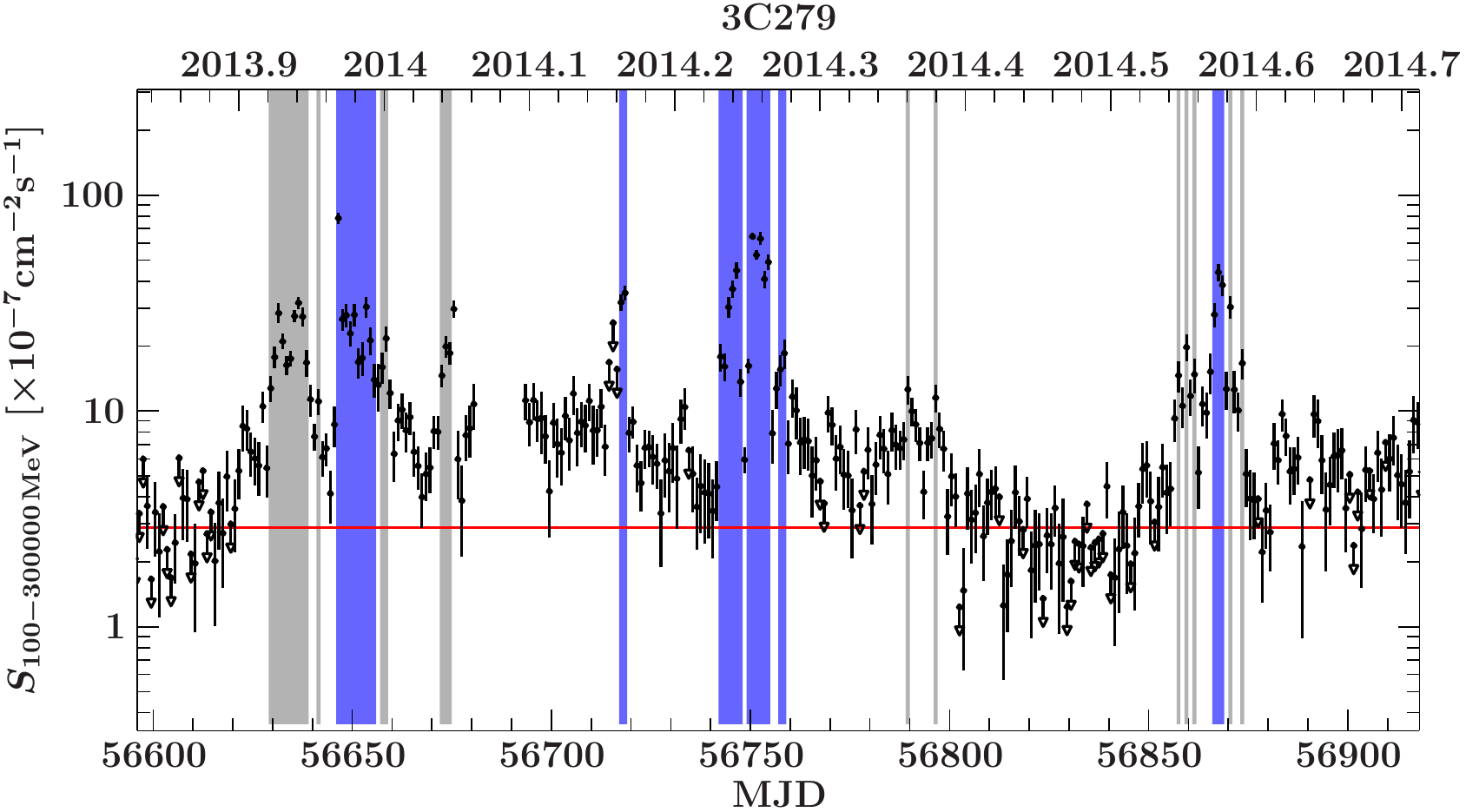}
\figsetgrpnote{Fermi/LAT  second zoomed γ-ray light curve for 3C\,279 in the energy range  $100$\,MeV -- $300$\,GeV using daily binning. Highlighted flares  (blue and gray) correspond to selected time intervals following the method of Sect.\,\ref{Flare_identification}, while blue intervals indicate the most promising flares, as listed in Table\,\ref{Tab_neutrino_expectation}. The red line corresponds to the identified ground level G.}
\figsetgrpend

\figsetgrpstart
\figsetgrpnum{1.10}
\figsetgrptitle{3C\,279}
\figsetplot{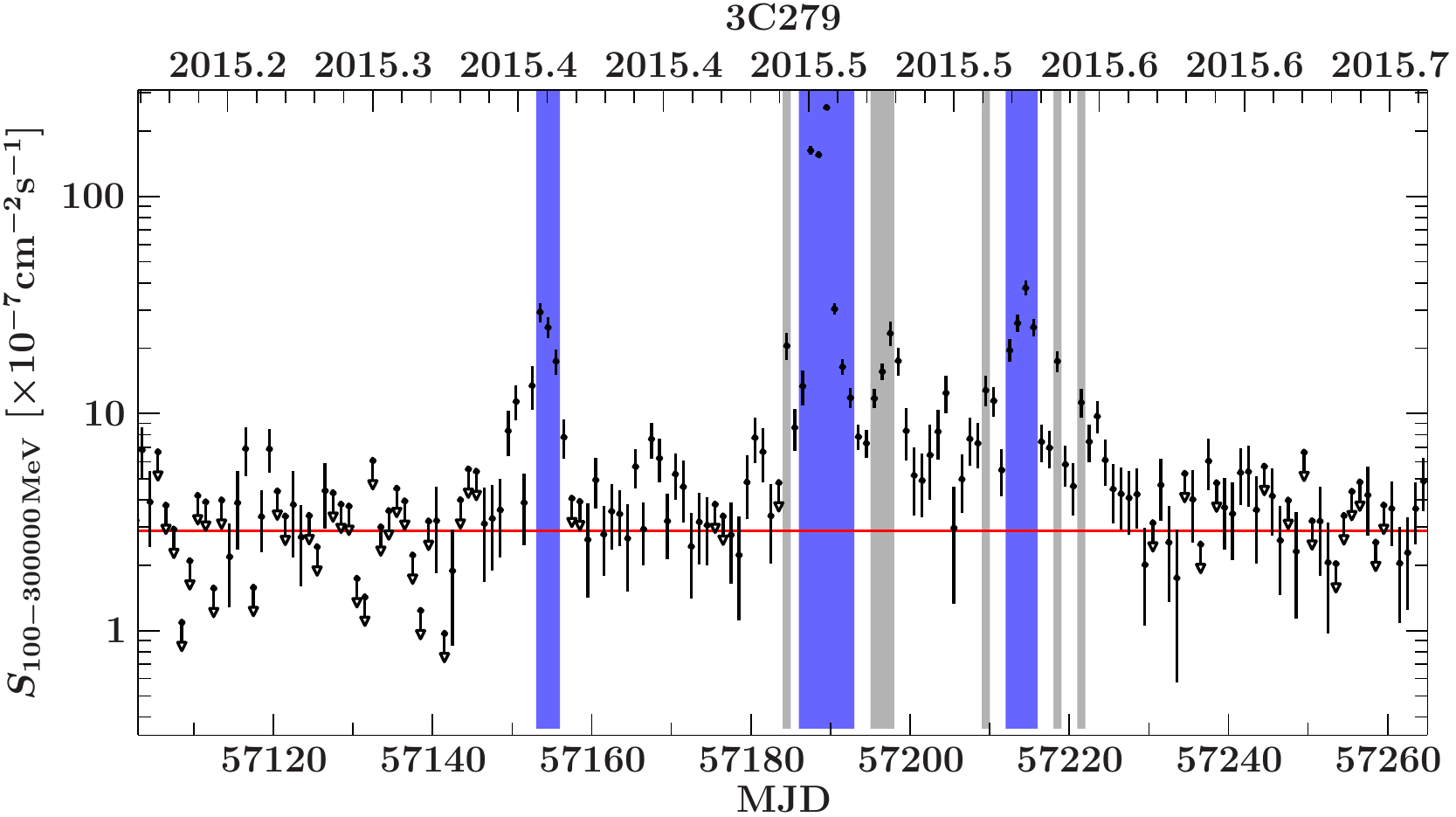}
\figsetgrpnote{Fermi/LAT  third zoomed γ-ray light curve for 3C\,279 in the energy range  $100$\,MeV -- $300$\,GeV using daily binning. Highlighted flares  (blue and gray) correspond to selected time intervals following the method of Sect.\,\ref{Flare_identification}, while blue intervals indicate the most promising flares, as listed in Table\,\ref{Tab_neutrino_expectation}. The red line corresponds to the identified ground level G.}
\figsetgrpend

\figsetgrpstart
\figsetgrpnum{1.11}
\figsetgrptitle{3C\,454.3}
\figsetplot{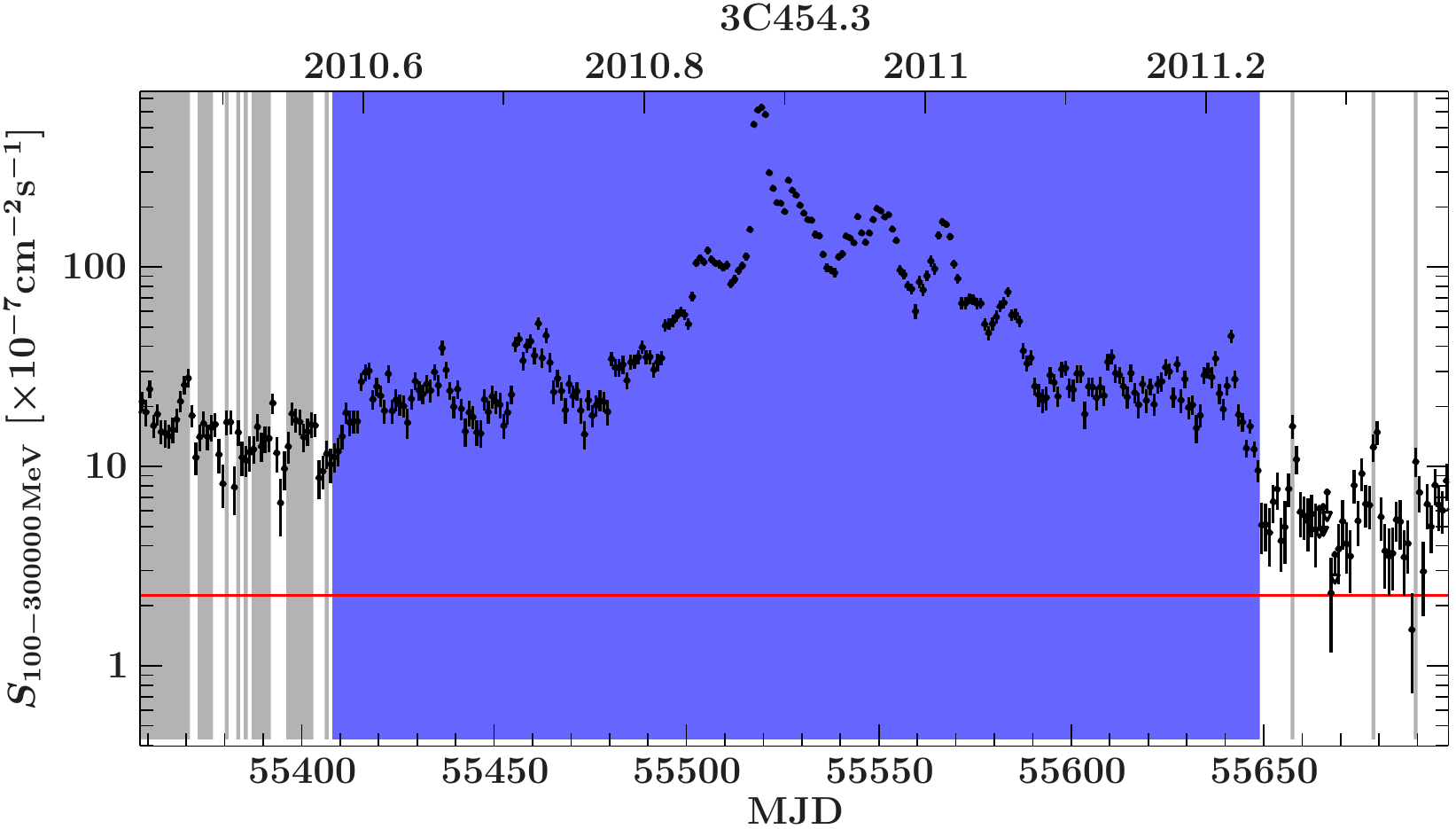}
\figsetgrpnote{Fermi/LAT first zoomed γ-ray light curve for 3C\,454.3 in the energy range  $100$\,MeV -- $300$\,GeV using daily binning. Highlighted flares  (blue and gray) correspond to selected time intervals following the method of Sect.\,\ref{Flare_identification}, while blue intervals indicate the most promising flares, as listed in Table\,\ref{Tab_neutrino_expectation}. The red line corresponds to the identified ground level G.}
\figsetgrpend

\figsetgrpstart
\figsetgrpnum{1.12}
\figsetgrptitle{CTA\,102}
\figsetplot{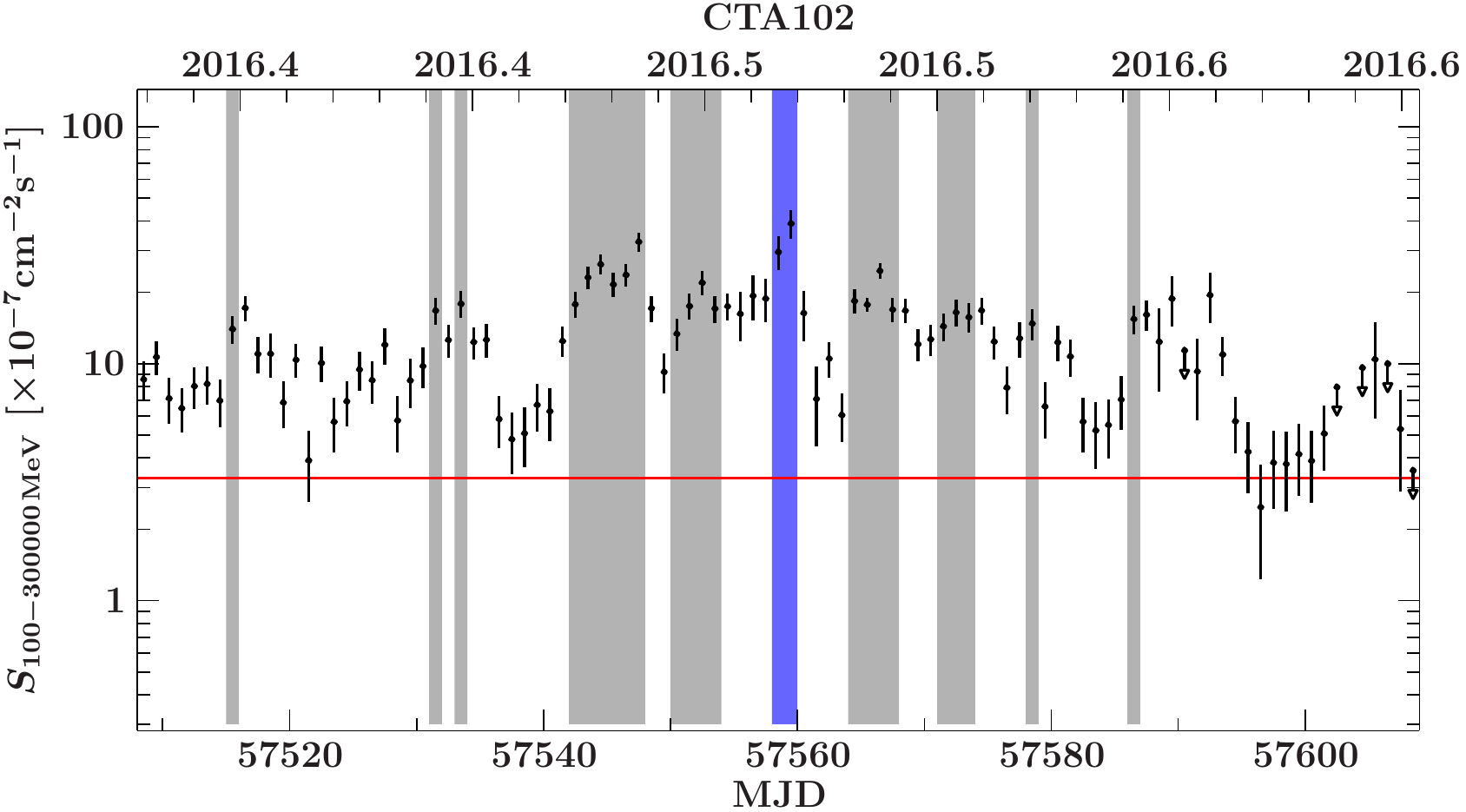}
\figsetgrpnote{Fermi/LAT  first zoomed γ-ray light curve for CTA\,102 in the energy range  $100$\,MeV -- $300$\,GeV using daily binning. Highlighted flares  (blue and gray) correspond to selected time intervals following the method of Sect.\,\ref{Flare_identification}, while blue intervals indicate the most promising flares, as listed in Table\,\ref{Tab_neutrino_expectation}. The red line corresponds to the identified ground level G.}
\figsetgrpend

\figsetgrpstart
\figsetgrpnum{1.13}
\figsetgrptitle{PKS\,0402$-$362}
\figsetplot{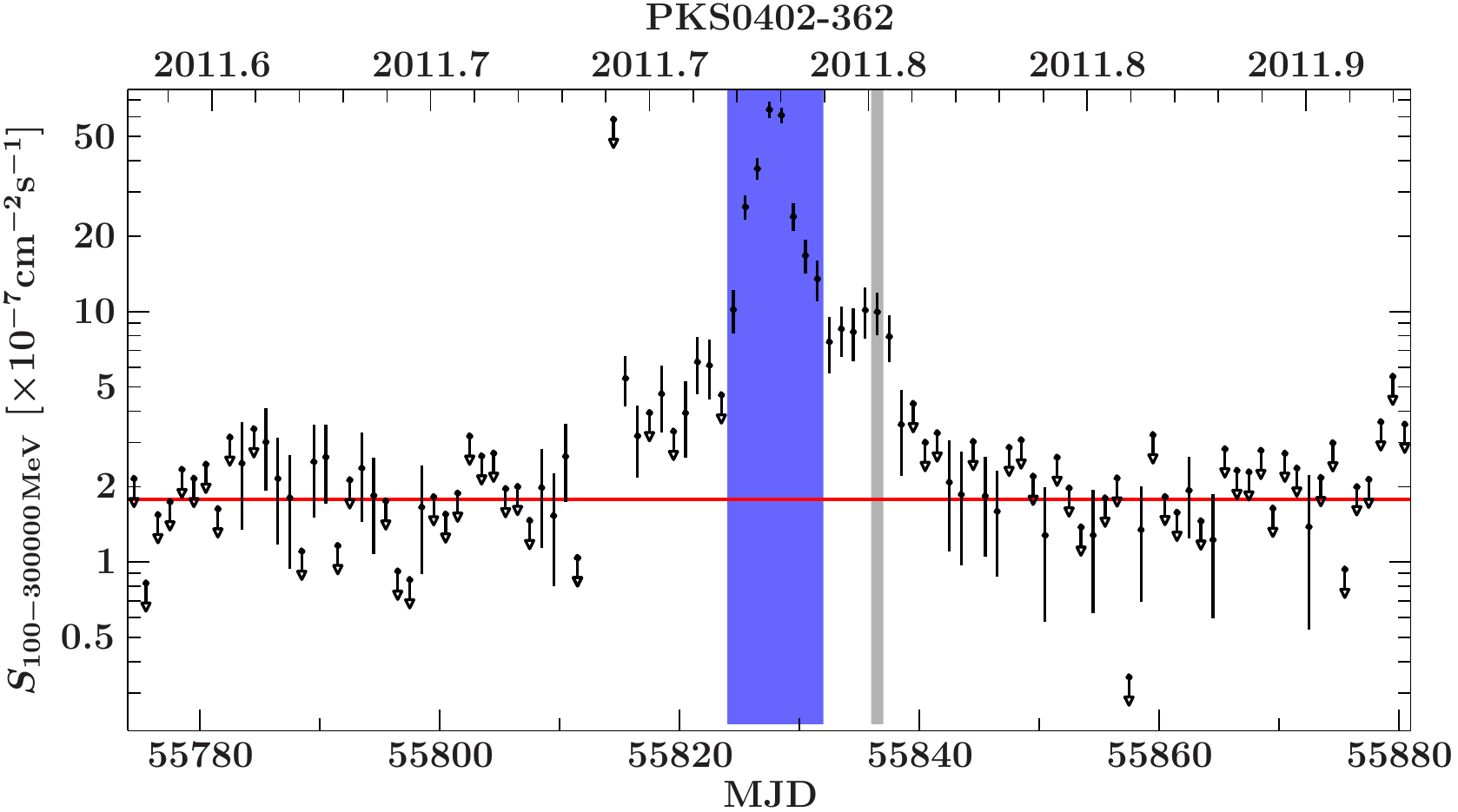}
\figsetgrpnote{Fermi/LAT first zoomed γ-ray light curve for PKS\,0402$-$362 in the energy range  $100$\,MeV -- $300$\,GeV using daily binning. Highlighted flares  (blue and gray) correspond to selected time intervals following the method of Sect.\,\ref{Flare_identification}, while blue intervals indicate the most promising flares, as listed in Table\,\ref{Tab_neutrino_expectation}. The red line corresponds to the identified ground level G.}
\figsetgrpend

\figsetgrpstart
\figsetgrpnum{1.14}
\figsetgrptitle{PKS\,1329$-$049}
\figsetplot{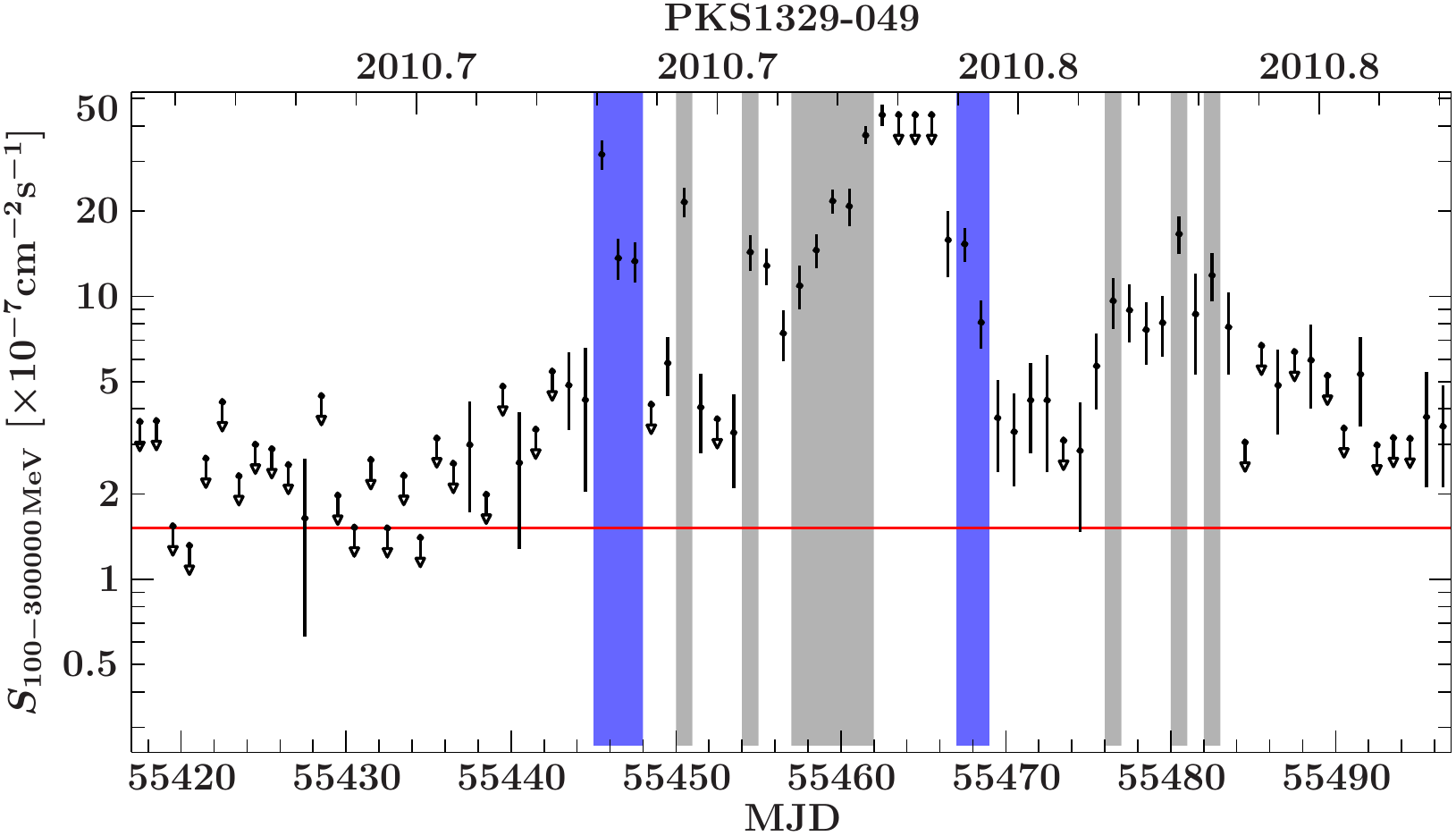}
\figsetgrpnote{Fermi/LAT first zoomed γ-ray light curve for PKS\,1329$-$049 in the energy range  $100$\,MeV -- $300$\,GeV using daily binning. Highlighted flares  (blue and gray) correspond to selected time intervals following the method of Sect.\,\ref{Flare_identification}, while blue intervals indicate the most promising flares, as listed in Table\,\ref{Tab_neutrino_expectation}. The red line corresponds to the identified ground level G.}
\figsetgrpend

\figsetgrpstart
\figsetgrpnum{1.15}
\figsetgrptitle{PKS\,1424$-$41}
\figsetplot{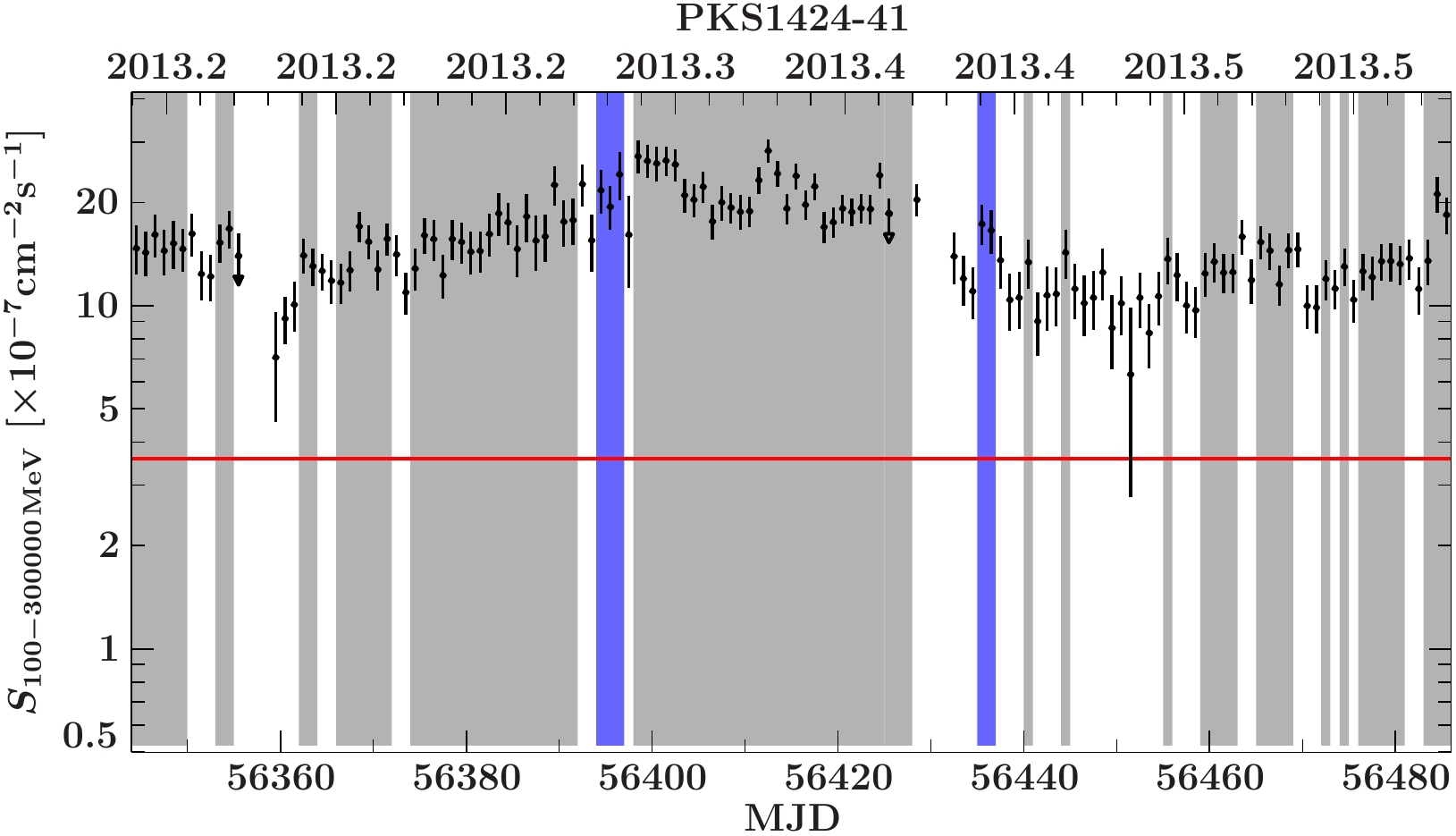}
\figsetgrpnote{Fermi/LAT first zoomed γ-ray light curve for PKS\,1424$-$41 in the energy range  $100$\,MeV -- $300$\,GeV using daily binning. Highlighted flares  (blue and gray) correspond to selected time intervals following the method of Sect.\,\ref{Flare_identification}, while blue intervals indicate the most promising flares, as listed in Table\,\ref{Tab_neutrino_expectation}. The red line corresponds to the identified ground level G.}
\figsetgrpend

\figsetgrpstart
\figsetgrpnum{1.16}
\figsetgrptitle{PKS\,1424$-$41}
\figsetplot{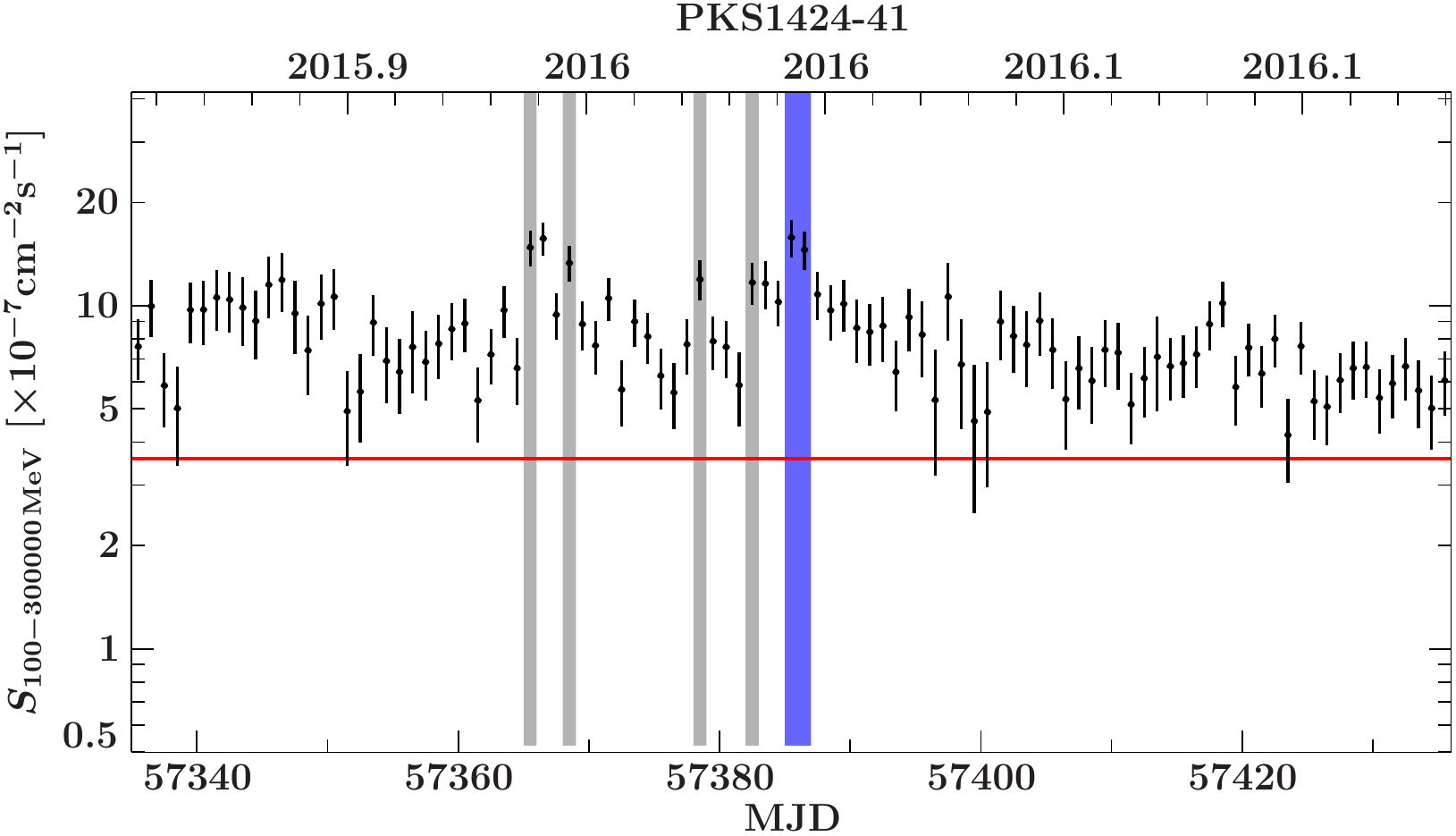}
\figsetgrpnote{Fermi/LAT second zoomed γ-ray light curve for PKS\,1424$-$41 in the energy range  $100$\,MeV -- $300$\,GeV using daily binning. Highlighted flares  (blue and gray) correspond to selected time intervals following the method of Sect.\,\ref{Flare_identification}, while blue intervals indicate the most promising flares, as listed in Table\,\ref{Tab_neutrino_expectation}. The red line corresponds to the identified ground level G.}
\figsetgrpend

\figsetgrpstart
\figsetgrpnum{1.17}
\figsetgrptitle{PKS\,1510$-$089}
\figsetplot{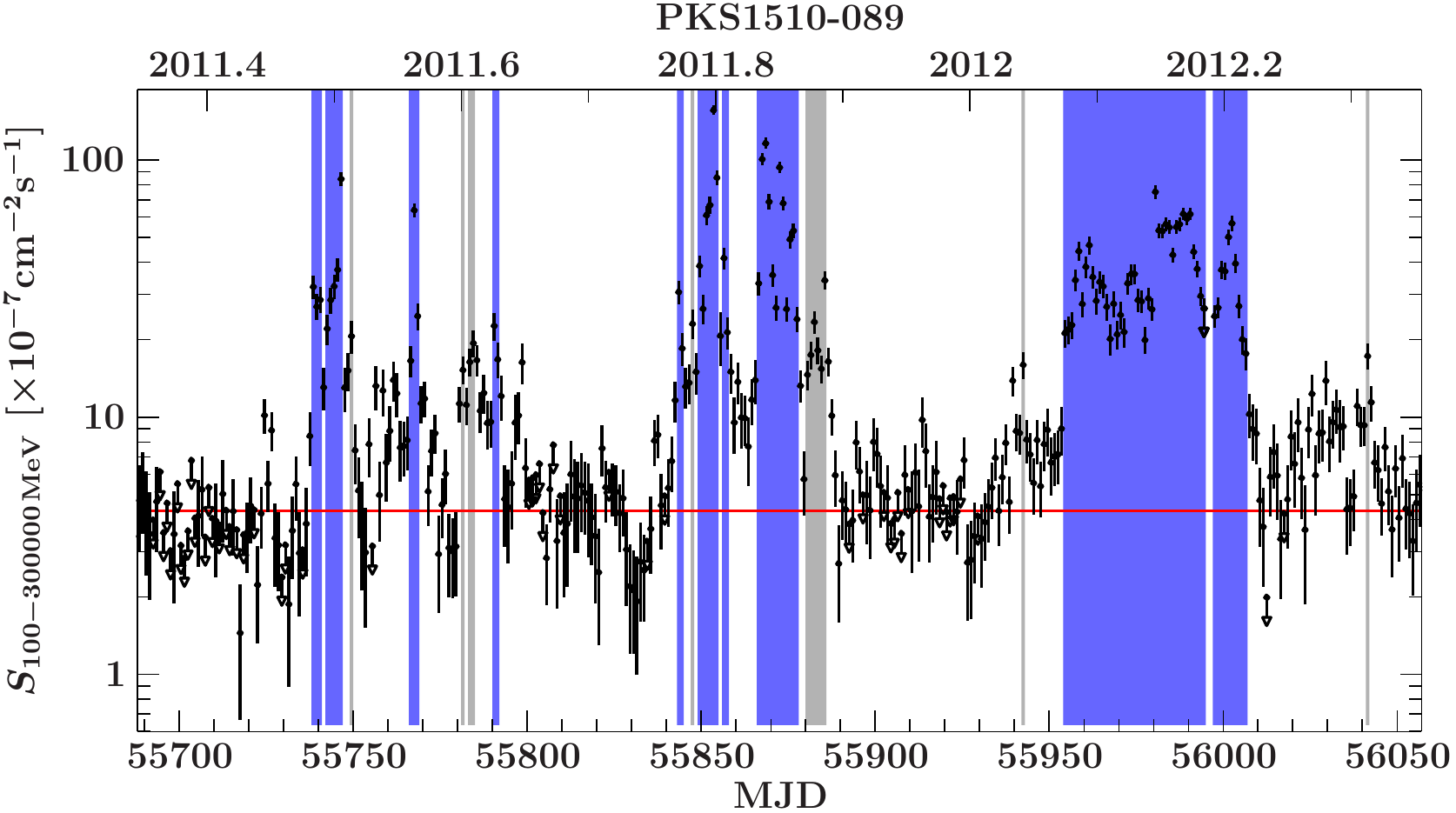}
\figsetgrpnote{Fermi/LAT first zoomed γ-ray light curve for PKS\,1510$-$089 in the energy range  $100$\,MeV -- $300$\,GeV using daily binning. Highlighted flares  (blue and gray) correspond to selected time intervals following the method of Sect.\,\ref{Flare_identification}, while blue intervals indicate the most promising flares, as listed in Table\,\ref{Tab_neutrino_expectation}. The red line corresponds to the identified ground level G.}
\figsetgrpend

\figsetgrpstart
\figsetgrpnum{1.18}
\figsetgrptitle{PKS\,1510$-$089}
\figsetplot{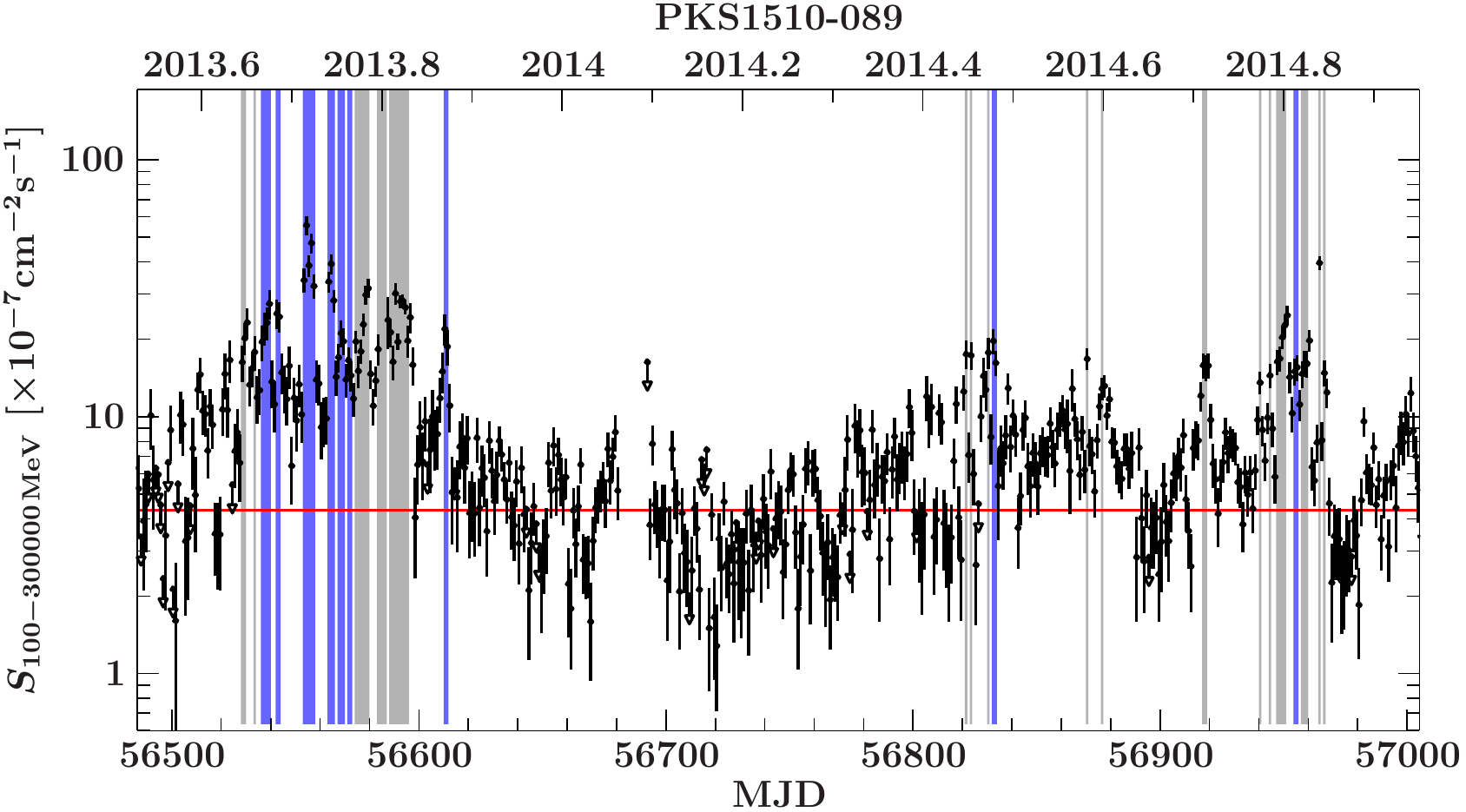}
\figsetgrpnote{Fermi/LAT second zoomed γ-ray light curve for PKS\,1510$-$089 in the energy range  $100$\,MeV -- $300$\,GeV using daily binning. Highlighted flares  (blue and gray) correspond to selected time intervals following the method of Sect.\,\ref{Flare_identification}, while blue intervals indicate the most promising flares, as listed in Table\,\ref{Tab_neutrino_expectation}. The red line corresponds to the identified ground level G.}
\figsetgrpend

\figsetgrpstart
\figsetgrpnum{1.19}
\figsetgrptitle{PKS\,1510$-$089}
\figsetplot{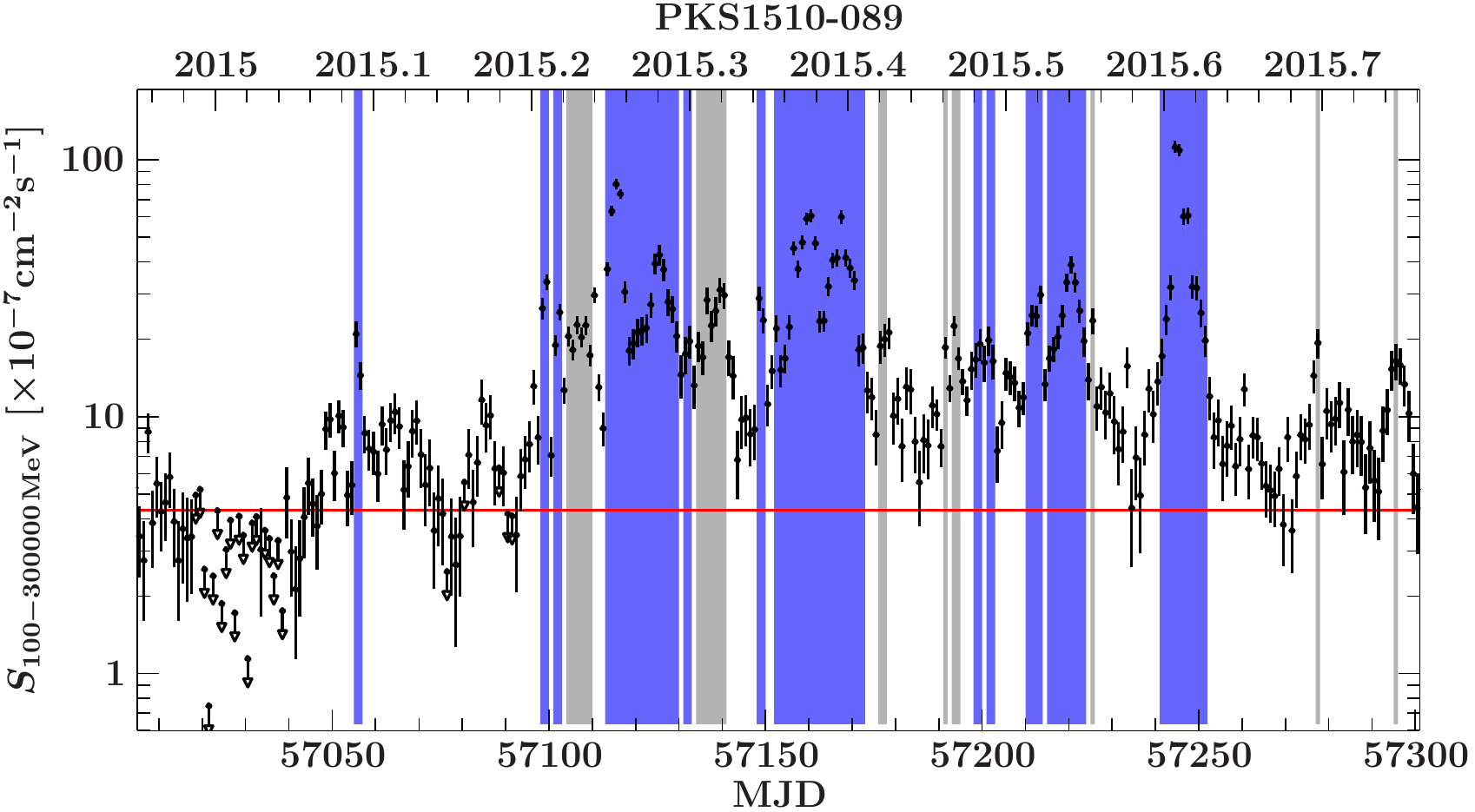}
\figsetgrpnote{Fermi/LAT third zoomed γ-ray light curve for PKS\,1510$-$089 in the energy range  $100$\,MeV -- $300$\,GeV using daily binning. Highlighted flares  (blue and gray) correspond to selected time intervals following the method of Sect.\,\ref{Flare_identification}, while blue intervals indicate the most promising flares, as listed in Table\,\ref{Tab_neutrino_expectation}. The red line corresponds to the identified ground level G.}
\figsetgrpend

\figsetgrpstart
\figsetgrpnum{1.20}
\figsetgrptitle{PKS\,1510$-$089}
\figsetplot{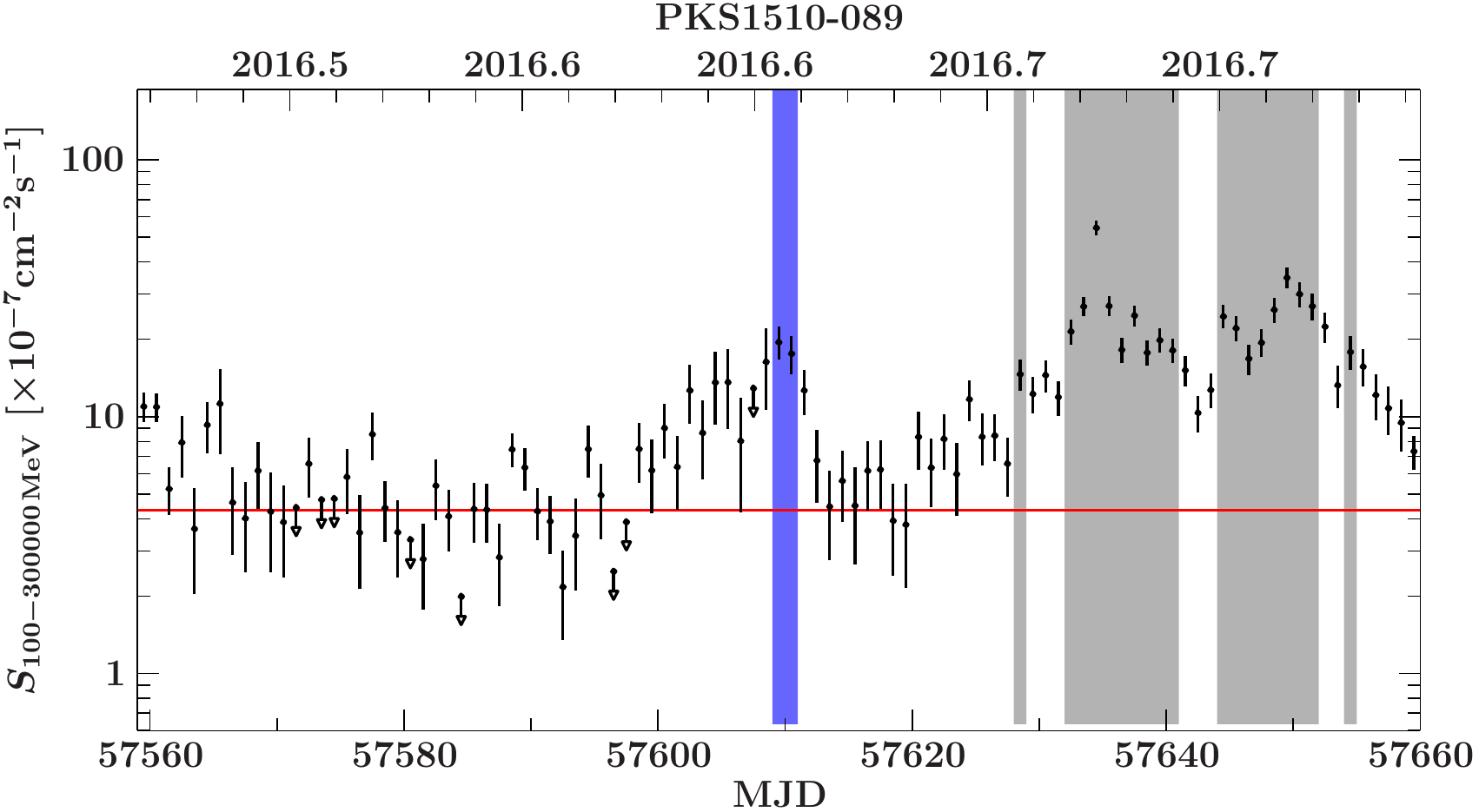}
\figsetgrpnote{Fermi/LAT fourth zoomed γ-ray light curve for PKS\,1510$-$089 in the energy range  $100$\,MeV -- $300$\,GeV using daily binning. Highlighted flares  (blue and gray) correspond to selected time intervals following the method of Sect.\,\ref{Flare_identification}, while blue intervals indicate the most promising flares, as listed in Table\,\ref{Tab_neutrino_expectation}. The red line corresponds to the identified ground level G.}
\figsetgrpend

\begin{figure}[pt]
  \centering
         \includegraphics[width=\hsize]{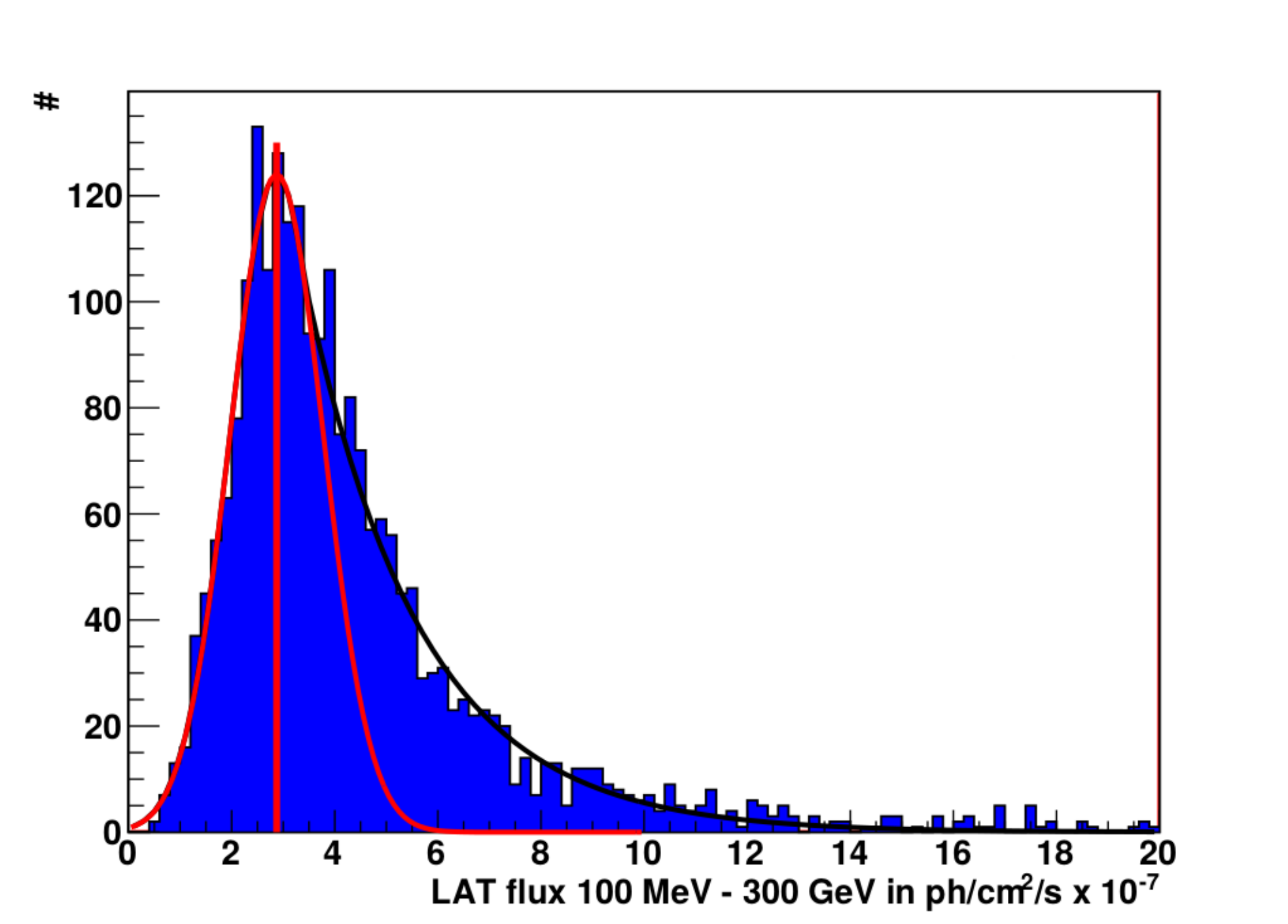} 
     \caption{Histogram of the \textit{Fermi}/LAT $\gamma$-ray flux of 3C\,279 in the energy range  $100$\,MeV - $300$\,GeV and a time range of $55317\,\text{MJD}$ (2010-05-01) -  $57662\,\text{MJD}$ (2016-10-01), using daily binning.
     The histogram is fitted with a Gaussian and exponential function. The red line represents the mean of the Gaussian fit that serves as ground level flux $\text{G}$. The black curve indicates the combined fit. }
\label{3C279 Gauss}
 \end{figure}

After pre-selecting flares based on the automated light curve sample, a dedicated analysis of identified blazars is performed following Sect.\,\ref{Fermi analysis} and the flare identification is repeated\footnote{As we search for the brightest flares, only 12 blazars are re-analyzed}.
Figure\,\ref{flare_selection_histo} displays the \textit{Fermi}/LAT $\gamma$-ray light curve of the blazar 3C\,279, indicating selected periods of enhanced activity. 
The ground level $\text{G}$ is derived by creating a histogram of \textit{Fermi}/LAT fluxes from the entire 6.5 year light curve and fitting the distribution with a Gaussian and exponential function. From the mean of the Gaussian fit we derive the ground level flux G. The tail of this flux distribution corresponds to flux values within flaring states. Figure\,\ref{3C279 Gauss} displays the corresponding histogram for the blazar 3C\,279.
This approach assumes that the studied blazar stays in the ground-level state  most of the time, which seems to be a reasonable assumption for the brightest blazars monitored by \textit{Fermi}/LAT.  
The IceCube effective area $\text{A}_{\text{Eff}}$ ensures a weighting of identified flares according to the IceCube sensitivity.
Short flares on daily to weekly time scales are favored, as no additional assumption about the flaring behavior is 
required. Longer outburst periods that are interrupted by non-detections are therefore split into several individual flares, because in this selection a flare corresponds to an uninterrupted period of activity. 

 
 \subsection{Multiwavelength Observations}\label{Multiwavelength observations}
\hspace{-13pt}
Quasi-Simultaneous multiwavelength observations are necessary to parameterize the spectral behavior of blazars during flaring states. We use \textit{Swift}/XRT and \textit{Fermi}/LAT data to characterize the high-energy emission hump of the blazar spectral energy distribution (SED) during bright flares selected in Sec.\,\ref{Flare_identification}.

\subsubsection{\textit{Swift}/XRT data analysis}
\hspace{-13pt}
\textit{Swift}/XRT data are reduced with standard methods, using the software package HEASOFT 6.15.1. The data are reduced, calibrated and cleaned by means of the XRT-PIPELINE script using standard filtering criteria. We selected 0-12 grades for observations in photon counting (pc) and grade 0 for those in window-timing mode (wt). 
Spectral fitting was performed with ISIS 1.6.2 \citep{Houck2000}.
The X-ray data were fitted with the \textit{Fermi}/LAT data using an empirical log parabolic model \citep{Massaro2004}.
The model also includes the X-ray photoelectric absorption tbnew\footnote{\url{https://pulsar.sternwarte.uni-erlangen.de/wilms/research/tbabs/}}
with the wilm abundances and vern cross-sections \citep{Wilms2000,Verner1996}.
The SEDs were fit in detector space with a diagonal response for the LAT data \citep{Krauss2016}.
We used a reduced Chi$^2$ method to find the best fit. Given 
the limitations of the multiwavelength data, this approach is not 
indicative of an absolute goodness-of-fit, but gives a good estimate of 
the SED shape and allows us to measure the bolometric flux well.
For flares where no simultaneous \mbox{X-ray} data were available, an averaged  \textit{Swift}/XRT spectrum over the entire considered time period of 6.5 years is used.
 
\subsubsection{\textit{Fermi}/LAT data analysis}\label{Fermi analysis}
\hspace{-13pt}
For the analysis of \textit{Fermi}/LAT data we performed a dedicated likelihood analysis for all blazar flares selected in Sect.\,\ref{Flare_identification}.
We used the \textit{Fermi} Science Tools (v10r0p5) together with the Pass 8 data and the $\text{P8R2}\_\text{SOURCE}\_\text{V6}$ instrument response functions.
We performed an unbinned likelihood analysis in a region of interest (ROI) of $10^{\circ}$ around the source of interest in an energy range of $100\,\text{MeV}$ to $300\,\text{GeV}$.
A zenith angle cut of $90^{\circ}$ was used together with the LAT EVENT\_CLASS = 128 and the LAT EVENT\_TYPE = 3, while the \emph{gtmktime} cuts DATA\_QUAL==1 $\&\&$ LAT\_CONFIG==1 were chosen.
For the modeling of the diffuse components\footnote{\url{https://fermi.gsfc.nasa.gov/ssc/data/access/lat/BackgroundModels.html}} we used the Galactic diffuse emission model gll\_iem\_v06.fits and the isotropic diffuse model iso\_P8R2\_SOURCE\_V6\_v06.txt \citep{Acero_2016}.
The model used in the likelihood analysis contains all sources of the third LAT source catalog \citep{Acero_2015} within a radius of 20 degrees from the source of interest.
Sources were fitted with either a power-law or logparabula fit following the 3FGL catalog.
In the inner 10 degree from the source of interest, the parameters PREFACTOR and INDEX were kept free to vary, while for sources further outside all parameters have been fixed to the 3FGL values.
\textit{Fermi}/LAT $\gamma$-ray light curves are calculated for a time interval of 6.5 years, using daily binning.
In case no significant detection can be achieved, our upper limit calculation is based on the method of \cite{Feldman_1998}. 
We used a threshold in test statistic of  $TS = 9$ for the spectra and  light curves calculation. In $\gamma$-ray astronomy, the test statistic is widely used as a measurement of the source detection significance following Wilks' theorem \citep{Wilks_1938,Mattox_1996}. 
 
\subsection{Calorimetric Neutrino Expectation} \label{Neutrino expectation}
\begin{figure}[htbp]
  \centering
      \includegraphics[width=\hsize]{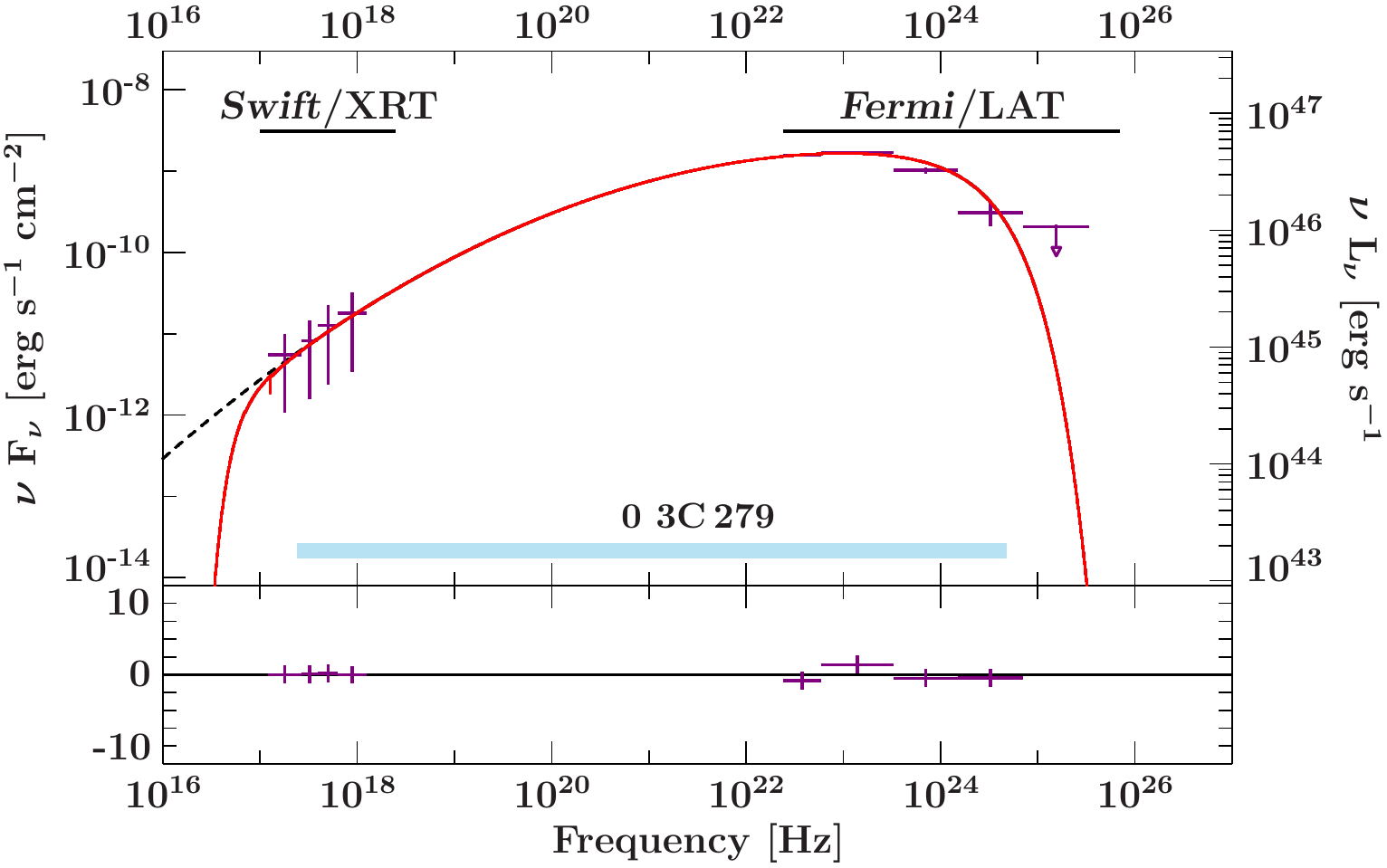}
       \put(-137,59){\tiny{$\mycbox{white}$}}
       \put(-8,73){\text{\begin{turn}{90}[ \end{turn}}}\\
    \caption{Quasi-simultaneous SED of the bright flare of 3C\,279 in 2015 (flare 1 in Table\,\ref{Tab_neutrino_expectation}). The red line indicates a log-parabola parametrization, including intrinsic absorption in X-rays and an exponential cut-off. The dashed black line shows the same parametrization without the absorption. The shadowed 
    area represents the integration range from $10\,\text{keV}$ to $20\,\text{GeV}$.\\~\\
    (The complete figure set (50 images) is available in the online journal.)}
  \label{SED_flare_95}
\end{figure}

\hspace{-13pt}
While neutrinos are created exclusively in hadronic interactions, high-energy $\gamma$ rays can originate from both leptonic and hadronic interactions.
We use our calorimetric approach developed in \cite{Krauss_2014} and \cite{Kadler_2016} in order to estimate the maximum number\footnote{This approach deliberately over-estimates values that could be derived from different possible physical modeling approaches \citep[e.g.][]{Bottcher_2013,Stone_2008}. The fundamental benefit of our approach is that it is model independent and suited to derive strict upper limits.}
of monoenergetic high-energy neutrino events a particular flare would be able to generate, assuming photopion production.
In a purely hadronic scenario, this leads to a balance between the integrated $\gamma$-ray energy flux $\Phi_{\gamma}$ and the integrated neutrino energy flux $\Phi_{\nu}$, following \cite{Krauss_2018}:
  \begin{align}
 \Phi_{\nu} \approx \Phi_{\gamma}
   \end{align}
Numerical simulations of pion photoproduction indicate that this balance implies a hard limit on the maximum number of neutrino events. Furthermore, this limit seems to be valid in most astrophysical scenarios, especially blazar jets \citep{Mucke_2000}.
In this maximum case, the total number of neutrinos an individual blazar flare can create is given by:
\begin{align}
 \label{N_max_formula}
 N_{\nu}^{\text{max}} = \text{A}_{\text{Eff}} ~\cdotp \left( \frac{\Phi_{\nu}}{E_{\nu}} \right) \cdotp \Delta t.
\end{align}
where $\text{A}_{\text{Eff}}$ corresponds to the energy-dependent IceCube effective area, $\Phi_{\nu}$ to the maximum integrated neutrino energy flux and $\Delta t$ to the duration of the selected flare, assuming neutrino and photon production to happen only within the identified time interval. 
We assume a flux of monoenergetic neutrinos following a $\delta$-peaked spectrum and thus select a neutrino energy of $E_{\nu} = 2$\,PeV to be comparable to \citet{Kadler_2016}. 
Following our previous work, the maximum neutrino expectation can be scaled to a more realistic estimate: 
\begin{align}
 N_{\nu}^{\text{pred}} = \mathnormal{f} \times N_{\nu}^{\text{max}}.
  \label{N_pred_formula}
\end{align}
Following \citet{Kadler_2016}, we use an empirical scaling factor of $\mathnormal{f} = 0.025$, which accounts for a combination of several factors such as neutrino flavor and blazar class. Please refer to Sect.\,\ref{Limitations} for further constraints on the  method.
To account for the maximum integrated neutrino energy flux $\Phi_{\nu}$ a given flare is generating, we integrate over the high-energy emission of the blazar SED in an energy range  $10\,\text{keV}$ to $20\,\text{GeV}$.
This energy range covers the entire range of measured \textit{Swift}/XRT and \textit{Fermi}/LAT data and ensures that the total X-ray to $\gamma$-ray emission is considered in our integrated flux estimate.
Quasi-simultaneous \textit{Swift}/XRT\footnote{including intrinsic absorption in  X-rays} and \textit{Fermi}/LAT observations of each flare are used to build SEDs, whose high-energy hump can be parameterized by a log-parabola model.
Figure\,\ref{SED_flare_95} displays a typical SED of a bright flare from 3C\,279 (flare number 1 in Table\,\ref{Tab_neutrino_expectation}), together with the fitted log-parabola and integration range. Light curves and SEDs of all studied flares can be found as supplementary material online.
\figsetgrpstart
\figsetgrpnum{3.1}
\figsetgrptitle{Flare 1 from 3C\,279}
\figsetplot{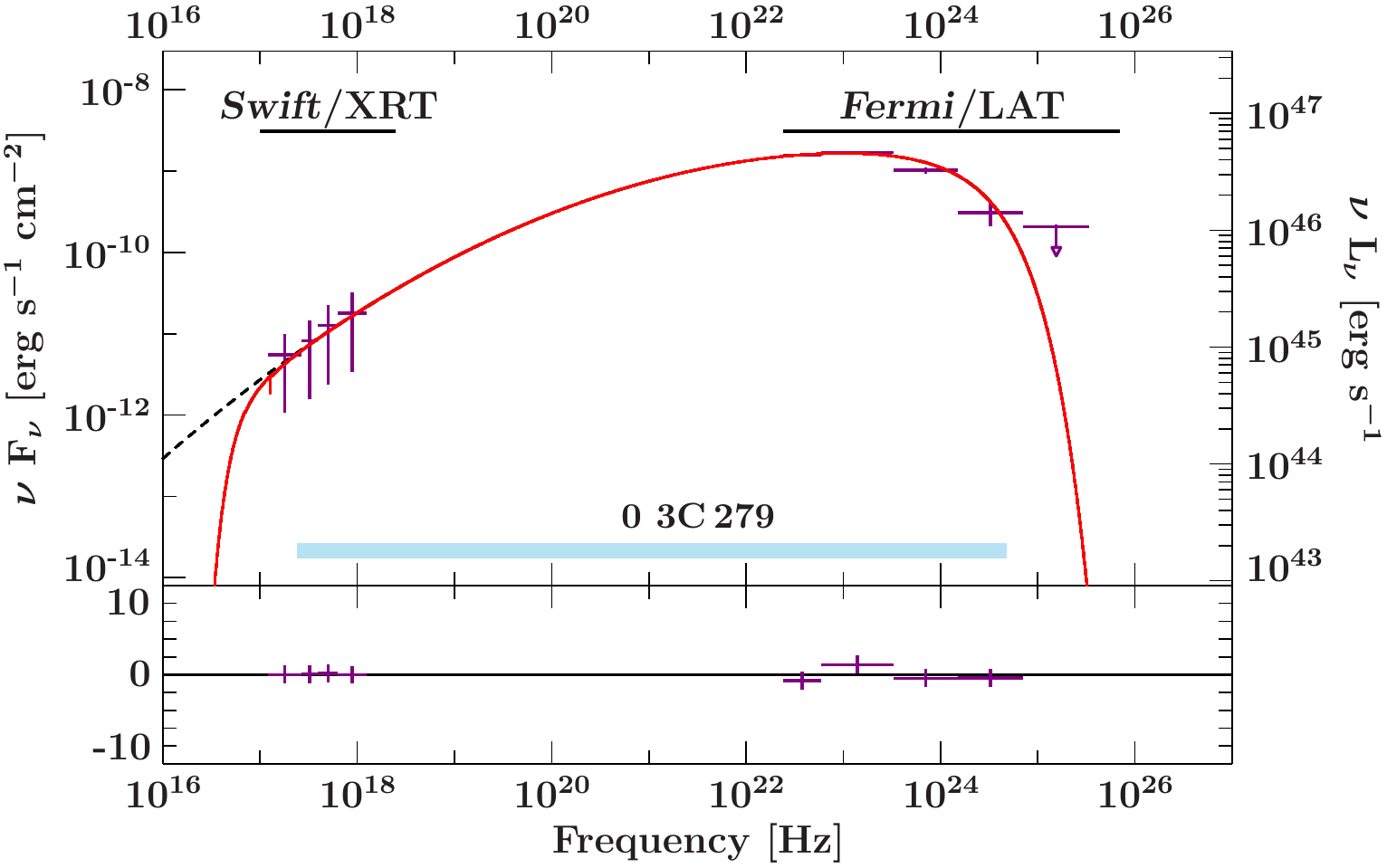}
\figsetgrpnote{Quasi-simultaneous SEDs of flares listed in Table\/\ref{Tab_neutrino_expectation}.
The red line indicates a log-parabola parametrization, including intrinsic absorption in X-rays and an exponential cut-off. The dashed black line shows the same parametrization without the absorption. The shadowed      area represents the integration range from $10\,\text{keV}$ to $20\,\text{GeV}$.}
\figsetgrpend

\figsetgrpstart
\figsetgrpnum{3.2}
\figsetgrptitle{Flare 2 from PKS\,1510$-$089}
\figsetplot{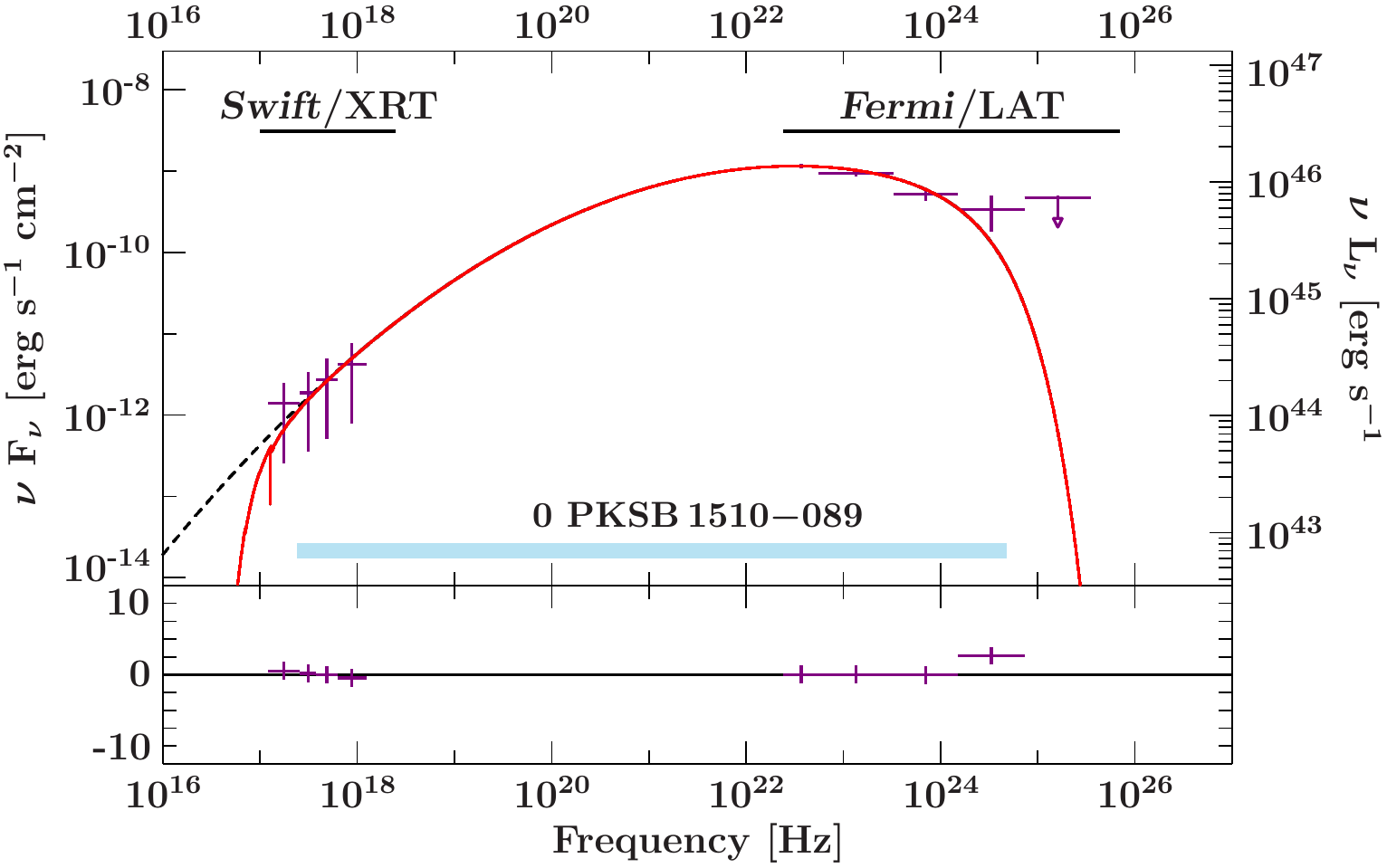}
\figsetgrpnote{Quasi-simultaneous SEDs of flares listed in Table\/\ref{Tab_neutrino_expectation}.
The red line indicates a log-parabola parametrization, including intrinsic absorption in X-rays and an exponential cut-off. The dashed black line shows the same parametrization without the absorption. The shadowed      area represents the integration range from $10\,\text{keV}$ to $20\,\text{GeV}$.}
\figsetgrpend

\figsetgrpstart
\figsetgrpnum{3.3}
\figsetgrptitle{Flare 3 from PKS\,1510$-$089}
\figsetplot{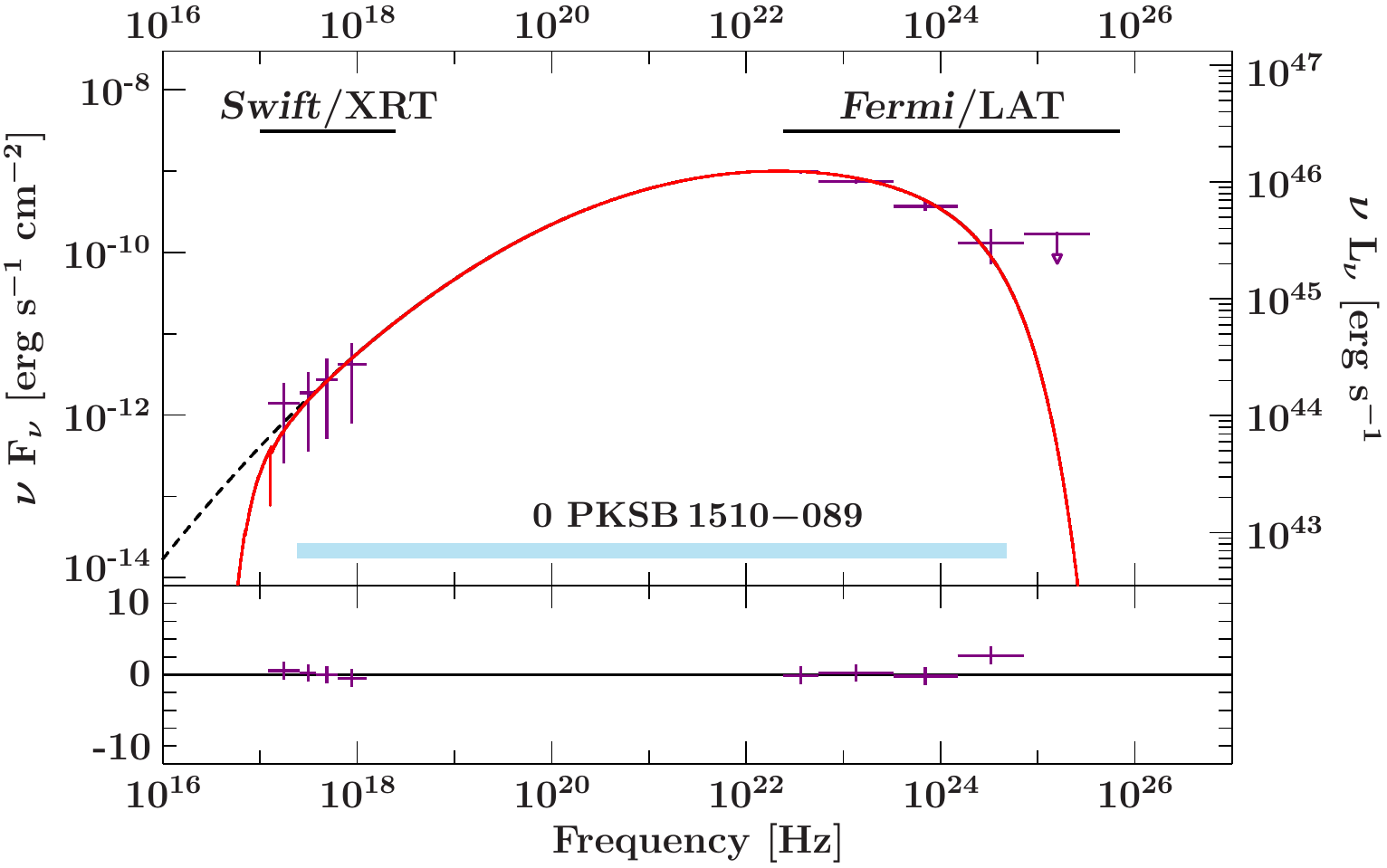}
\figsetgrpnote{Quasi-simultaneous SEDs of flares listed in Table\/\ref{Tab_neutrino_expectation}
. The red line indicates a log-parabola parametrization, including intrinsic absorption in X-rays and an exponential cut-off. The dashed black line shows the same parametrization without the absorption. The shadowed      area represents the integration range from $10\,\text{keV}$ to $20\,\text{GeV}$.}
\figsetgrpend

\figsetgrpstart
\figsetgrpnum{3.4}
\figsetgrptitle{Flare 4 from PKS\,1510$-$089}
\figsetplot{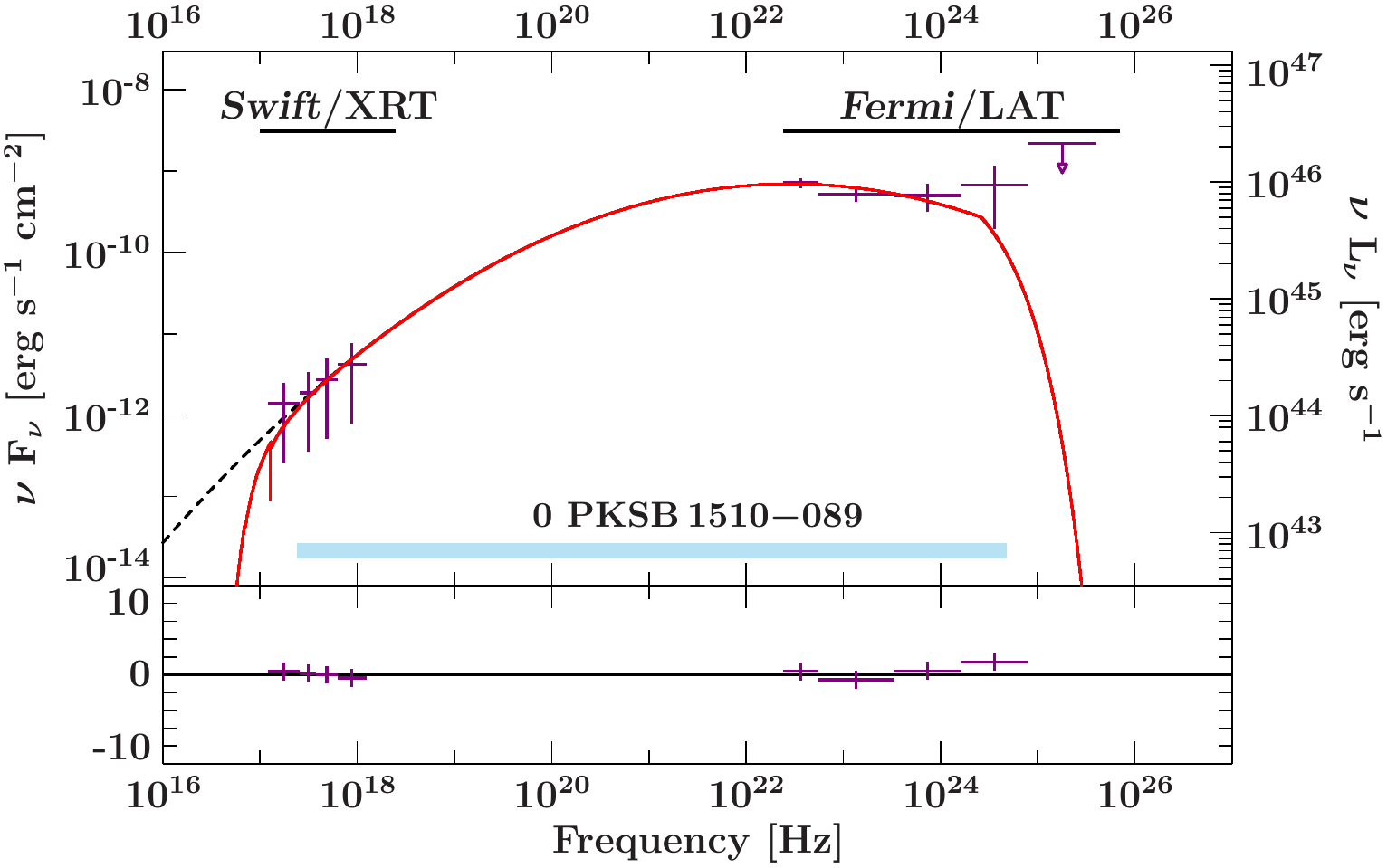}
\figsetgrpnote{Quasi-simultaneous SEDs of flares listed in Table\/\ref{Tab_neutrino_expectation}
. The red line indicates a log-parabola parametrization, including intrinsic absorption in X-rays and an exponential cut-off. The dashed black line shows the same parametrization without the absorption. The shadowed      area represents the integration range from $10\,\text{keV}$ to $20\,\text{GeV}$.}
\figsetgrpend

\figsetgrpstart
\figsetgrpnum{3.5}
\figsetgrptitle{Flare 5 from 3C\,279}
\figsetplot{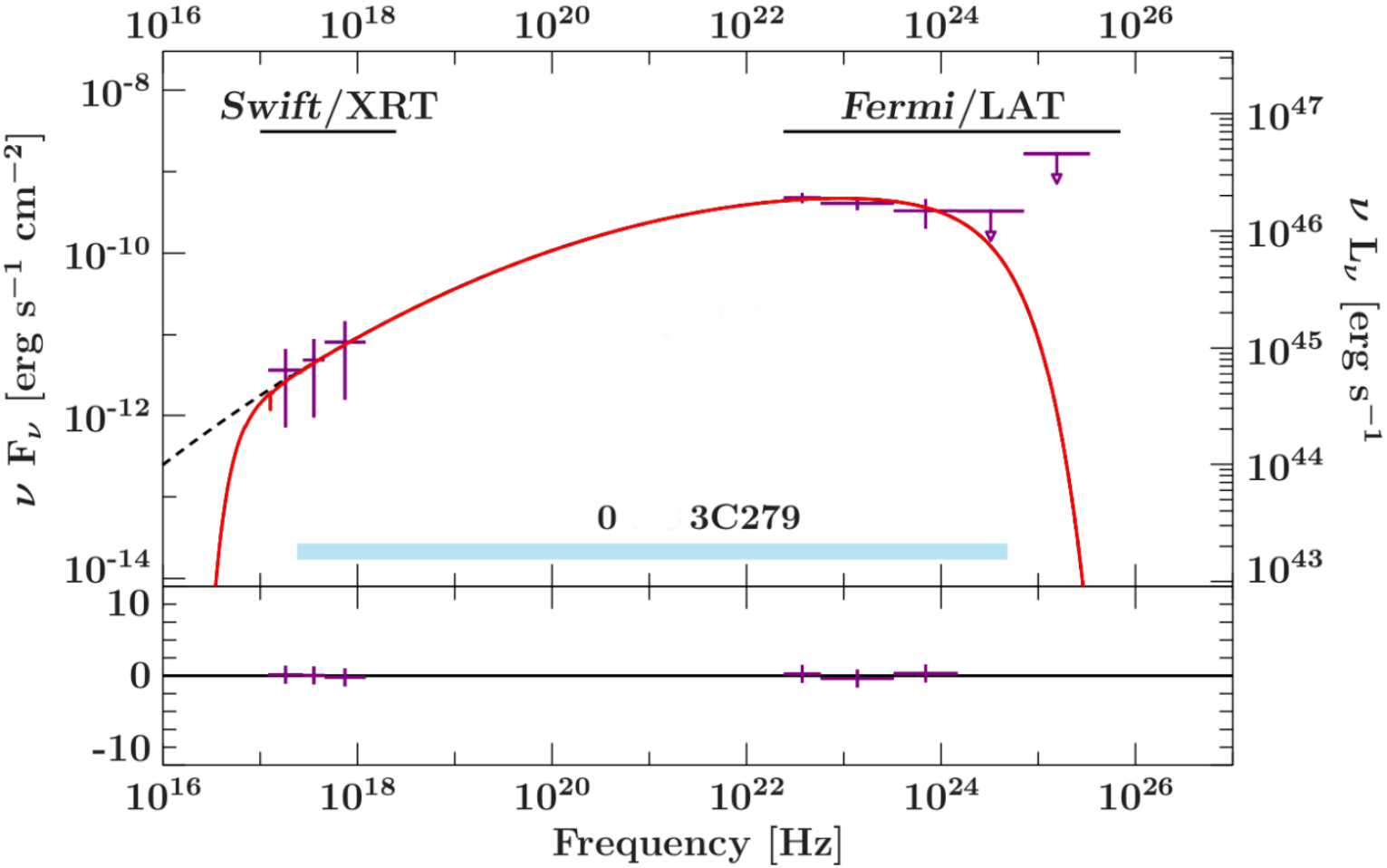}
\figsetgrpnote{Quasi-simultaneous SEDs of flares listed in Table\/\ref{Tab_neutrino_expectation}
. The red line indicates a log-parabola parametrization, including intrinsic absorption in X-rays and an exponential cut-off. The dashed black line shows the same parametrization without the absorption. The shadowed      area represents the integration range from $10\,\text{keV}$ to $20\,\text{GeV}$.}
\figsetgrpend

\figsetgrpstart
\figsetgrpnum{3.6}
\figsetgrptitle{Flare 6 from 3C\,279}
\figsetplot{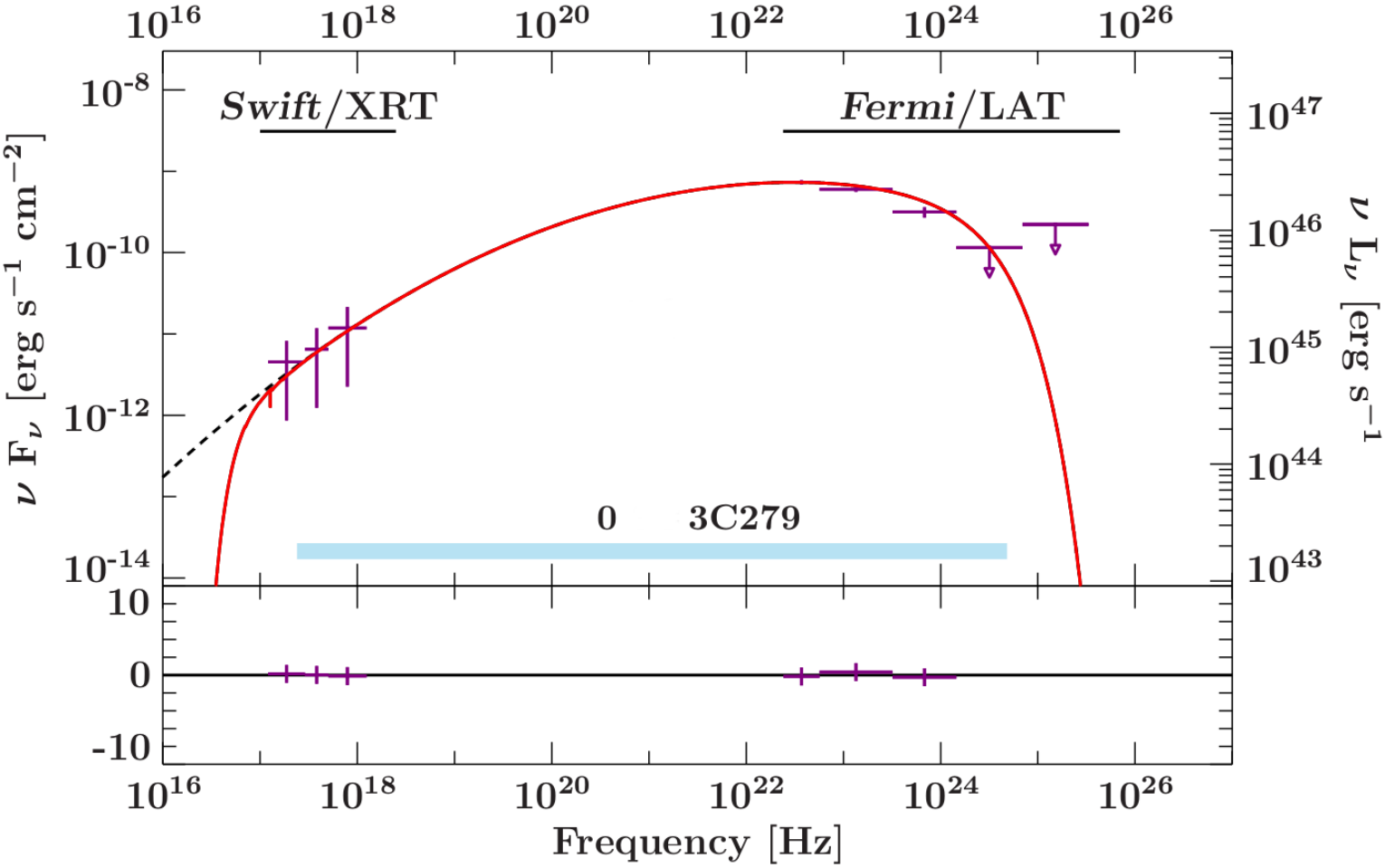}
\figsetgrpnote{Quasi-simultaneous SEDs of flares listed in Table\/\ref{Tab_neutrino_expectation}
. The red line indicates a log-parabola parametrization, including intrinsic absorption in X-rays and an exponential cut-off. The dashed black line shows the same parametrization without the absorption. The shadowed      area represents the integration range from $10\,\text{keV}$ to $20\,\text{GeV}$.}
\figsetgrpend

\figsetgrpstart
\figsetgrpnum{3.7}
\figsetgrptitle{Flare 7 from PKS\,1510$-$089}
\figsetplot{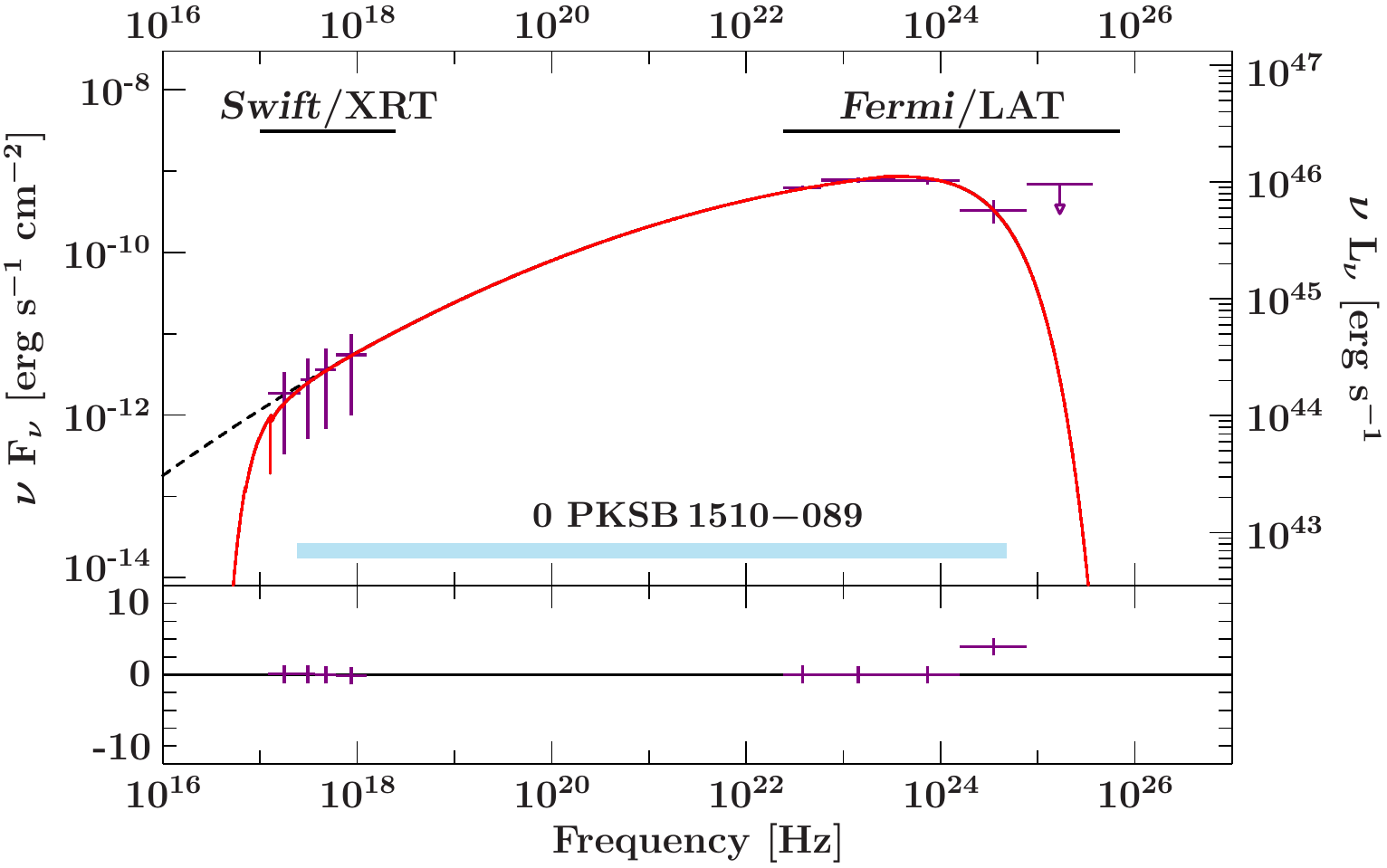}
\figsetgrpnote{Quasi-simultaneous SEDs of flares listed in Table\/\ref{Tab_neutrino_expectation}
. The red line indicates a log-parabola parametrization, including intrinsic absorption in X-rays and an exponential cut-off. The dashed black line shows the same parametrization without the absorption. The shadowed      area represents the integration range from $10\,\text{keV}$ to $20\,\text{GeV}$.}
\figsetgrpend

\figsetgrpstart
\figsetgrpnum{3.8}
\figsetgrptitle{Flare 8 from 3C\,279}
\figsetplot{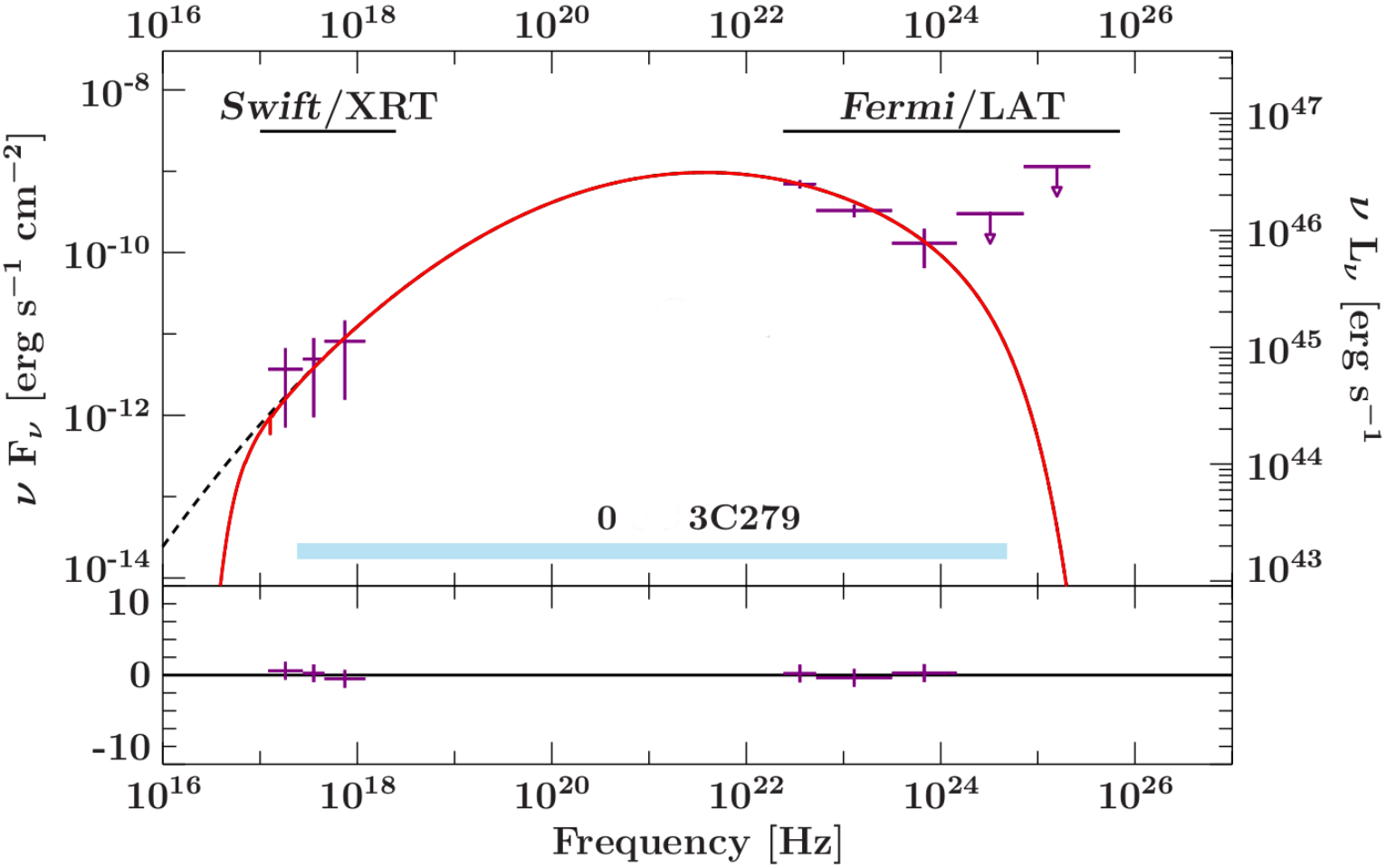}
\figsetgrpnote{Quasi-simultaneous SEDs of flares listed in Table\/\ref{Tab_neutrino_expectation}
. The red line indicates a log-parabola parametrization, including intrinsic absorption in X-rays and an exponential cut-off. The dashed black line shows the same parametrization without the absorption. The shadowed      area represents the integration range from $10\,\text{keV}$ to $20\,\text{GeV}$.}
\figsetgrpend

\figsetgrpstart
\figsetgrpnum{3.9}
\figsetgrptitle{Flare 9 from PKS\,1510$-$089}
\figsetplot{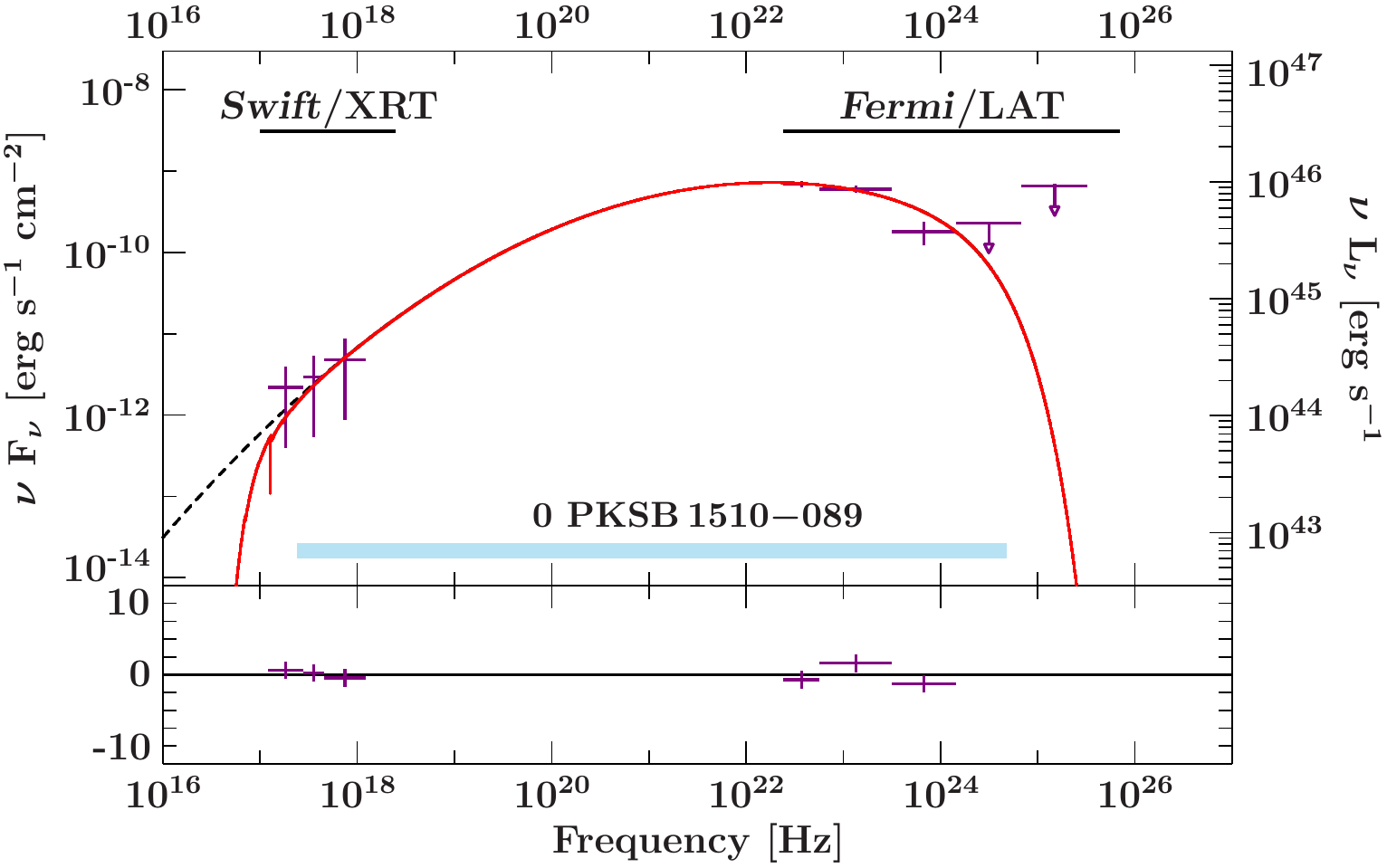}
\figsetgrpnote{Quasi-simultaneous SEDs of flares listed in Table\/\ref{Tab_neutrino_expectation}
. The red line indicates a log-parabola parametrization, including intrinsic absorption in X-rays and an exponential cut-off. The dashed black line shows the same parametrization without the absorption. The shadowed      area represents the integration range from $10\,\text{keV}$ to $20\,\text{GeV}$.}
\figsetgrpend

\figsetgrpstart
\figsetgrpnum{3.10}
\figsetgrptitle{Flare 10 from PKS\,1510$-$089}
\figsetplot{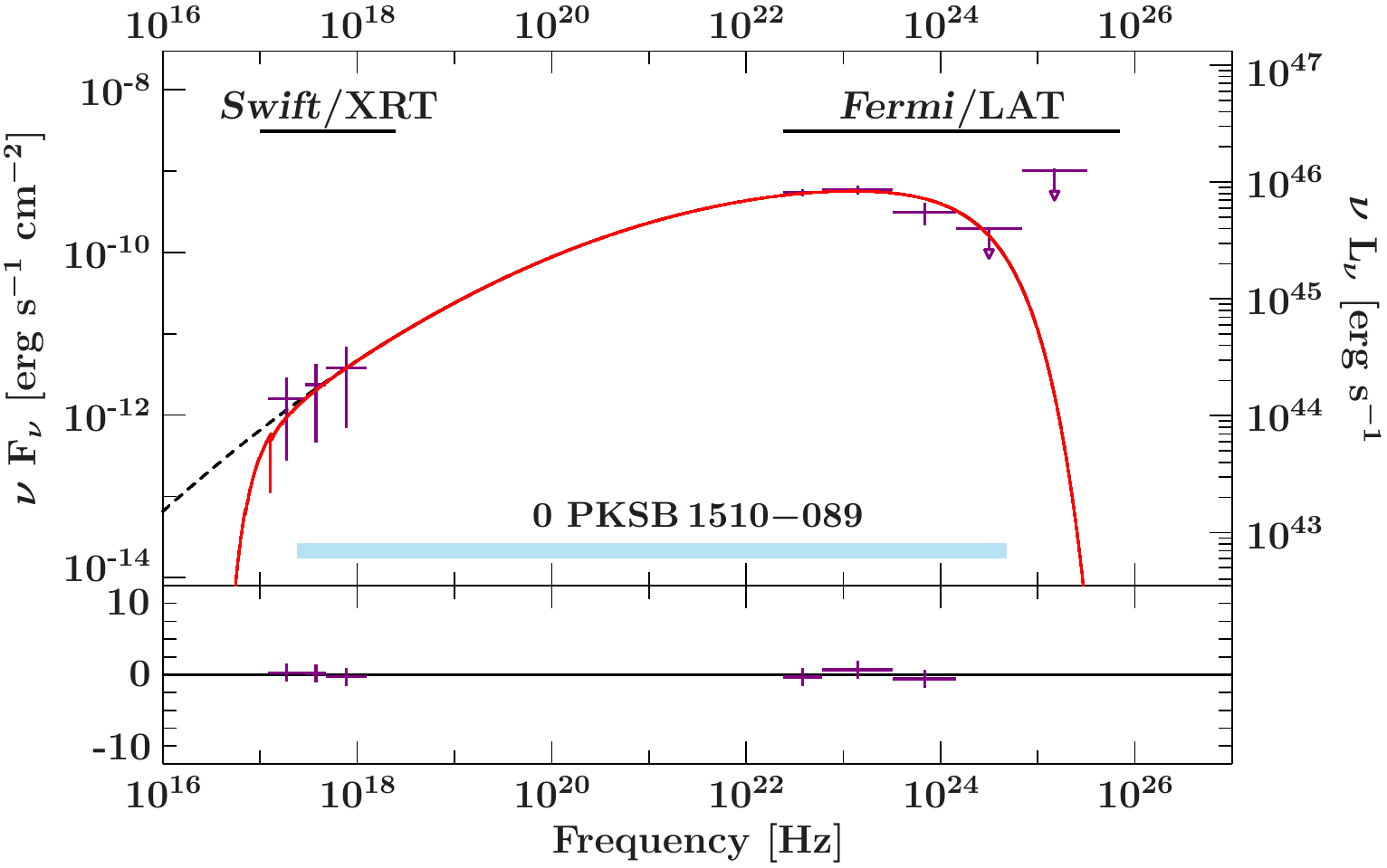}
\figsetgrpnote{Quasi-simultaneous SEDs of flares listed in Table\/\ref{Tab_neutrino_expectation}
. The red line indicates a log-parabola parametrization, including intrinsic absorption in X-rays and an exponential cut-off. The dashed black line shows the same parametrization without the absorption. The shadowed      area represents the integration range from $10\,\text{keV}$ to $20\,\text{GeV}$.}
\figsetgrpend

\figsetgrpstart
\figsetgrpnum{3.11}
\figsetgrptitle{Flare 11 from PKS\,1510$-$089}
\figsetplot{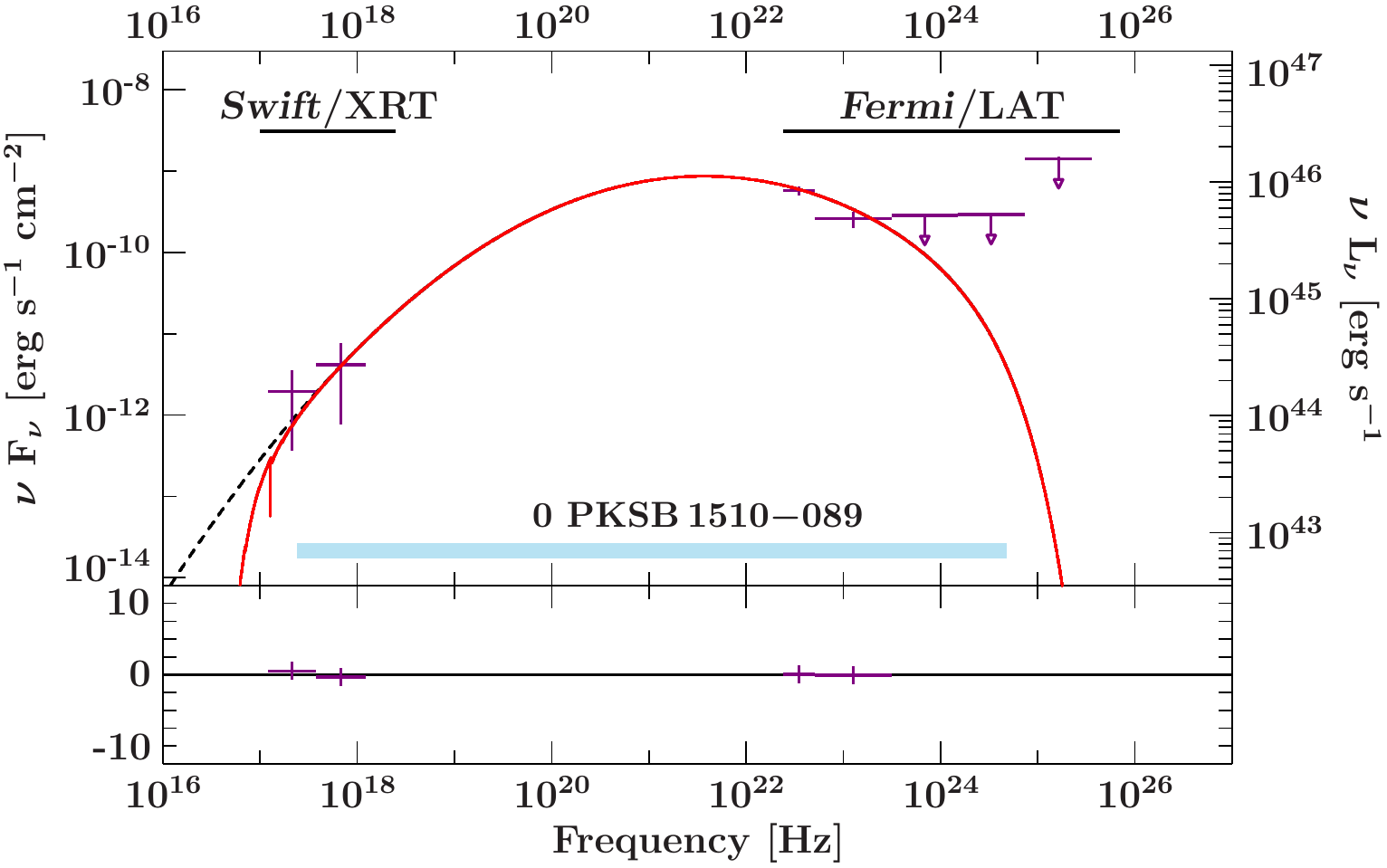}
\figsetgrpnote{Quasi-simultaneous SEDs of flares listed in Table\/\ref{Tab_neutrino_expectation}
. The red line indicates a log-parabola parametrization, including intrinsic absorption in X-rays and an exponential cut-off. The dashed black line shows the same parametrization without the absorption. The shadowed      area represents the integration range from $10\,\text{keV}$ to $20\,\text{GeV}$.}
\figsetgrpend

\figsetgrpstart
\figsetgrpnum{3.12}
\figsetgrptitle{Flare 12 from PKS\,1510$-$089}
\figsetplot{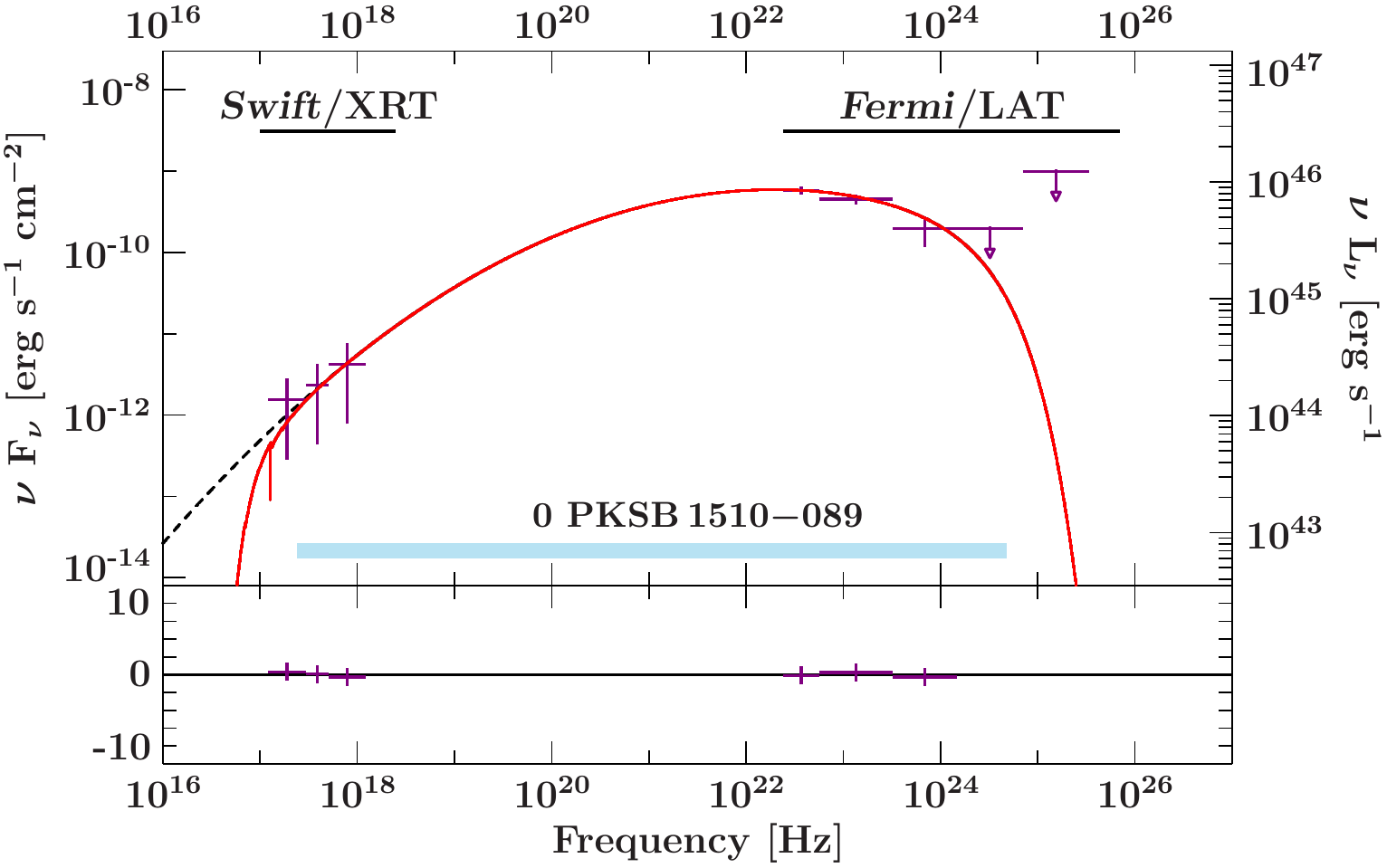}
\figsetgrpnote{Quasi-simultaneous SEDs of flares listed in Table\/\ref{Tab_neutrino_expectation}
. The red line indicates a log-parabola parametrization, including intrinsic absorption in X-rays and an exponential cut-off. The dashed black line shows the same parametrization without the absorption. The shadowed      area represents the integration range from $10\,\text{keV}$ to $20\,\text{GeV}$.}
\figsetgrpend

\figsetgrpstart
\figsetgrpnum{3.13}
\figsetgrptitle{Flare 13 from PKS\,1510$-$089}
\figsetplot{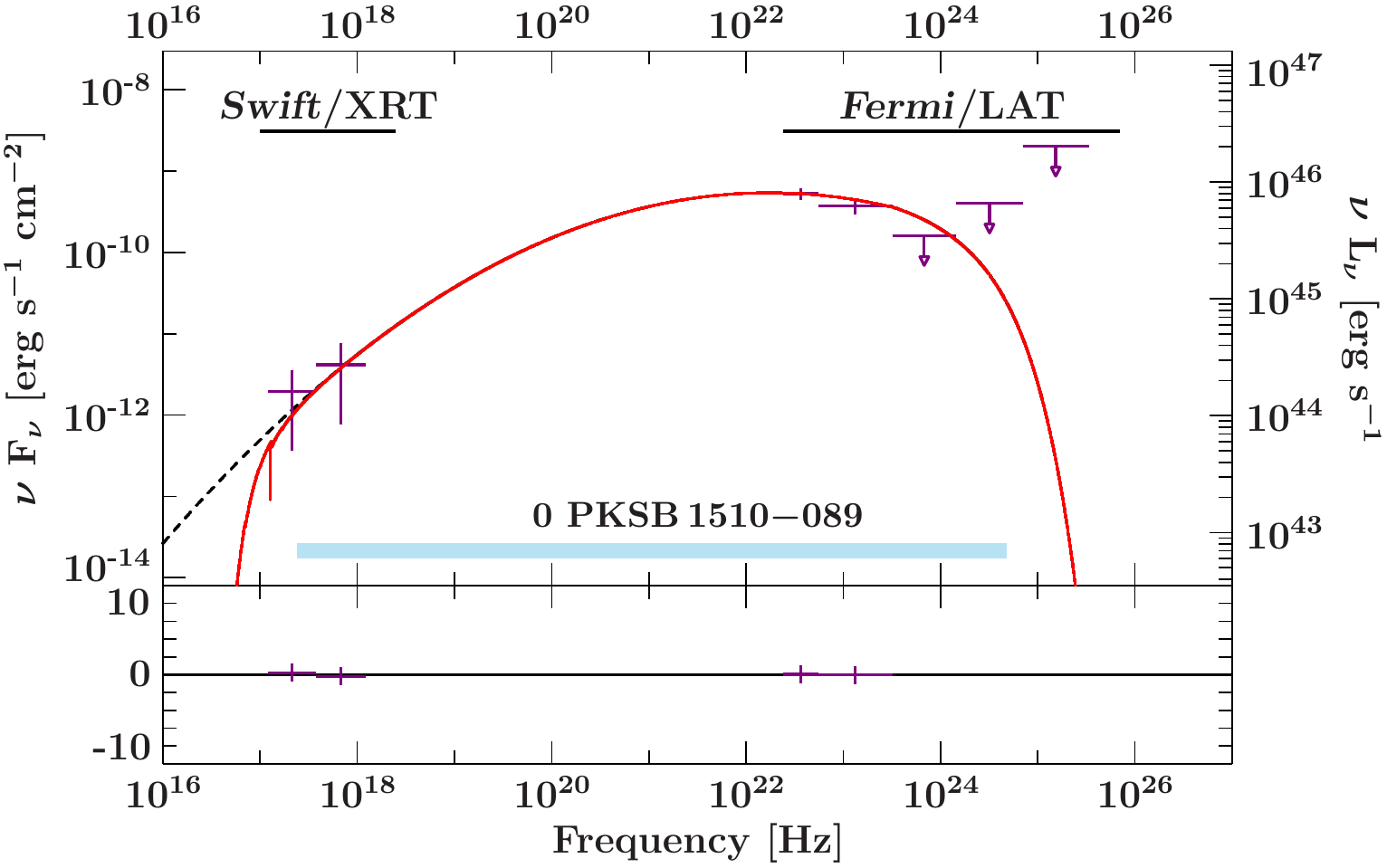}
\figsetgrpnote{Quasi-simultaneous SEDs of flares listed in Table\/\ref{Tab_neutrino_expectation}
. The red line indicates a log-parabola parametrization, including intrinsic absorption in X-rays and an exponential cut-off. The dashed black line shows the same parametrization without the absorption. The shadowed      area represents the integration range from $10\,\text{keV}$ to $20\,\text{GeV}$.}
\figsetgrpend

\figsetgrpstart
\figsetgrpnum{3.14}
\figsetgrptitle{Flare 14 from PKS\,1510$-$089}
\figsetplot{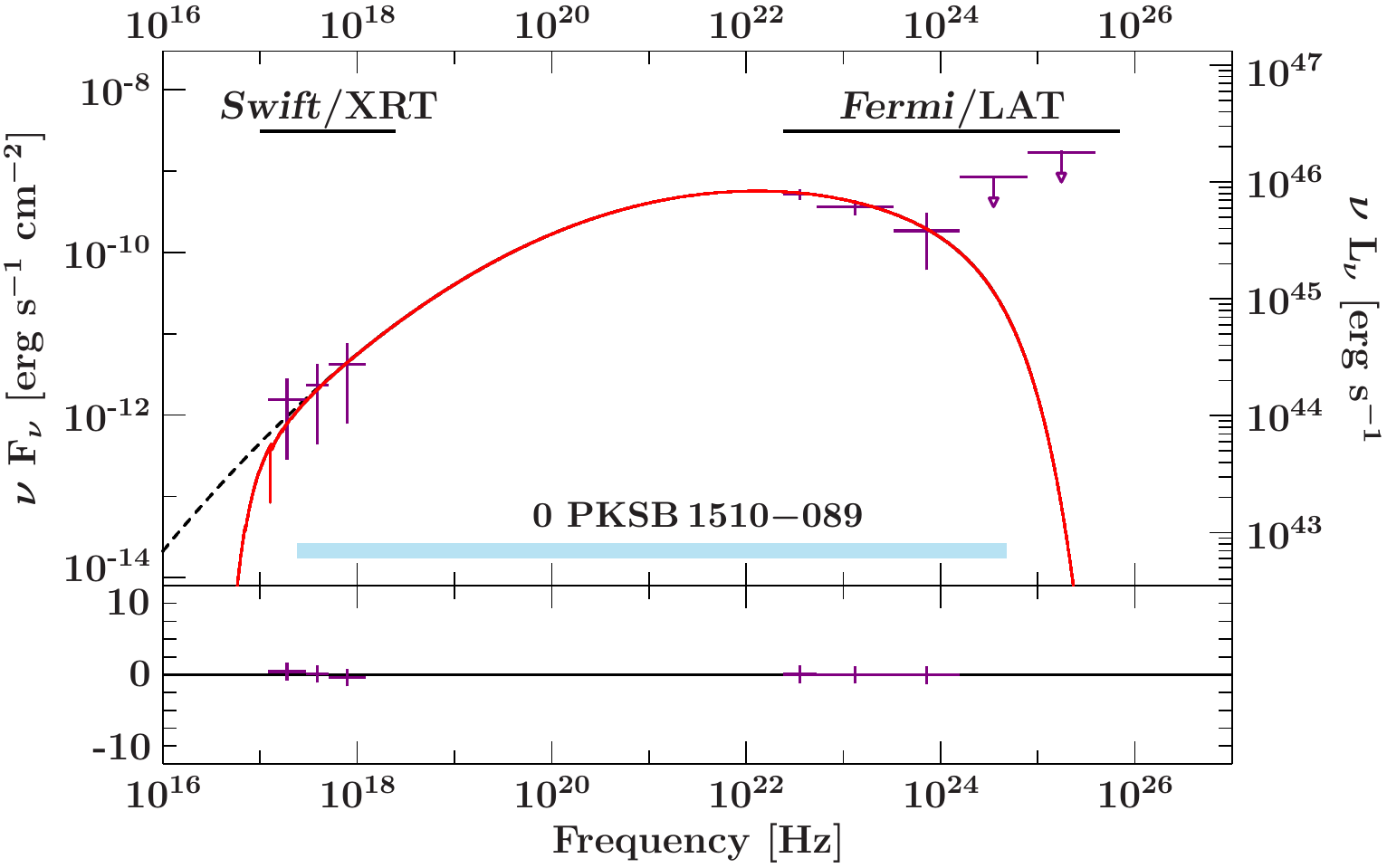}
\figsetgrpnote{Quasi-simultaneous SEDs of flares listed in Table\/\ref{Tab_neutrino_expectation}
. The red line indicates a log-parabola parametrization, including intrinsic absorption in X-rays and an exponential cut-off. The dashed black line shows the same parametrization without the absorption. The shadowed      area represents the integration range from $10\,\text{keV}$ to $20\,\text{GeV}$.}
\figsetgrpend

\figsetgrpstart
\figsetgrpnum{3.15}
\figsetgrptitle{Flare 15 from PKS\,1510$-$089}
\figsetplot{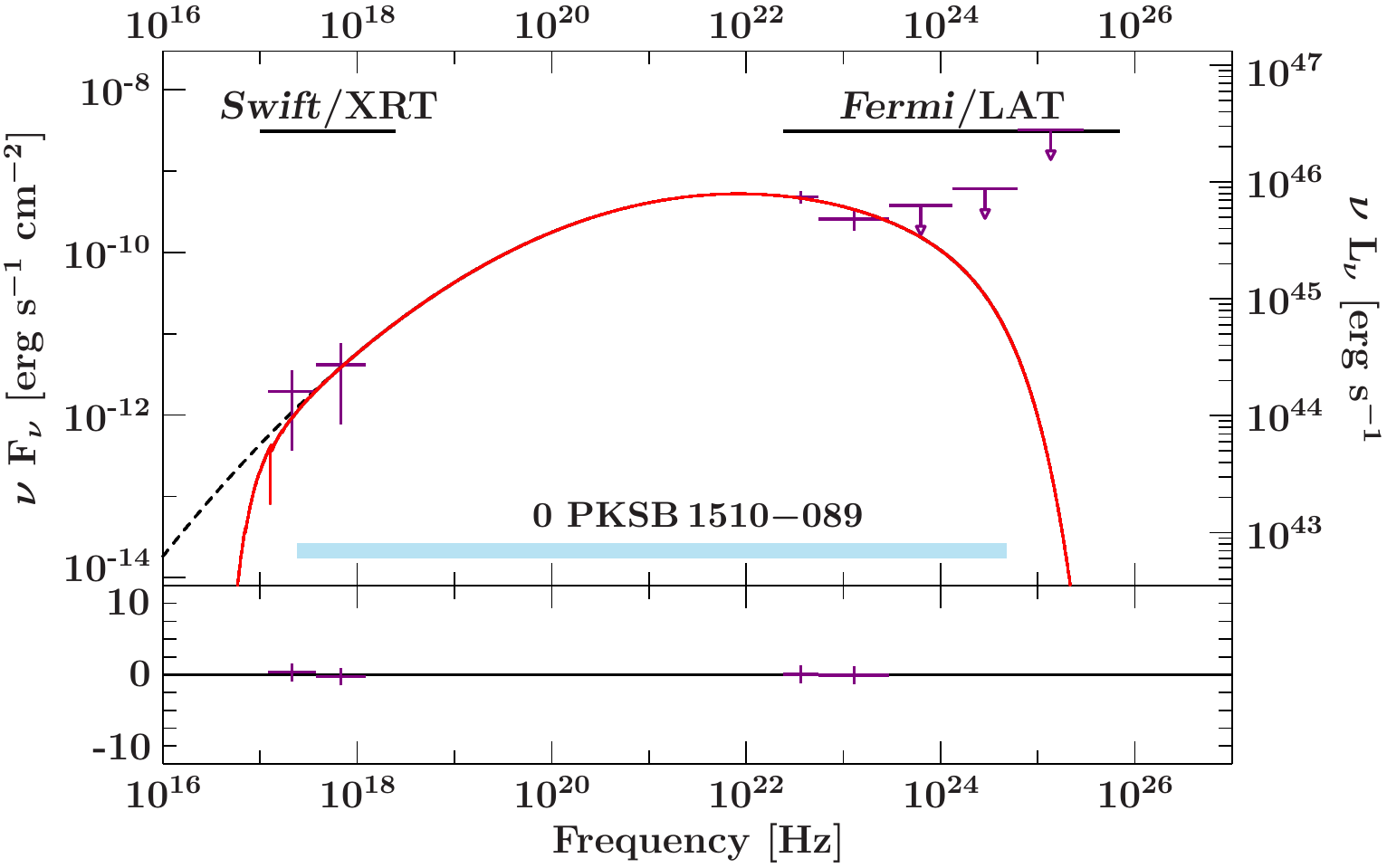}
\figsetgrpnote{Quasi-simultaneous SEDs of flares listed in Table\/\ref{Tab_neutrino_expectation}
. The red line indicates a log-parabola parametrization, including intrinsic absorption in X-rays and an exponential cut-off. The dashed black line shows the same parametrization without the absorption. The shadowed      area represents the integration range from $10\,\text{keV}$ to $20\,\text{GeV}$.}
\figsetgrpend

\figsetgrpstart
\figsetgrpnum{3.16}
\figsetgrptitle{Flare 16 from PKS\,1510$-$089}
\figsetplot{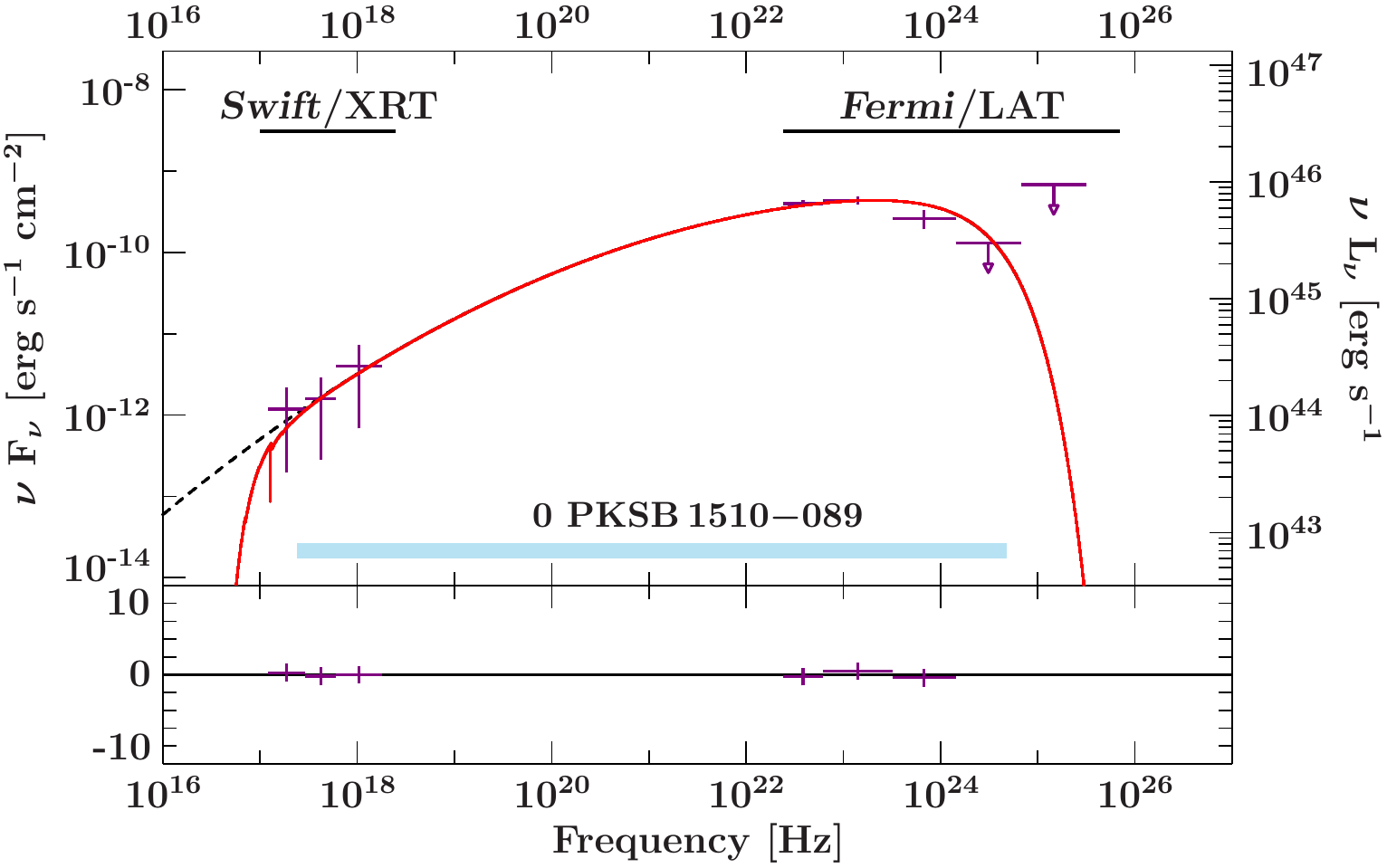}
\figsetgrpnote{Quasi-simultaneous SEDs of flares listed in Table\/\ref{Tab_neutrino_expectation}
. The red line indicates a log-parabola parametrization, including intrinsic absorption in X-rays and an exponential cut-off. The dashed black line shows the same parametrization without the absorption. The shadowed      area represents the integration range from $10\,\text{keV}$ to $20\,\text{GeV}$.}
\figsetgrpend

\figsetgrpstart
\figsetgrpnum{3.17}
\figsetgrptitle{Flare 17 from PKS\,1510$-$089}
\figsetplot{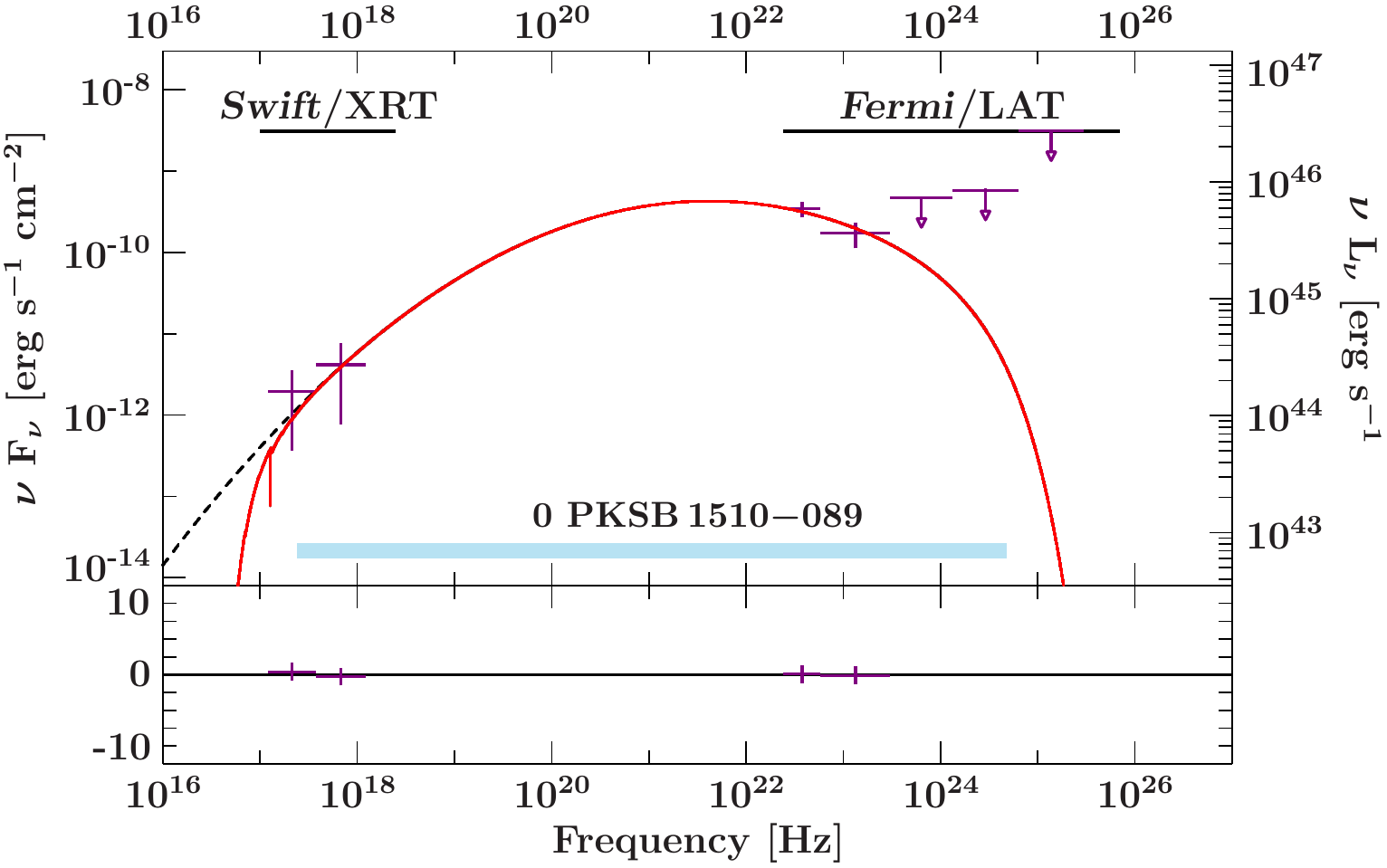}
\figsetgrpnote{Quasi-simultaneous SEDs of flares listed in Table\/\ref{Tab_neutrino_expectation}
. The red line indicates a log-parabola parametrization, including intrinsic absorption in X-rays and an exponential cut-off. The dashed black line shows the same parametrization without the absorption. The shadowed      area represents the integration range from $10\,\text{keV}$ to $20\,\text{GeV}$.}
\figsetgrpend

\figsetgrpstart
\figsetgrpnum{3.18}
\figsetgrptitle{Flare 18 from PKS\,1510$-$089}
\figsetplot{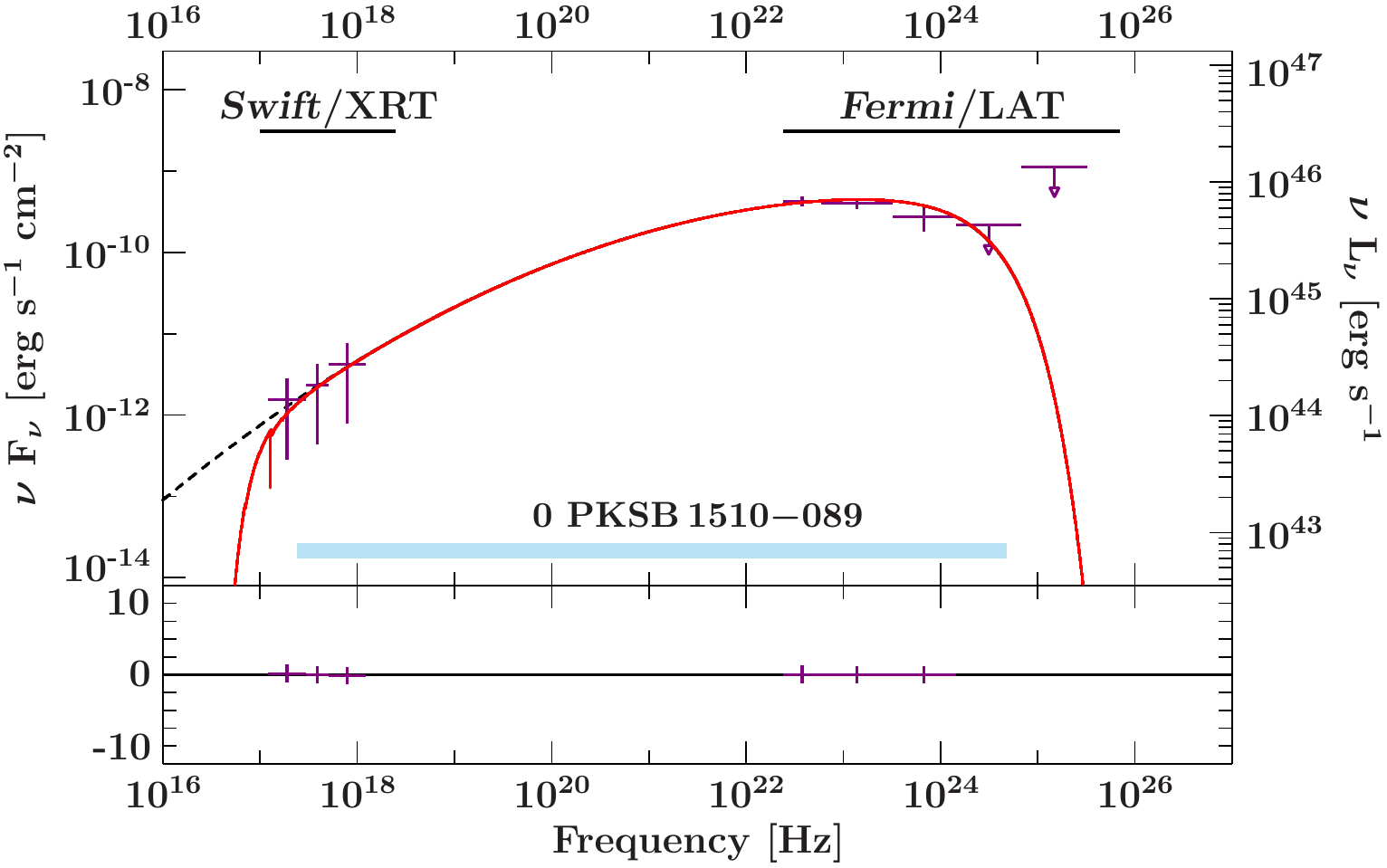}
\figsetgrpnote{Quasi-simultaneous SEDs of flares listed in Table\/\ref{Tab_neutrino_expectation}
. The red line indicates a log-parabola parametrization, including intrinsic absorption in X-rays and an exponential cut-off. The dashed black line shows the same parametrization without the absorption. The shadowed      area represents the integration range from $10\,\text{keV}$ to $20\,\text{GeV}$.}
\figsetgrpend

\figsetgrpstart
\figsetgrpnum{3.19}
\figsetgrptitle{Flare 19 from PKS\,1510$-$089}
\figsetplot{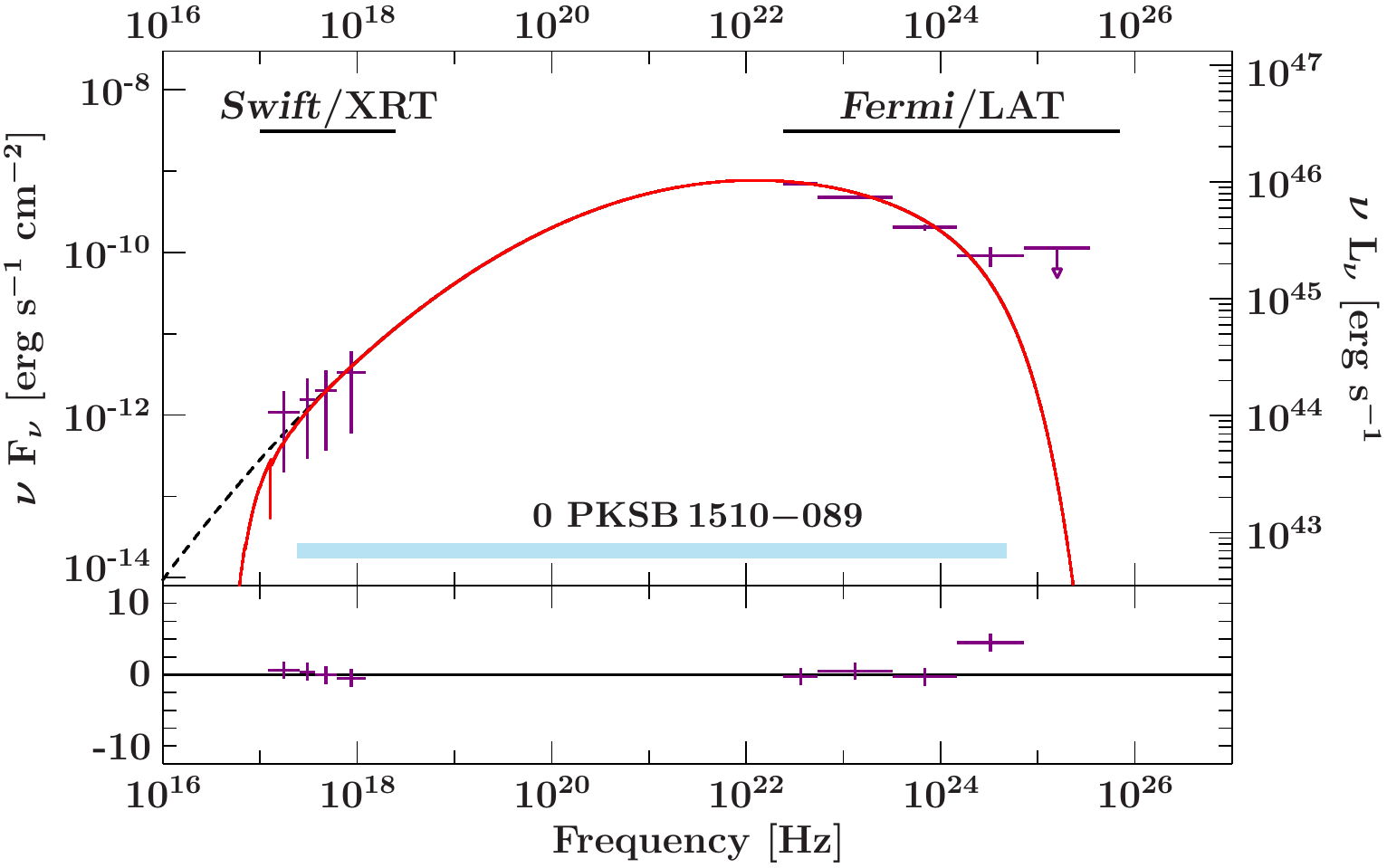}
\figsetgrpnote{Quasi-simultaneous SEDs of flares listed in Table\/\ref{Tab_neutrino_expectation}
. The red line indicates a log-parabola parametrization, including intrinsic absorption in X-rays and an exponential cut-off. The dashed black line shows the same parametrization without the absorption. The shadowed      area represents the integration range from $10\,\text{keV}$ to $20\,\text{GeV}$.}
\figsetgrpend

\figsetgrpstart
\figsetgrpnum{3.20}
\figsetgrptitle{Flare 20 from PKS\,1510$-$089}
\figsetplot{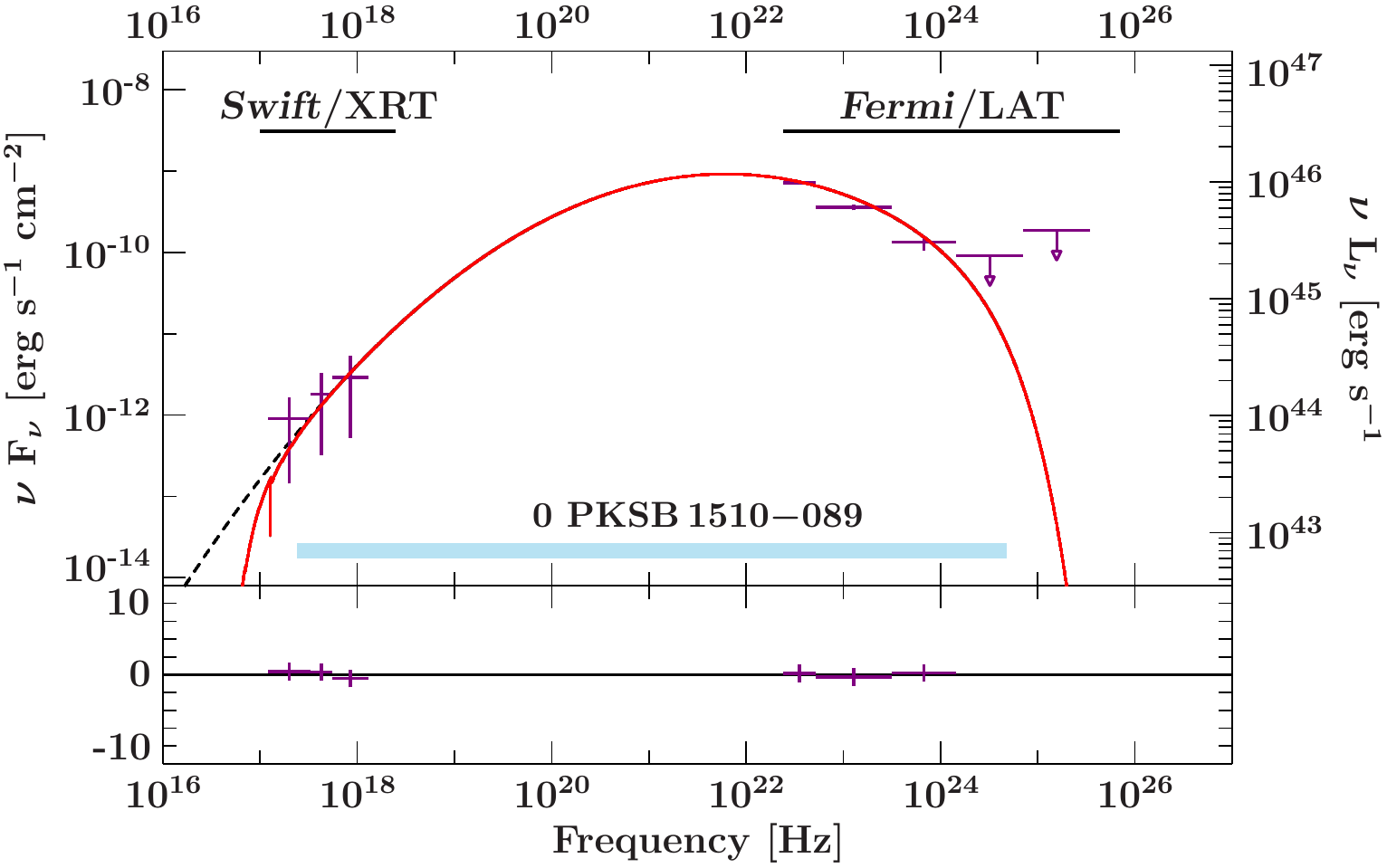}
\figsetgrpnote{Quasi-simultaneous SEDs of flares listed in Table\/\ref{Tab_neutrino_expectation}
. The red line indicates a log-parabola parametrization, including intrinsic absorption in X-rays and an exponential cut-off. The dashed black line shows the same parametrization without the absorption. The shadowed      area represents the integration range from $10\,\text{keV}$ to $20\,\text{GeV}$.}
\figsetgrpend

\figsetgrpstart
\figsetgrpnum{3.21}
\figsetgrptitle{Flare 21 from PKS\,1510$-$089}
\figsetplot{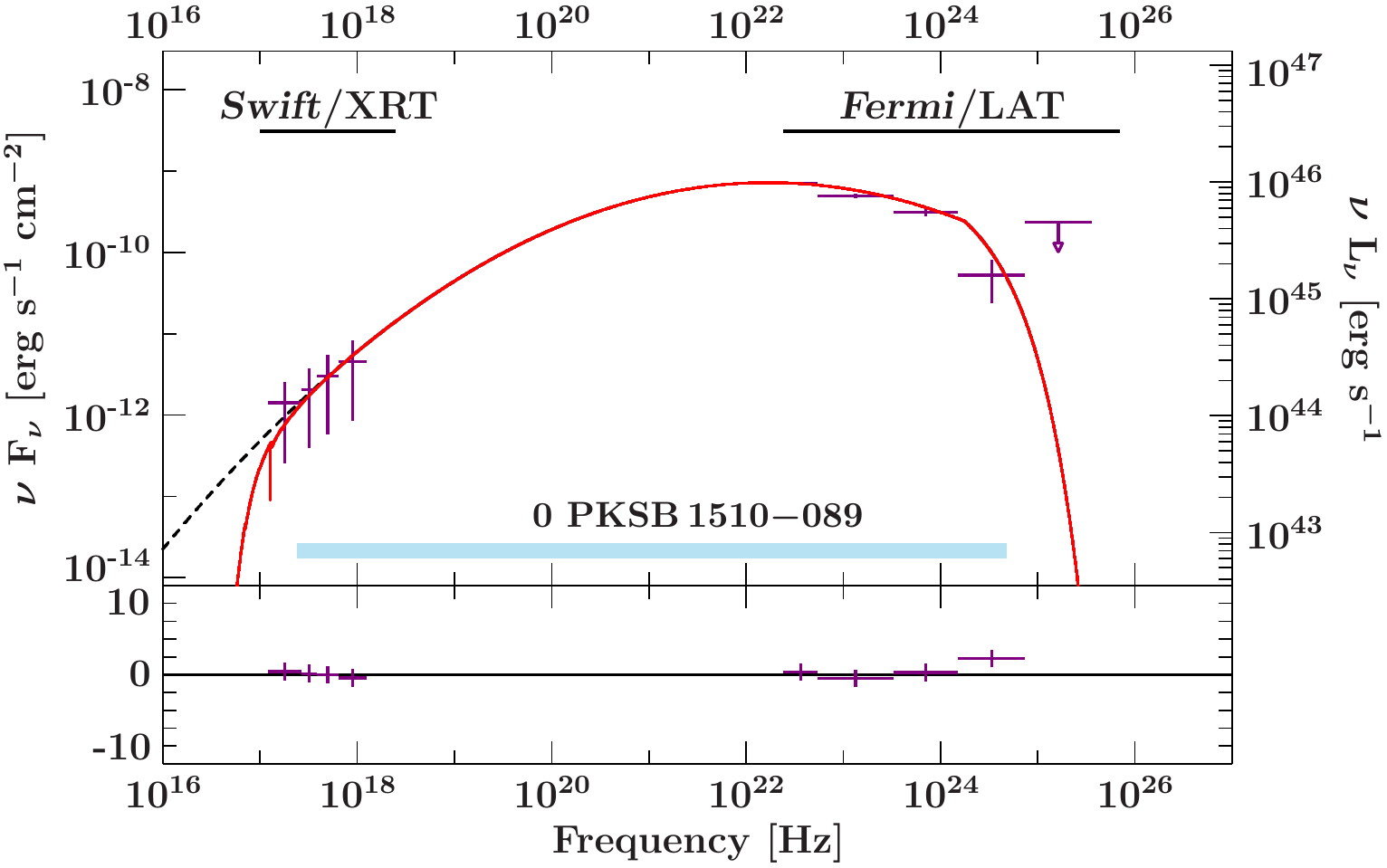}
\figsetgrpnote{Quasi-simultaneous SEDs of flares listed in Table\/\ref{Tab_neutrino_expectation}
. The red line indicates a log-parabola parametrization, including intrinsic absorption in X-rays and an exponential cut-off. The dashed black line shows the same parametrization without the absorption. The shadowed      area represents the integration range from $10\,\text{keV}$ to $20\,\text{GeV}$.}
\figsetgrpend

\figsetgrpstart
\figsetgrpnum{3.22}
\figsetgrptitle{Flare 22 from PKS\,1510$-$089}
\figsetplot{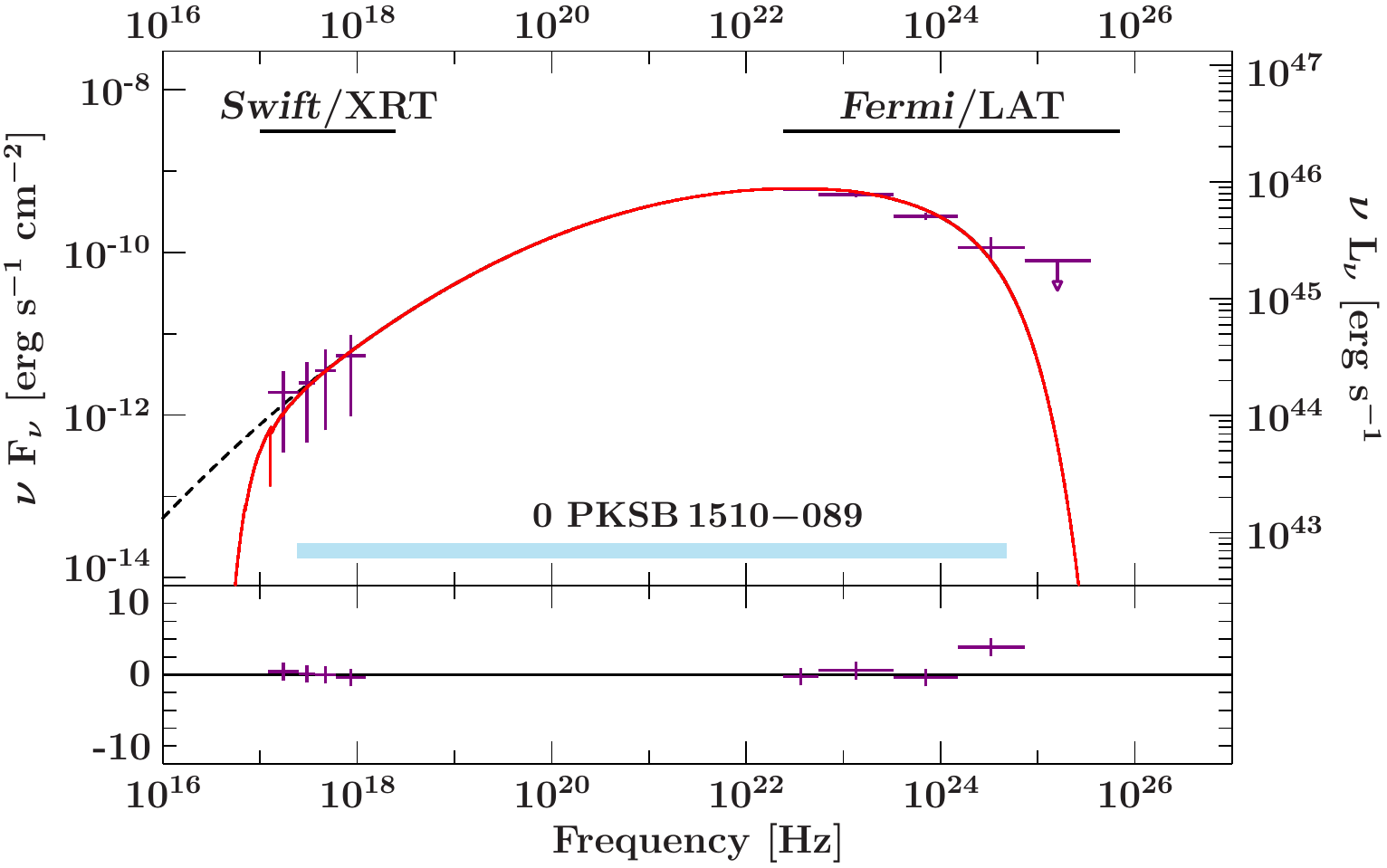}
\figsetgrpnote{Quasi-simultaneous SEDs of flares listed in Table\/\ref{Tab_neutrino_expectation}
. The red line indicates a log-parabola parametrization, including intrinsic absorption in X-rays and an exponential cut-off. The dashed black line shows the same parametrization without the absorption. The shadowed      area represents the integration range from $10\,\text{keV}$ to $20\,\text{GeV}$.}
\figsetgrpend

\figsetgrpstart
\figsetgrpnum{3.23}
\figsetgrptitle{Flare 23 from PKS\,1510$-$089}
\figsetplot{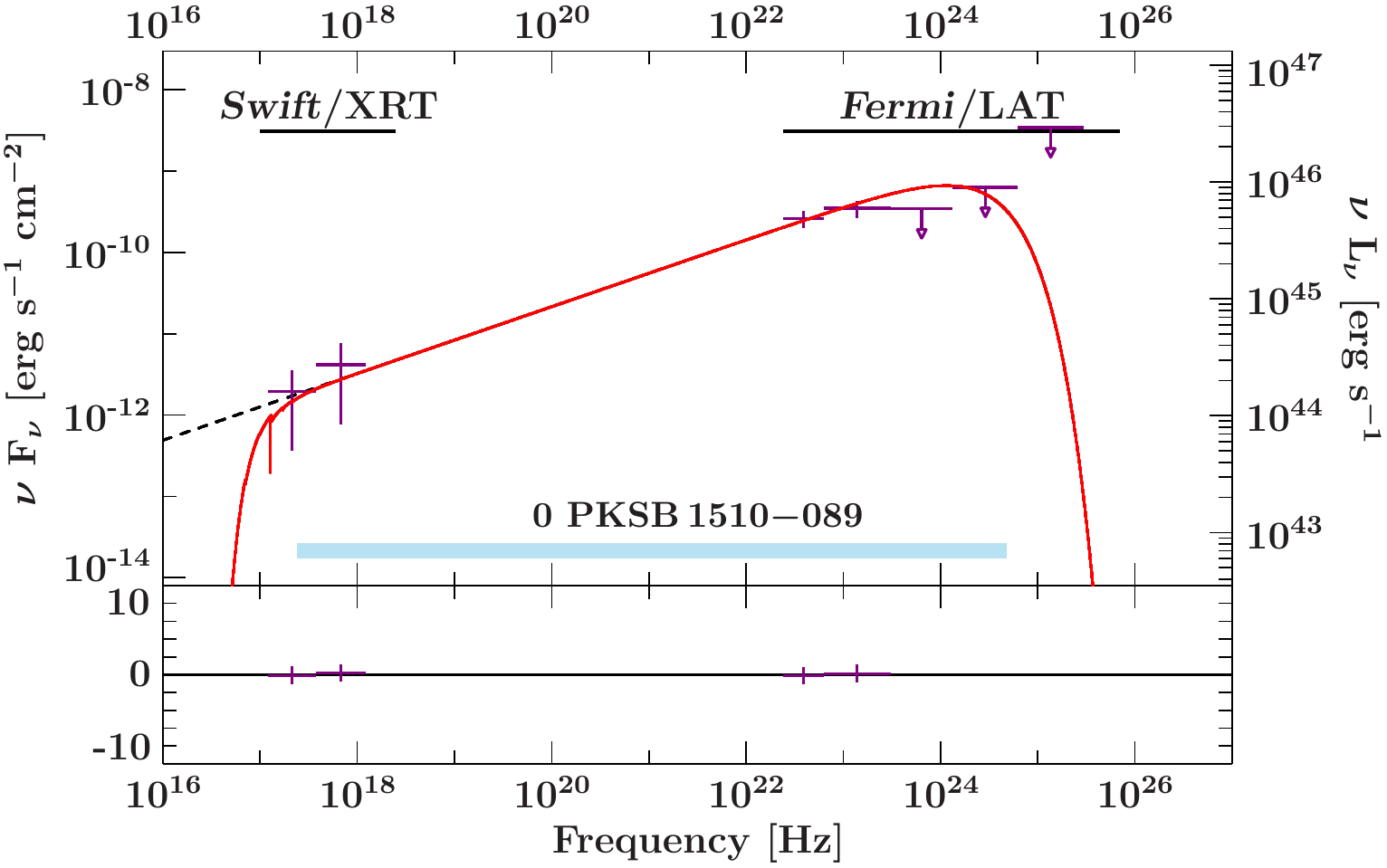}
\figsetgrpnote{Quasi-simultaneous SEDs of flares listed in Table\/\ref{Tab_neutrino_expectation}
. The red line indicates a log-parabola parametrization, including intrinsic absorption in X-rays and an exponential cut-off. The dashed black line shows the same parametrization without the absorption. The shadowed      area represents the integration range from $10\,\text{keV}$ to $20\,\text{GeV}$.}
\figsetgrpend

\figsetgrpstart
\figsetgrpnum{3.24}
\figsetgrptitle{Flare 24 from PKS\,1510$-$089}
\figsetplot{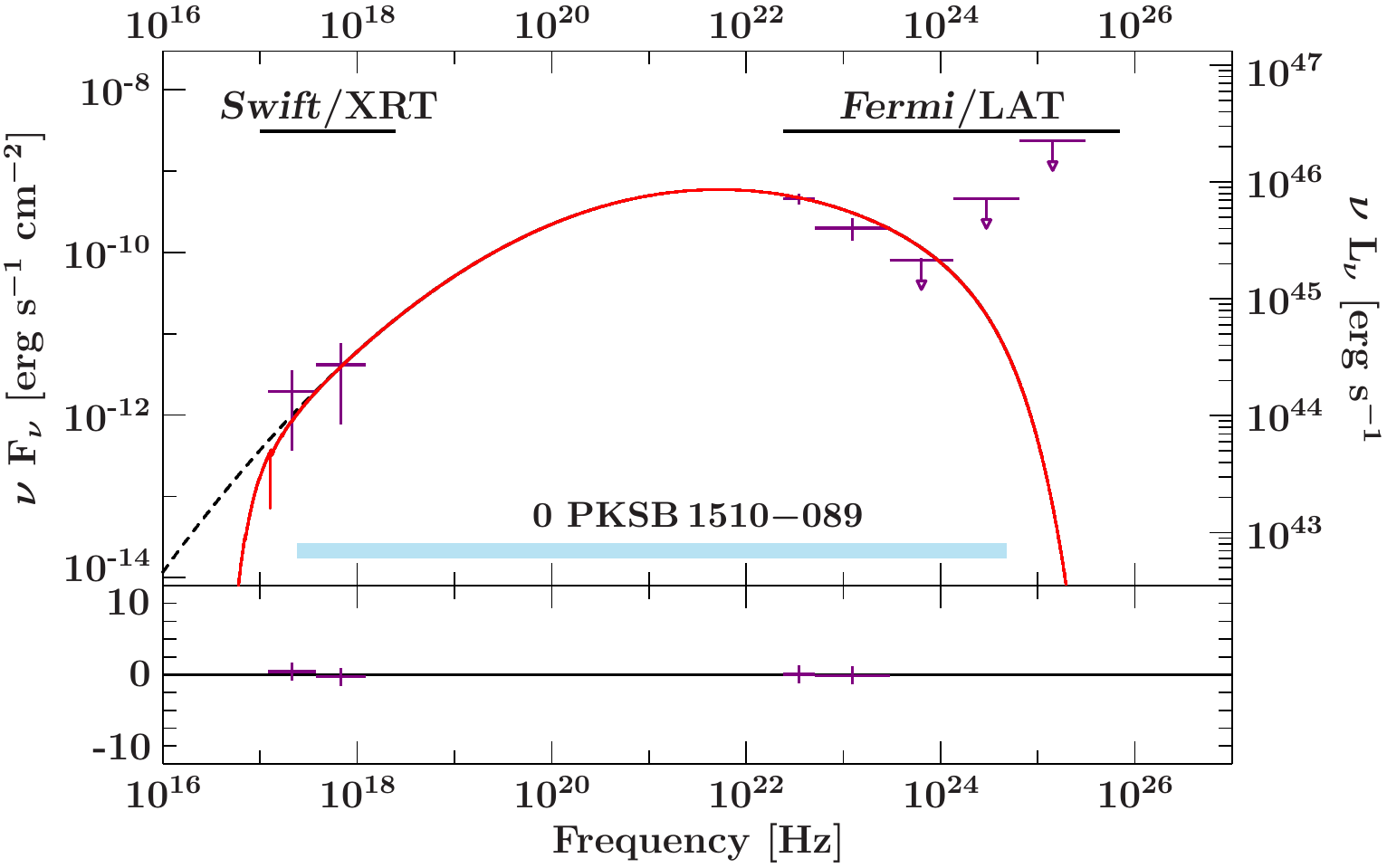}
\figsetgrpnote{Quasi-simultaneous SEDs of flares listed in Table\/\ref{Tab_neutrino_expectation}
. The red line indicates a log-parabola parametrization, including intrinsic absorption in X-rays and an exponential cut-off. The dashed black line shows the same parametrization without the absorption. The shadowed      area represents the integration range from $10\,\text{keV}$ to $20\,\text{GeV}$.}
\figsetgrpend

\figsetgrpstart
\figsetgrpnum{3.25}
\figsetgrptitle{Flare 25 from PKS\,1510$-$089}
\figsetplot{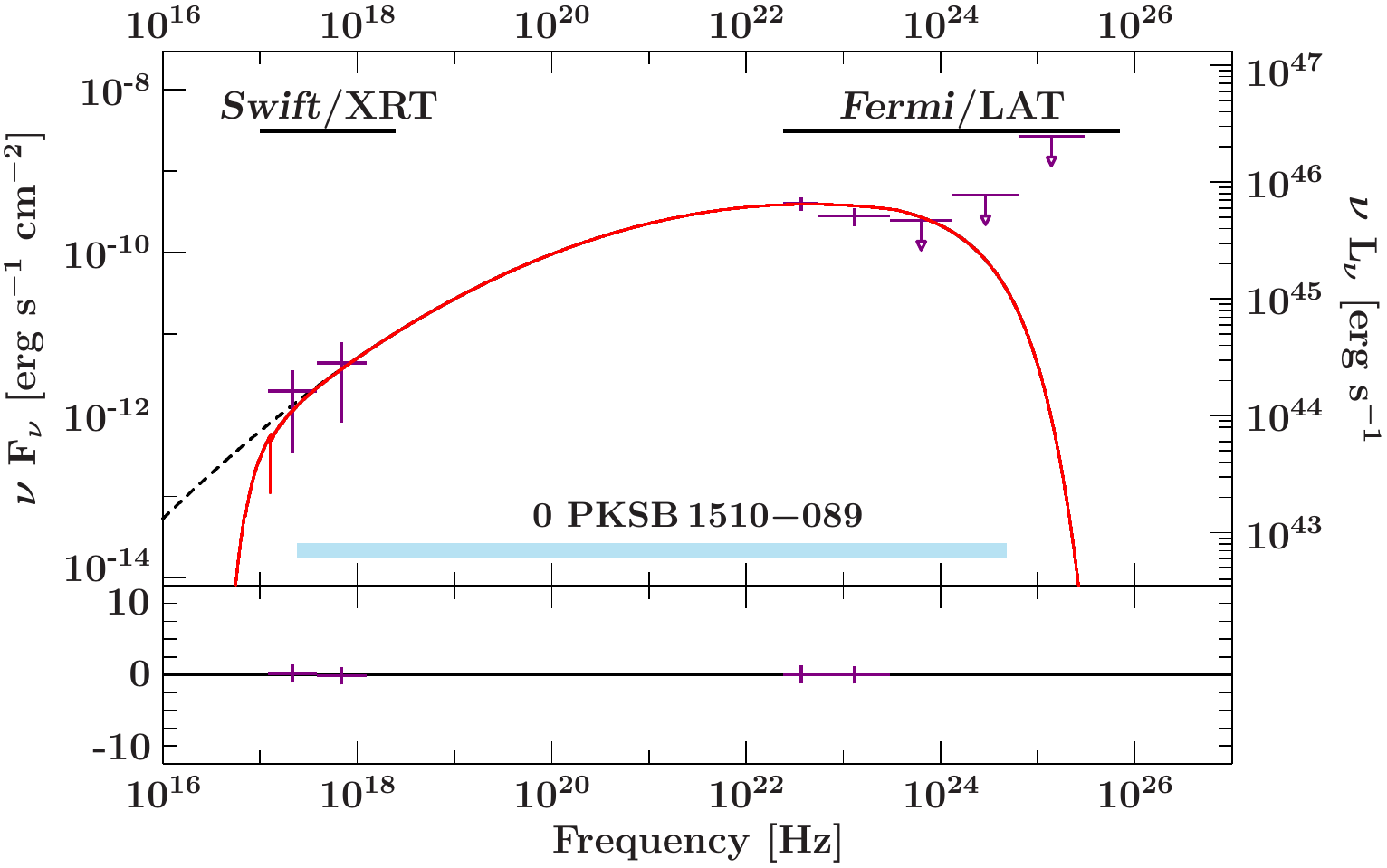}
\figsetgrpnote{Quasi-simultaneous SEDs of flares listed in Table\/\ref{Tab_neutrino_expectation}
. The red line indicates a log-parabola parametrization, including intrinsic absorption in X-rays and an exponential cut-off. The dashed black line shows the same parametrization without the absorption. The shadowed      area represents the integration range from $10\,\text{keV}$ to $20\,\text{GeV}$.}
\figsetgrpend

\figsetgrpstart
\figsetgrpnum{3.26}
\figsetgrptitle{Flare 26 from PKS\,1510$-$089}
\figsetplot{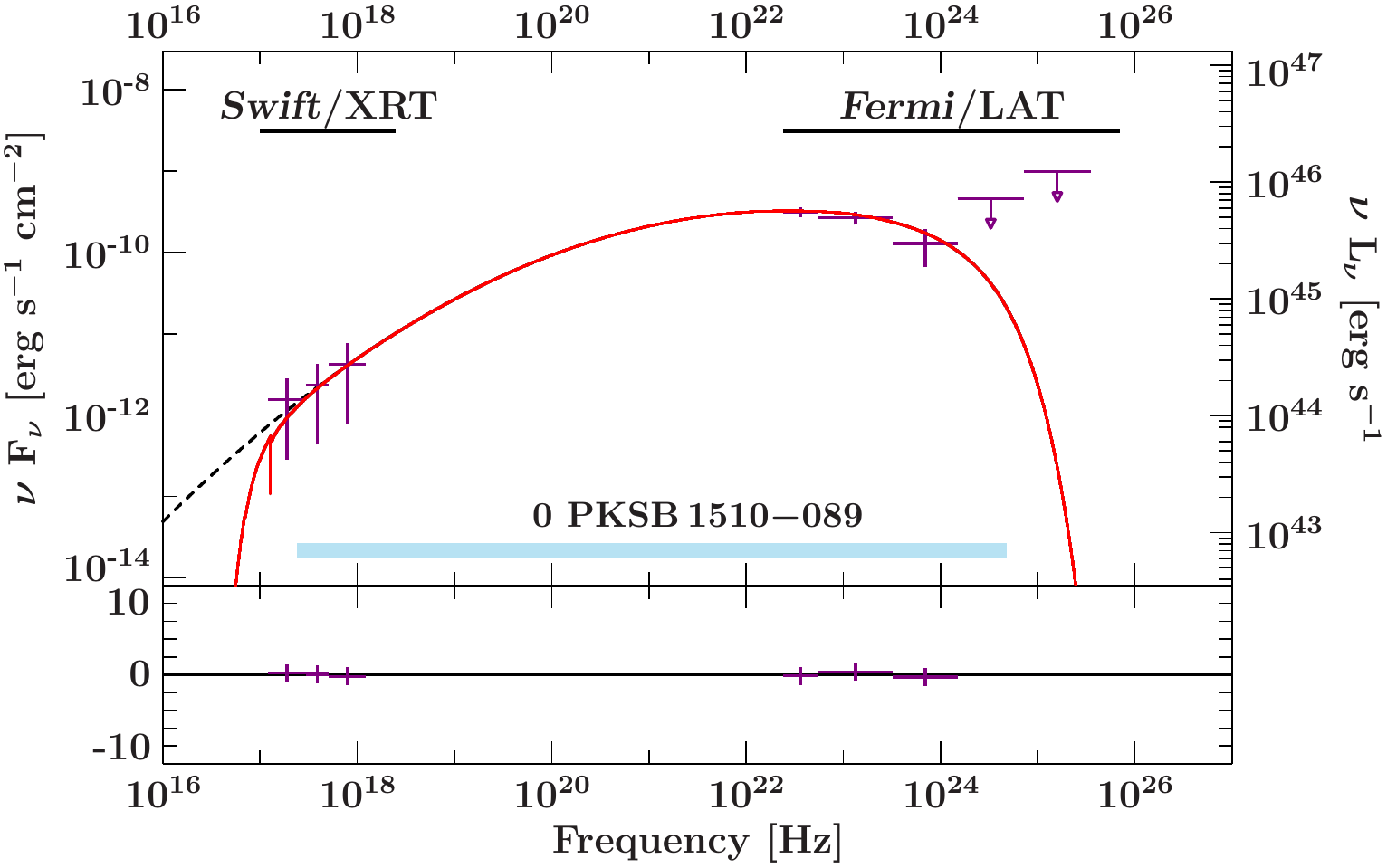}
\figsetgrpnote{Quasi-simultaneous SEDs of flares listed in Table\/\ref{Tab_neutrino_expectation}
. The red line indicates a log-parabola parametrization, including intrinsic absorption in X-rays and an exponential cut-off. The dashed black line shows the same parametrization without the absorption. The shadowed      area represents the integration range from $10\,\text{keV}$ to $20\,\text{GeV}$.}
\figsetgrpend

\figsetgrpstart
\figsetgrpnum{3.27}
\figsetgrptitle{Flare 27 from 3C\,279}
\figsetplot{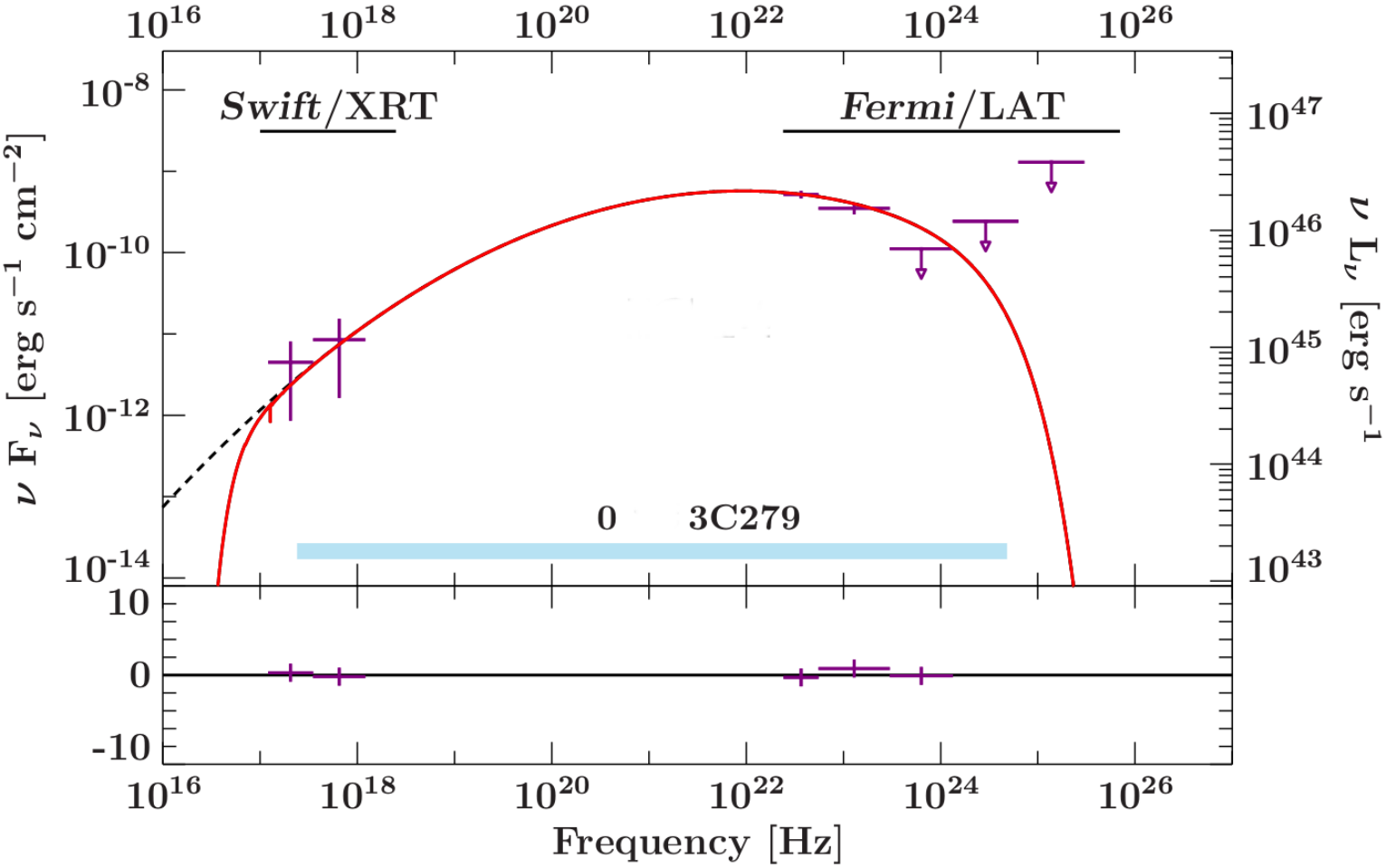}
\figsetgrpnote{Quasi-simultaneous SEDs of flares listed in Table\/\ref{Tab_neutrino_expectation}
. The red line indicates a log-parabola parametrization, including intrinsic absorption in X-rays and an exponential cut-off. The dashed black line shows the same parametrization without the absorption. The shadowed      area represents the integration range from $10\,\text{keV}$ to $20\,\text{GeV}$.}
\figsetgrpend

\figsetgrpstart
\figsetgrpnum{3.28}
\figsetgrptitle{Flare 28 from PKS\,0402$-$362}
\figsetplot{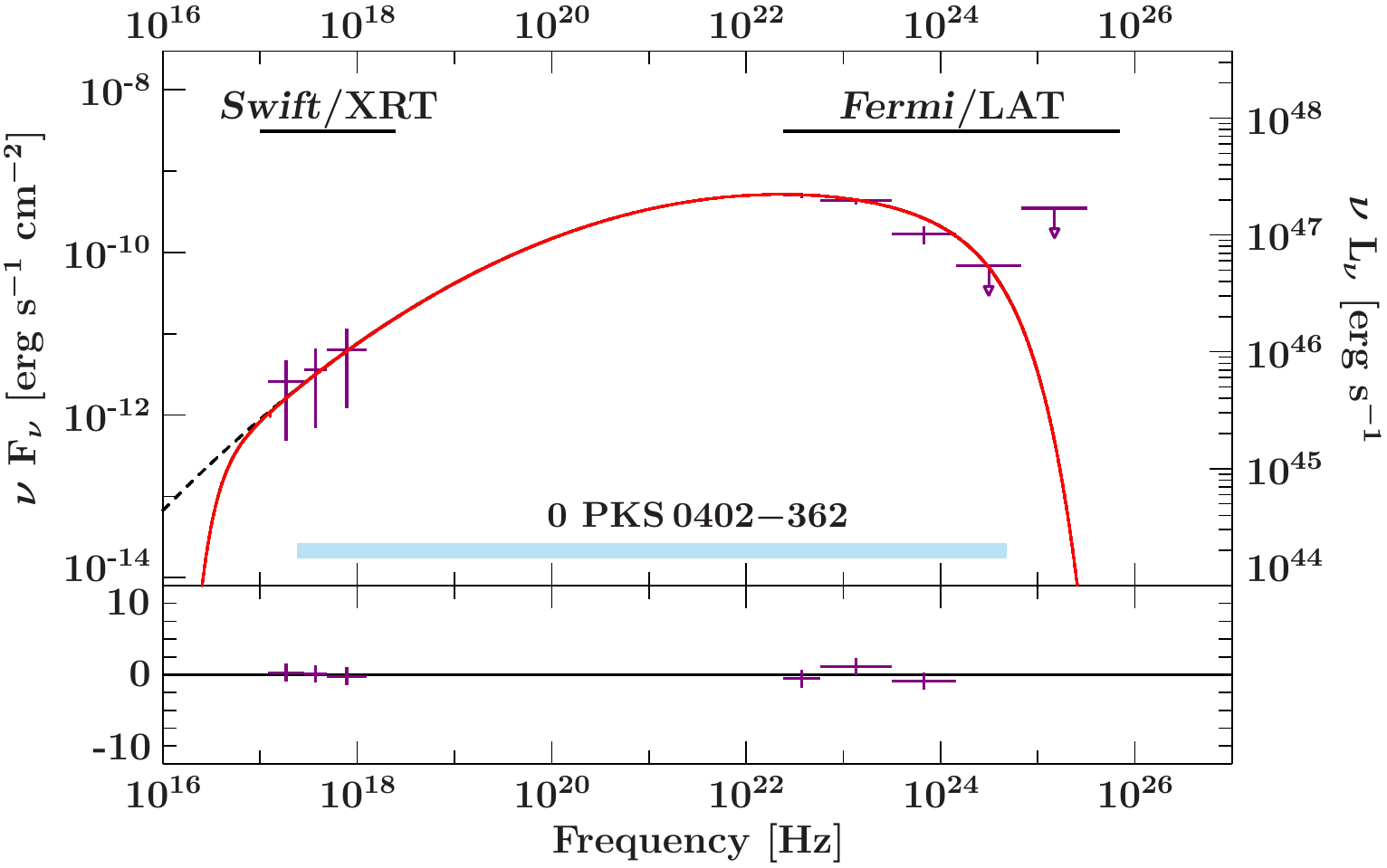}
\figsetgrpnote{Quasi-simultaneous SEDs of flares listed in Table\/\ref{Tab_neutrino_expectation}
. The red line indicates a log-parabola parametrization, including intrinsic absorption in X-rays and an exponential cut-off. The dashed black line shows the same parametrization without the absorption. The shadowed      area represents the integration range from $10\,\text{keV}$ to $20\,\text{GeV}$.}
\figsetgrpend

\figsetgrpstart
\figsetgrpnum{3.29}
\figsetgrptitle{Flare 29 from PKS\,1510$-$089}
\figsetplot{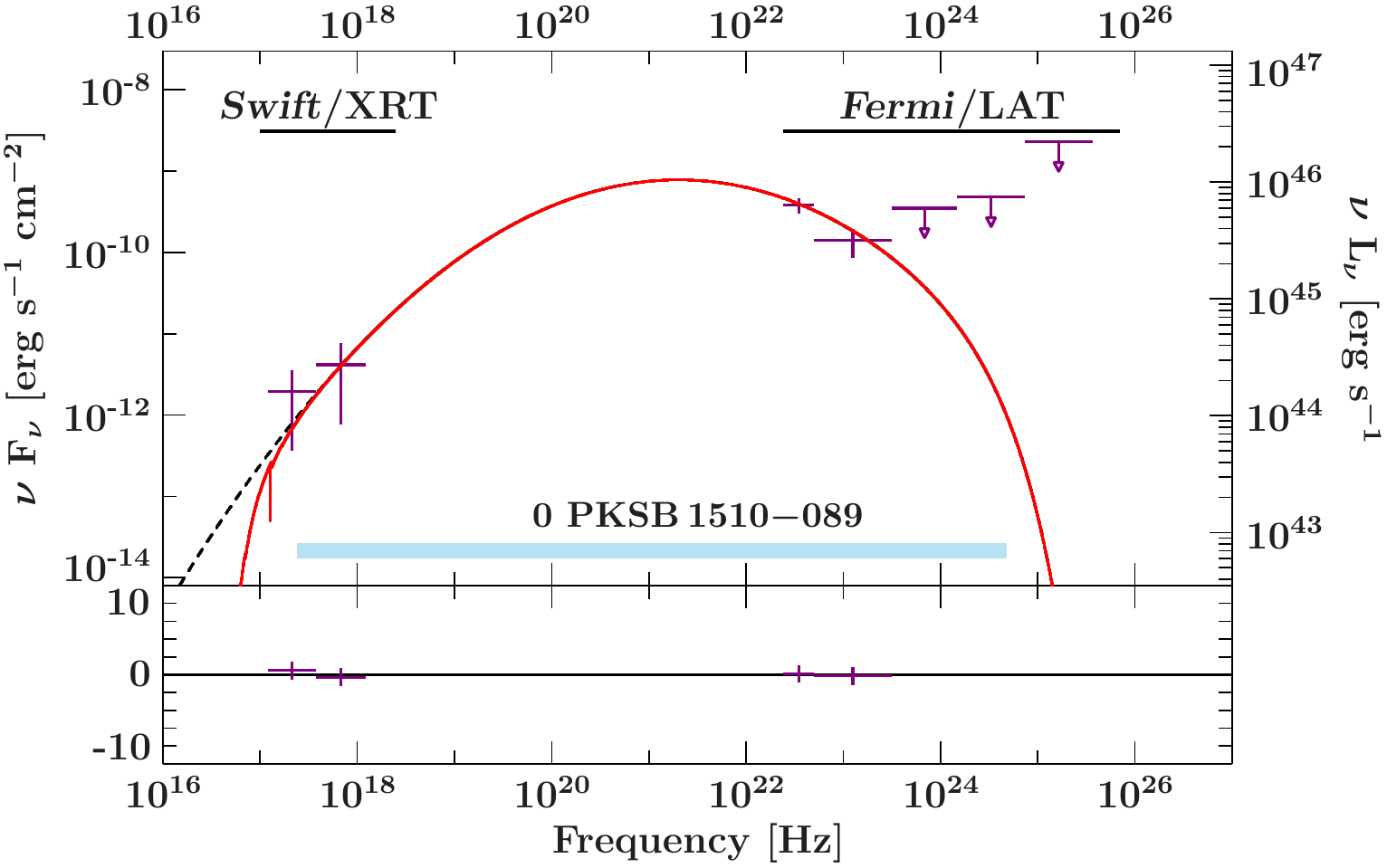}
\figsetgrpnote{Quasi-simultaneous SEDs of flares listed in Table\/\ref{Tab_neutrino_expectation}
. The red line indicates a log-parabola parametrization, including intrinsic absorption in X-rays and an exponential cut-off. The dashed black line shows the same parametrization without the absorption. The shadowed      area represents the integration range from $10\,\text{keV}$ to $20\,\text{GeV}$.}
\figsetgrpend

\figsetgrpstart
\figsetgrpnum{3.30}
\figsetgrptitle{Flare 30 from PKS\,1510$-$089}
\figsetplot{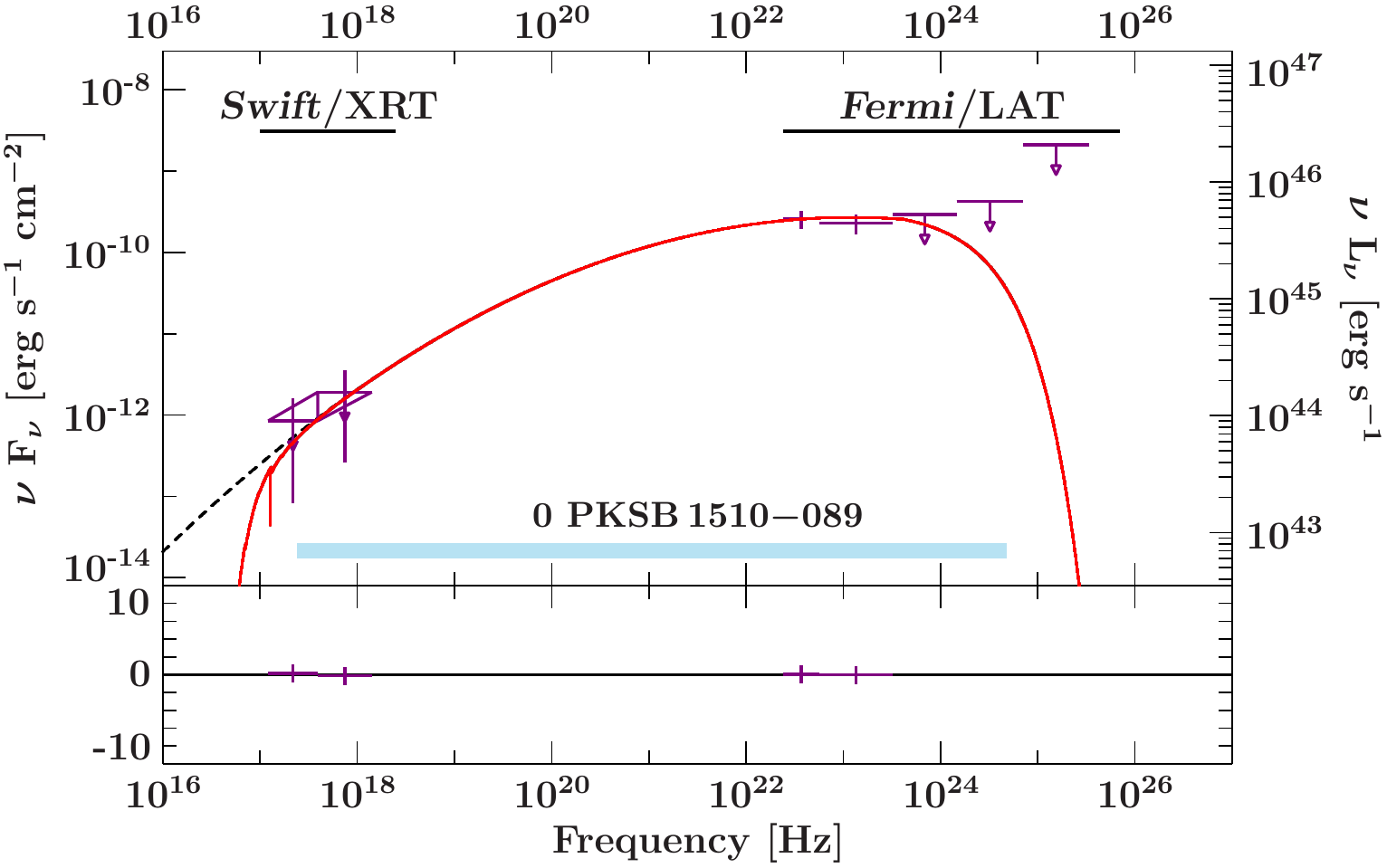}
\figsetgrpnote{Quasi-simultaneous SEDs of flares listed in Table\/\ref{Tab_neutrino_expectation}
. The red line indicates a log-parabola parametrization, including intrinsic absorption in X-rays and an exponential cut-off. The dashed black line shows the same parametrization without the absorption. The shadowed      area represents the integration range from $10\,\text{keV}$ to $20\,\text{GeV}$.}
\figsetgrpend

\figsetgrpstart
\figsetgrpnum{3.31}
\figsetgrptitle{Flare 31 from CTA\,102}
\figsetplot{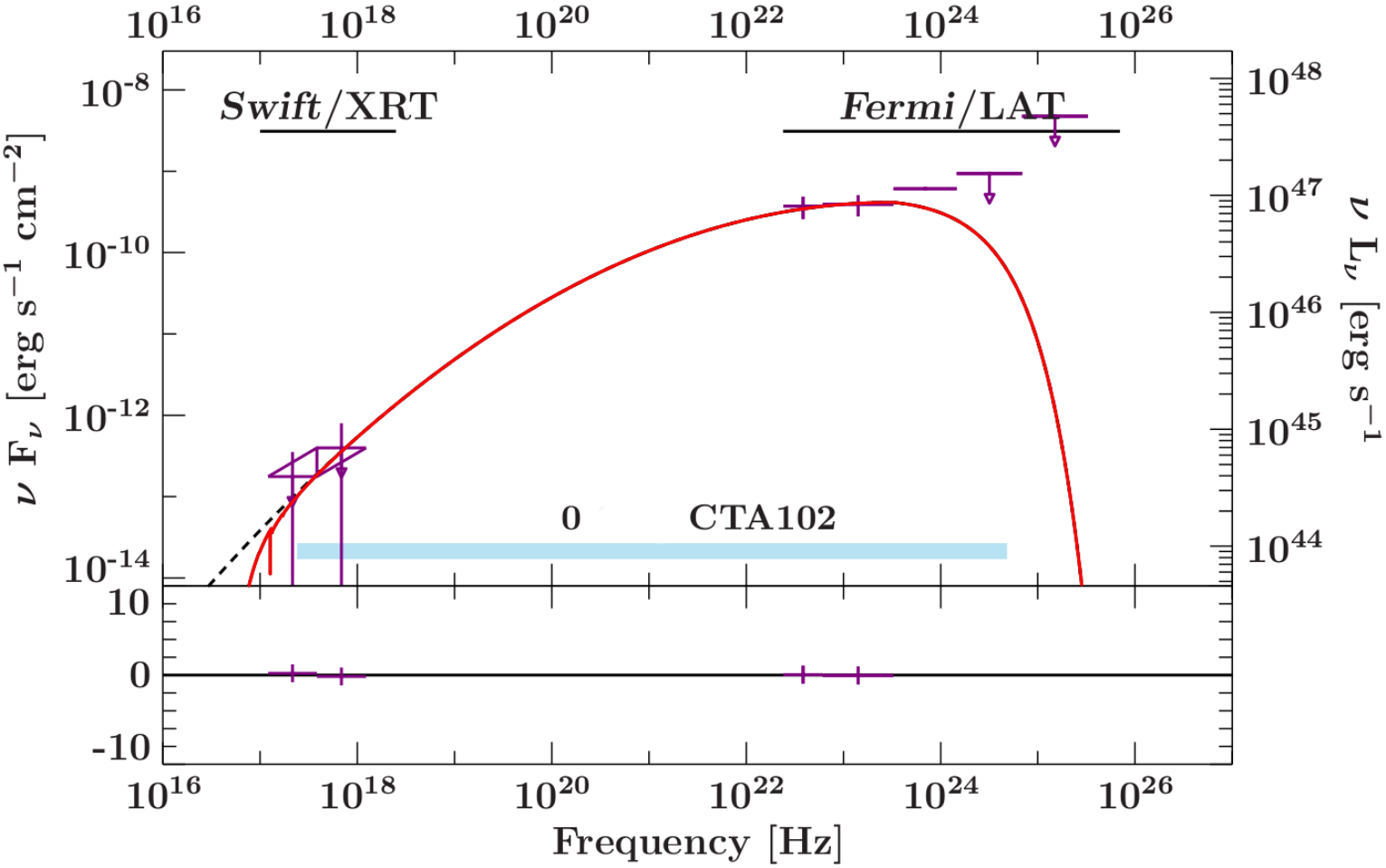}
\figsetgrpnote{Quasi-simultaneous SEDs of flares listed in Table\/\ref{Tab_neutrino_expectation}
. The red line indicates a log-parabola parametrization, including intrinsic absorption in X-rays and an exponential cut-off. The dashed black line shows the same parametrization without the absorption. The shadowed      area represents the integration range from $10\,\text{keV}$ to $20\,\text{GeV}$.}
\figsetgrpend

\figsetgrpstart
\figsetgrpnum{3.32}
\figsetgrptitle{Flare 32 from 3C\,279}
\figsetplot{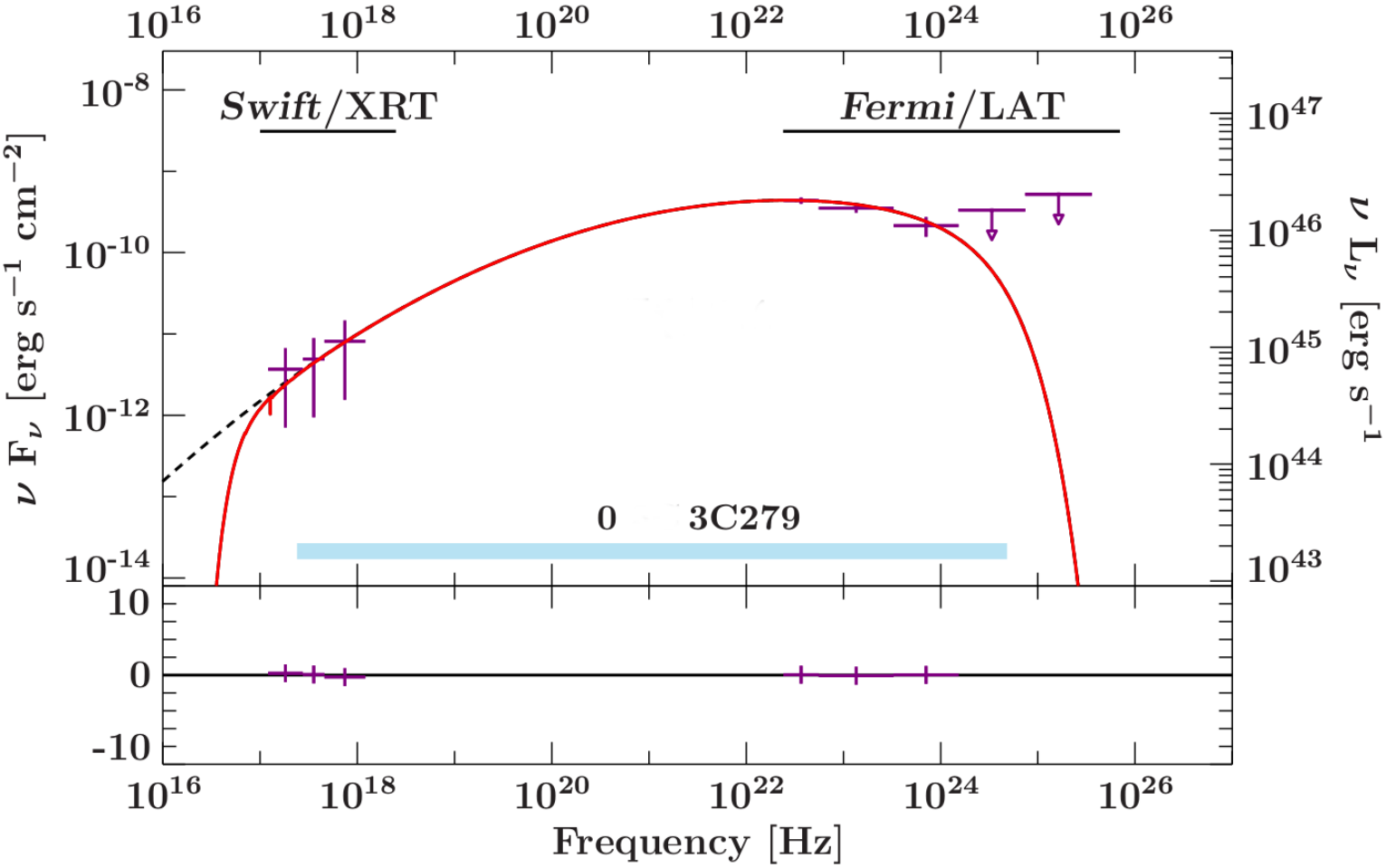}
\figsetgrpnote{Quasi-simultaneous SEDs of flares listed in Table\/\ref{Tab_neutrino_expectation}
. The red line indicates a log-parabola parametrization, including intrinsic absorption in X-rays and an exponential cut-off. The dashed black line shows the same parametrization without the absorption. The shadowed      area represents the integration range from $10\,\text{keV}$ to $20\,\text{GeV}$.}
\figsetgrpend

\figsetgrpstart
\figsetgrpnum{3.33}
\figsetgrptitle{Flare 33 from PKS\,1510$-$089}
\figsetplot{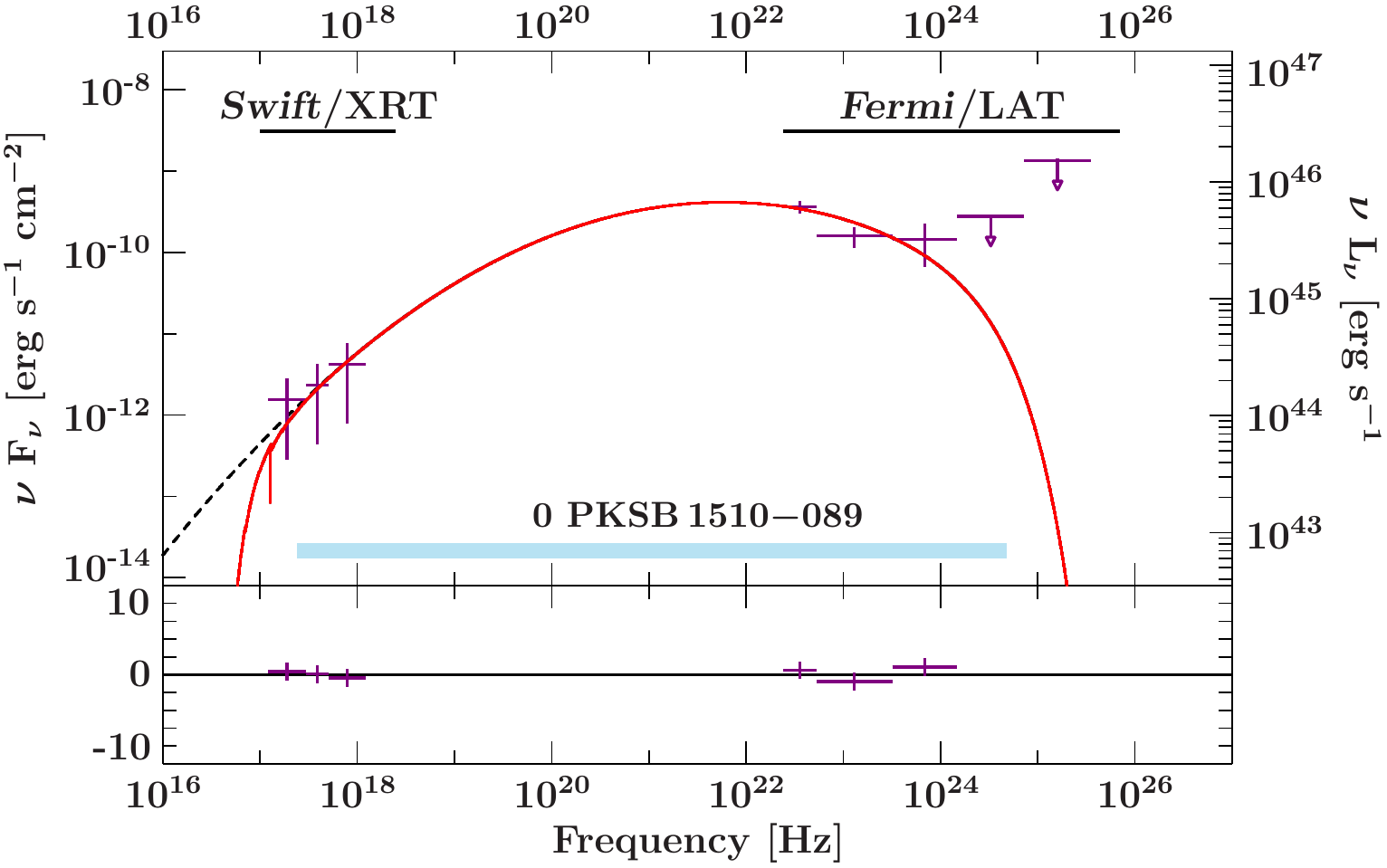}
\figsetgrpnote{Quasi-simultaneous SEDs of flares listed in Table\/\ref{Tab_neutrino_expectation}
. The red line indicates a log-parabola parametrization, including intrinsic absorption in X-rays and an exponential cut-off. The dashed black line shows the same parametrization without the absorption. The shadowed      area represents the integration range from $10\,\text{keV}$ to $20\,\text{GeV}$.}
\figsetgrpend

\figsetgrpstart
\figsetgrpnum{3.34}
\figsetgrptitle{Flare 34 from 3C\,279}
\figsetplot{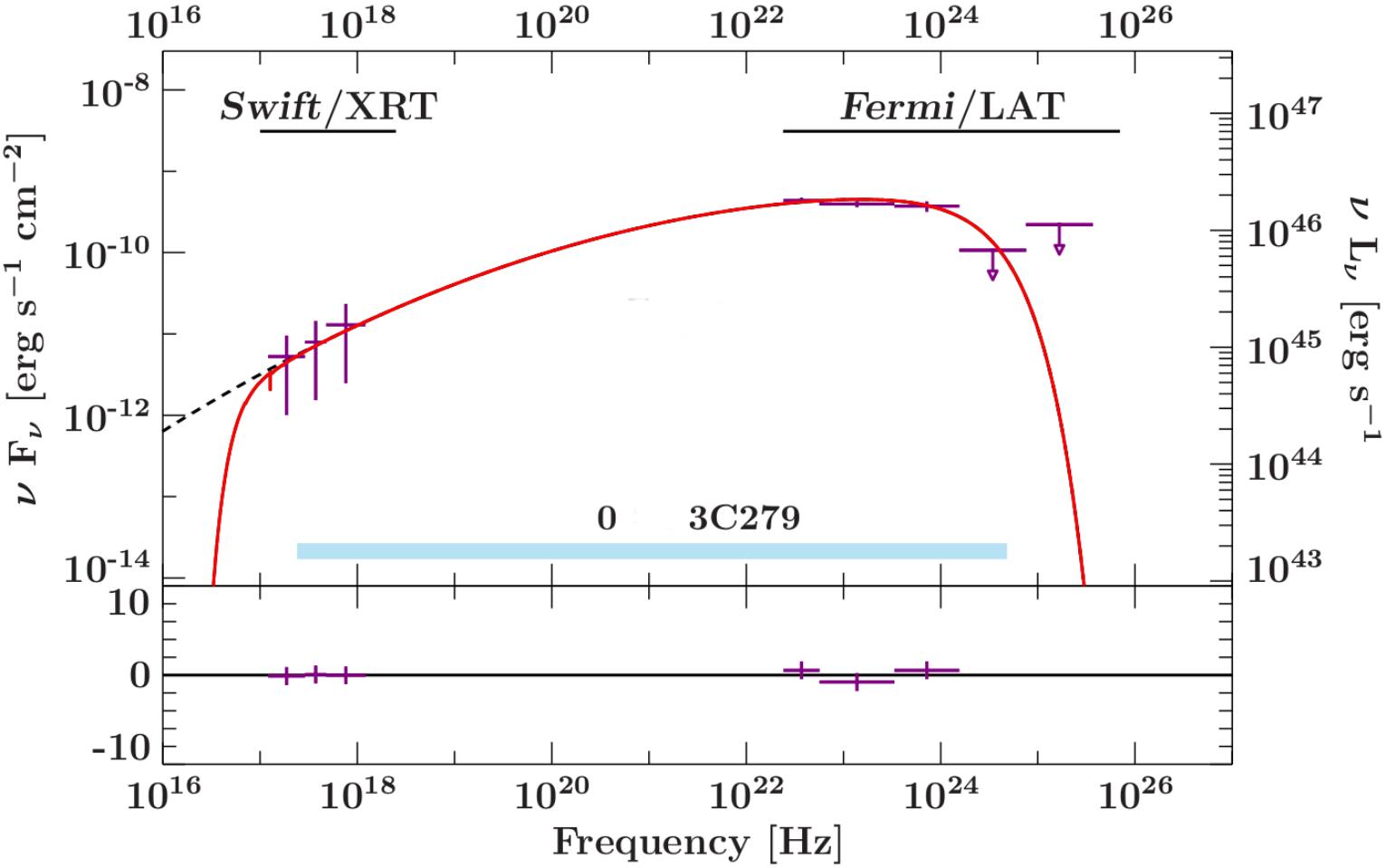}
\figsetgrpnote{Quasi-simultaneous SEDs of flares listed in Table\/\ref{Tab_neutrino_expectation}
. The red line indicates a log-parabola parametrization, including intrinsic absorption in X-rays and an exponential cut-off. The dashed black line shows the same parametrization without the absorption. The shadowed      area represents the integration range from $10\,\text{keV}$ to $20\,\text{GeV}$.}
\figsetgrpend

\figsetgrpstart
\figsetgrpnum{3.35}
\figsetgrptitle{Flare 35 from 3C\,279}
\figsetplot{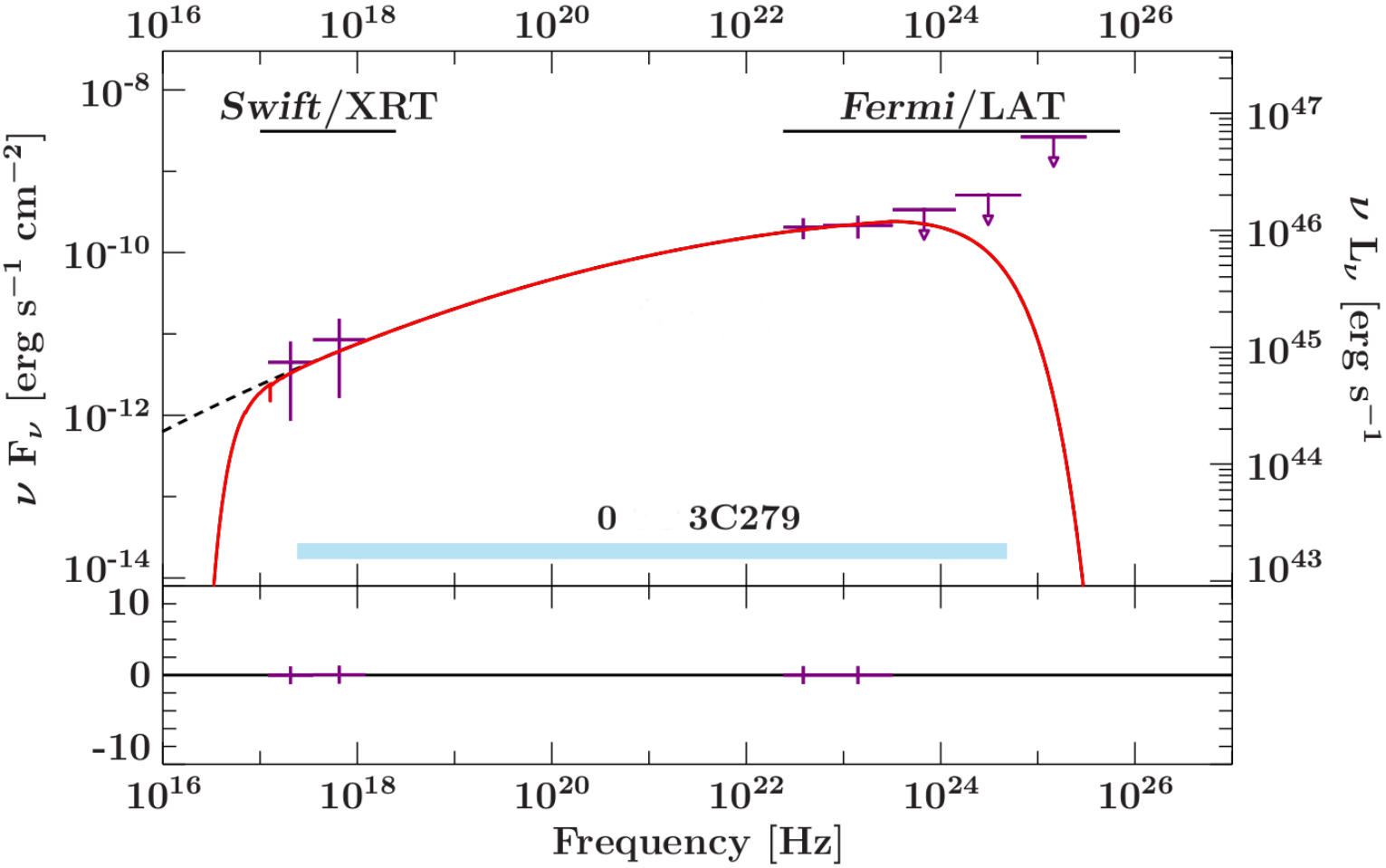}
\figsetgrpnote{Quasi-simultaneous SEDs of flares listed in Table\/\ref{Tab_neutrino_expectation}
. The red line indicates a log-parabola parametrization, including intrinsic absorption in X-rays and an exponential cut-off. The dashed black line shows the same parametrization without the absorption. The shadowed      area represents the integration range from $10\,\text{keV}$ to $20\,\text{GeV}$.}
\figsetgrpend

\figsetgrpstart
\figsetgrpnum{3.36}
\figsetgrptitle{Flare 36 from 3C\,279}
\figsetplot{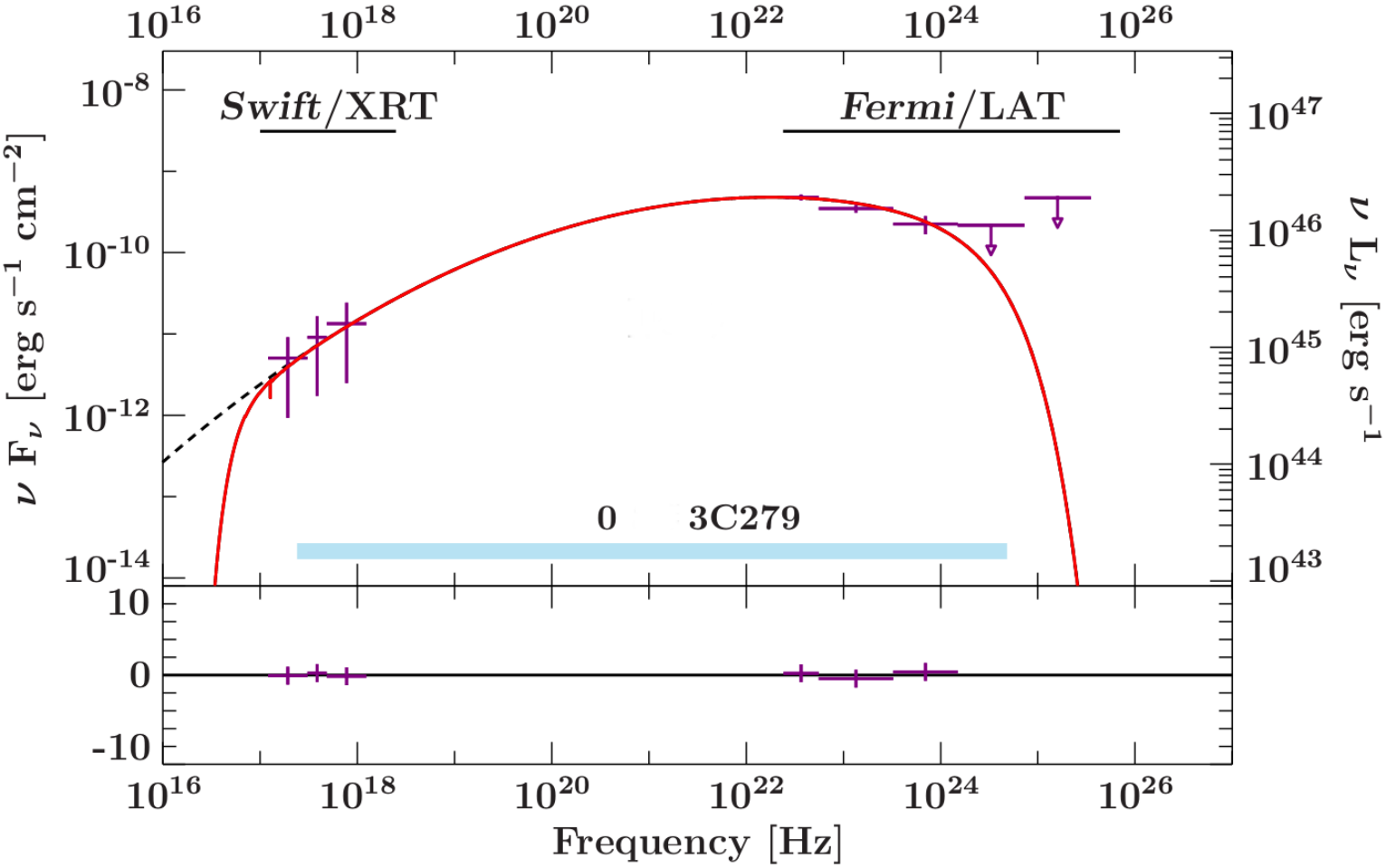}
\figsetgrpnote{Quasi-simultaneous SEDs of flares listed in Table\/\ref{Tab_neutrino_expectation}
. The red line indicates a log-parabola parametrization, including intrinsic absorption in X-rays and an exponential cut-off. The dashed black line shows the same parametrization without the absorption. The shadowed      area represents the integration range from $10\,\text{keV}$ to $20\,\text{GeV}$.}
\figsetgrpend

\figsetgrpstart
\figsetgrpnum{3.37}
\figsetgrptitle{Flare 37 from PKS\,1510$-$089}
\figsetplot{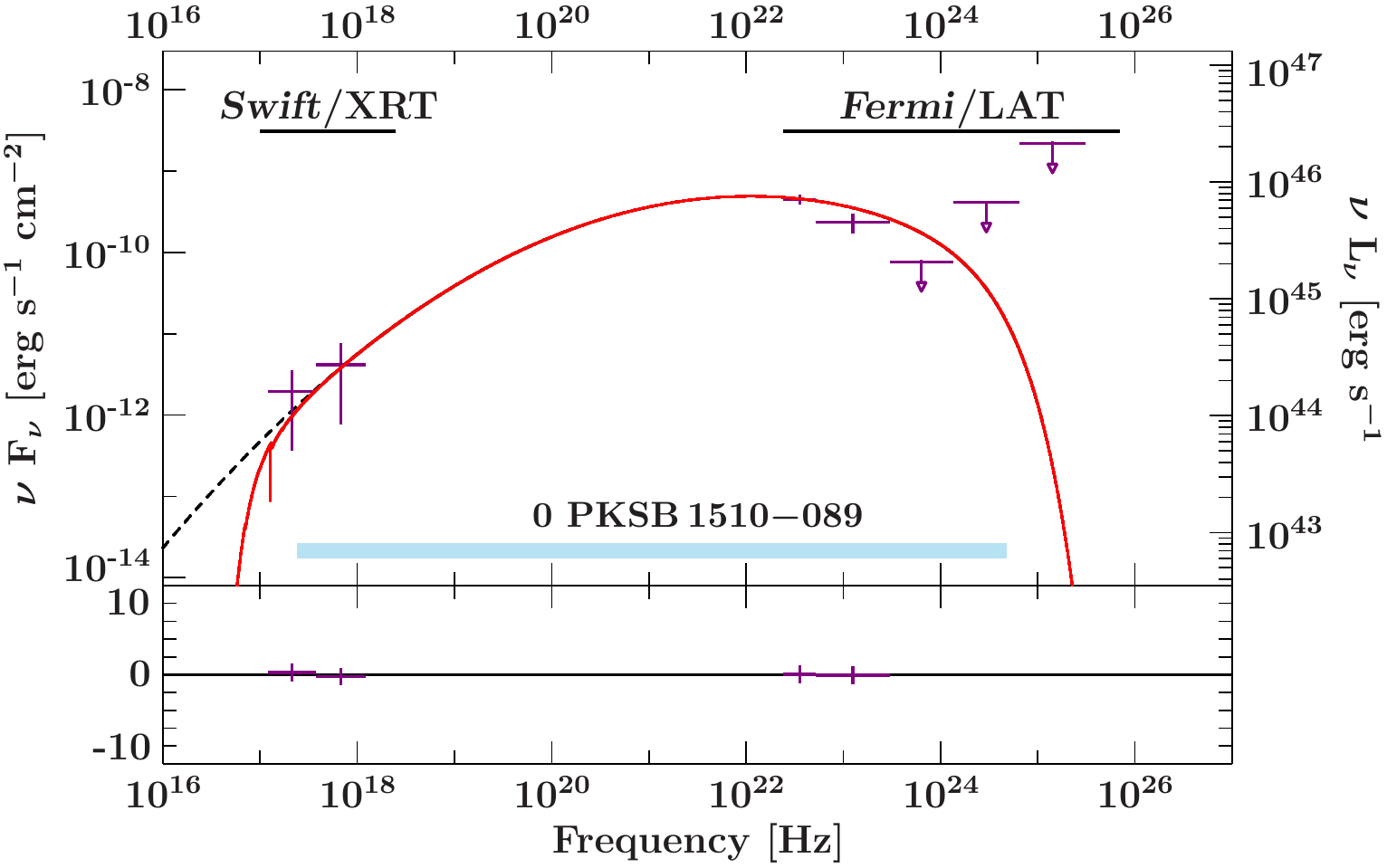}
\figsetgrpnote{Quasi-simultaneous SEDs of flares listed in Table\/\ref{Tab_neutrino_expectation}
. The red line indicates a log-parabola parametrization, including intrinsic absorption in X-rays and an exponential cut-off. The dashed black line shows the same parametrization without the absorption. The shadowed      area represents the integration range from $10\,\text{keV}$ to $20\,\text{GeV}$.}
\figsetgrpend

\figsetgrpstart
\figsetgrpnum{3.38}
\figsetgrptitle{Flare 38 from 3C\,454.3}
\figsetplot{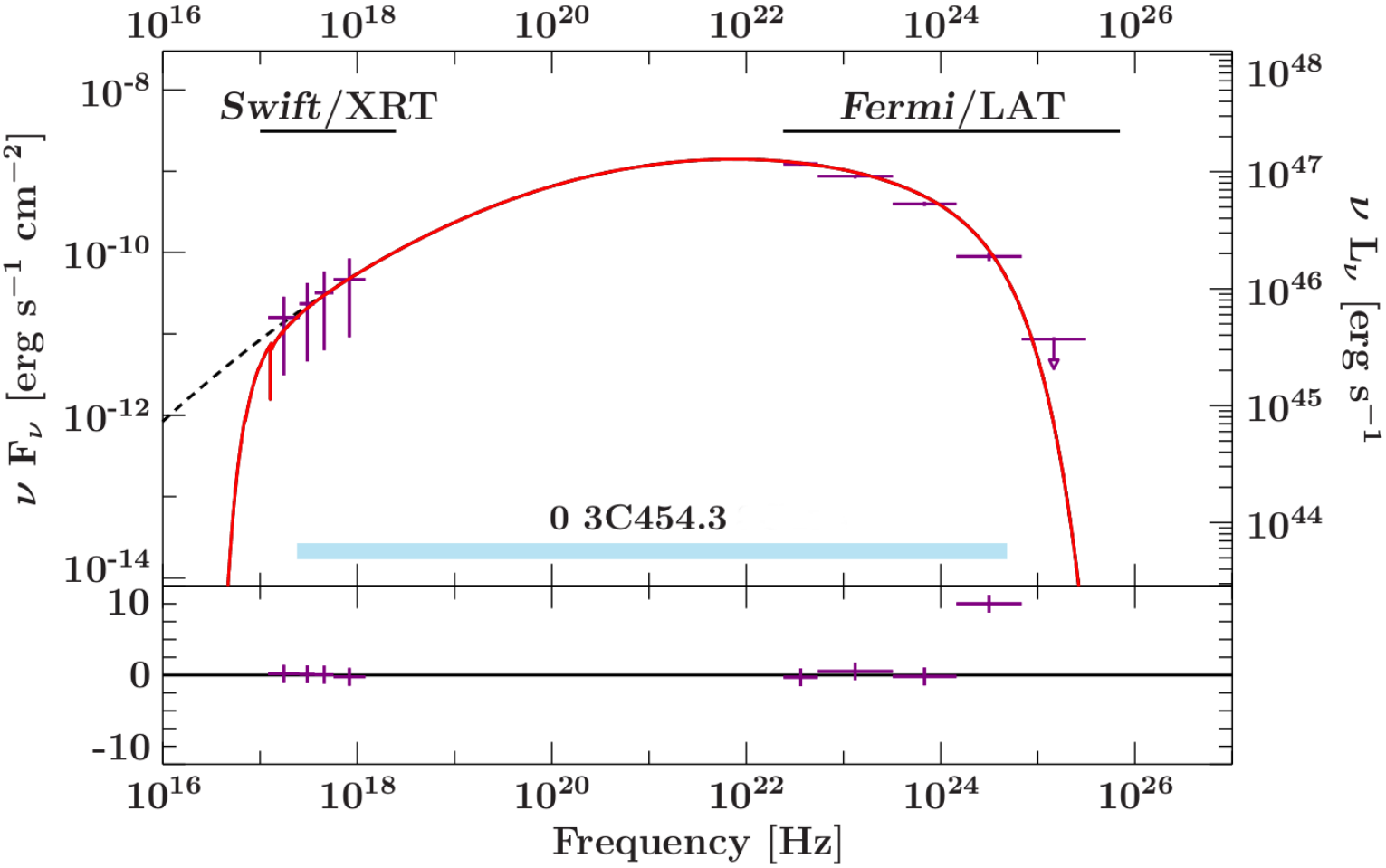}
\figsetgrpnote{Quasi-simultaneous SEDs of flares listed in Table\/\ref{Tab_neutrino_expectation}
. The red line indicates a log-parabola parametrization, including intrinsic absorption in X-rays and an exponential cut-off. The dashed black line shows the same parametrization without the absorption. The shadowed      area represents the integration range from $10\,\text{keV}$ to $20\,\text{GeV}$.}
\figsetgrpend

\figsetgrpstart
\figsetgrpnum{3.39}
\figsetgrptitle{Flare 39 from PKS\,1510$-$089}
\figsetplot{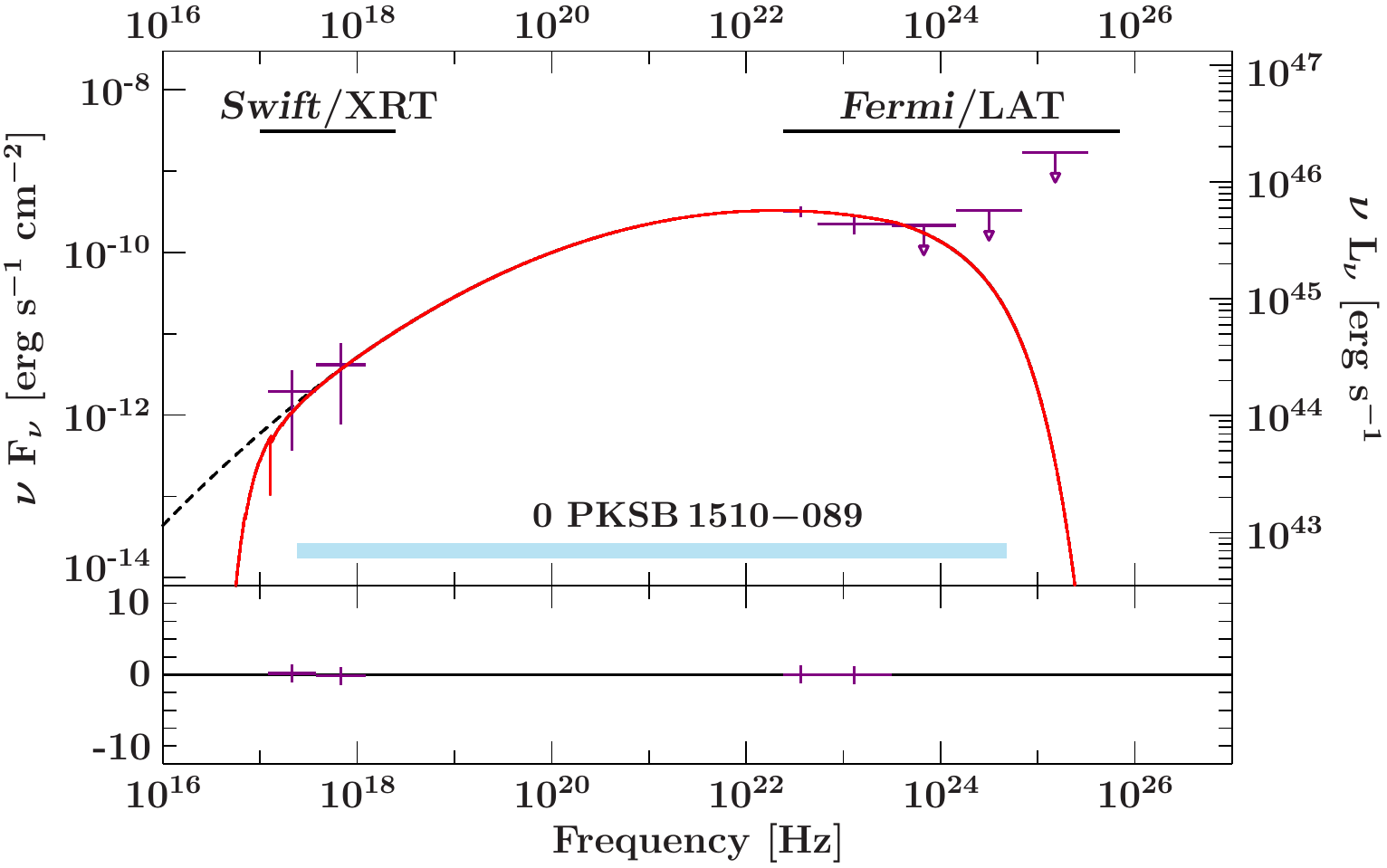}
\figsetgrpnote{Quasi-simultaneous SEDs of flares listed in Table\/\ref{Tab_neutrino_expectation}
. The red line indicates a log-parabola parametrization, including intrinsic absorption in X-rays and an exponential cut-off. The dashed black line shows the same parametrization without the absorption. The shadowed      area represents the integration range from $10\,\text{keV}$ to $20\,\text{GeV}$.}
\figsetgrpend

\figsetgrpstart
\figsetgrpnum{3.40}
\figsetgrptitle{Flare 40 from PKS\,1510$-$089}
\figsetplot{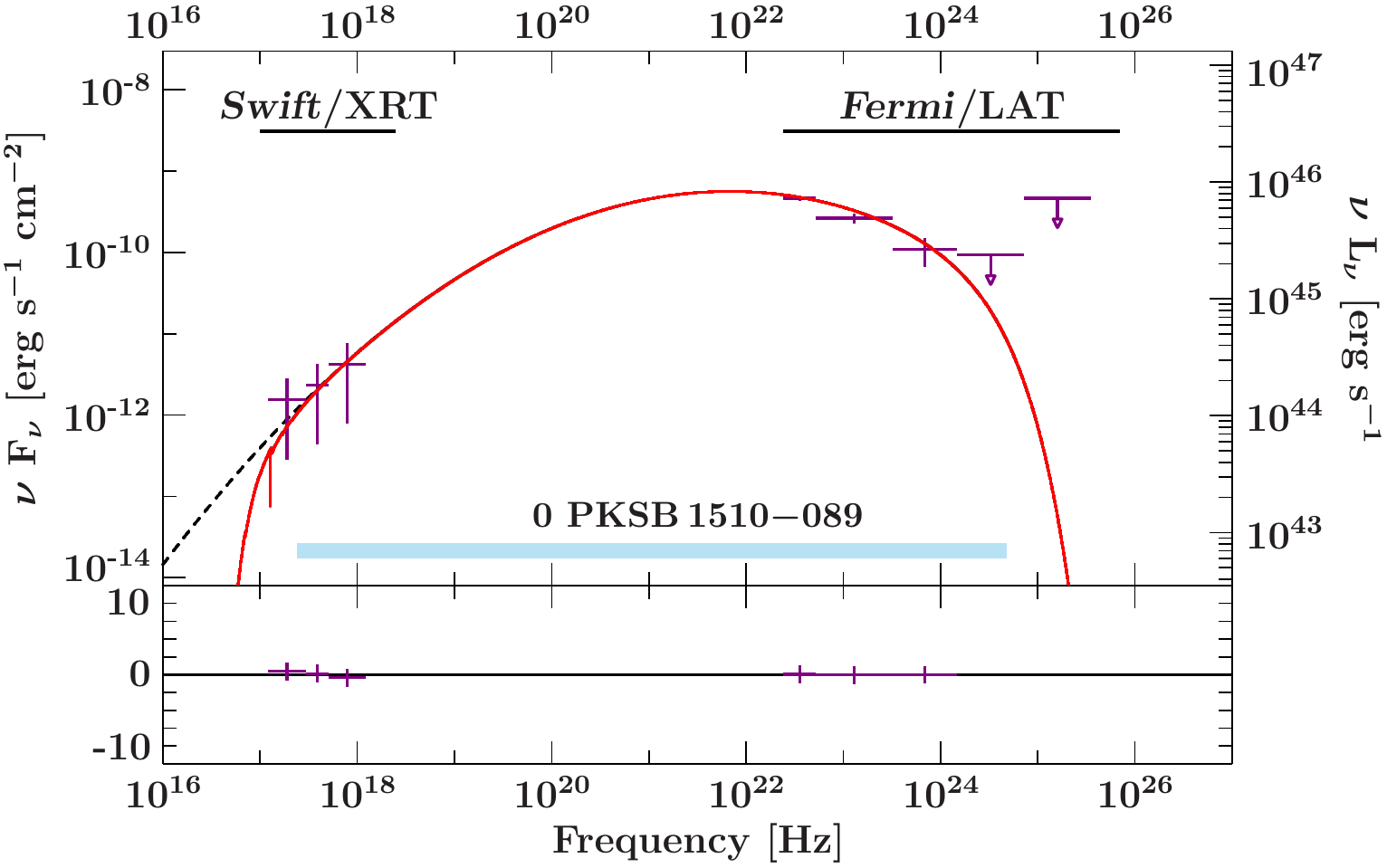}
\figsetgrpnote{Quasi-simultaneous SEDs of flares listed in Table\/\ref{Tab_neutrino_expectation}
. The red line indicates a log-parabola parametrization, including intrinsic absorption in X-rays and an exponential cut-off. The dashed black line shows the same parametrization without the absorption. The shadowed      area represents the integration range from $10\,\text{keV}$ to $20\,\text{GeV}$.}
\figsetgrpend

\figsetgrpstart
\figsetgrpnum{3.41}
\figsetgrptitle{Flare 41 from PKS\,1424$-$418}
\figsetplot{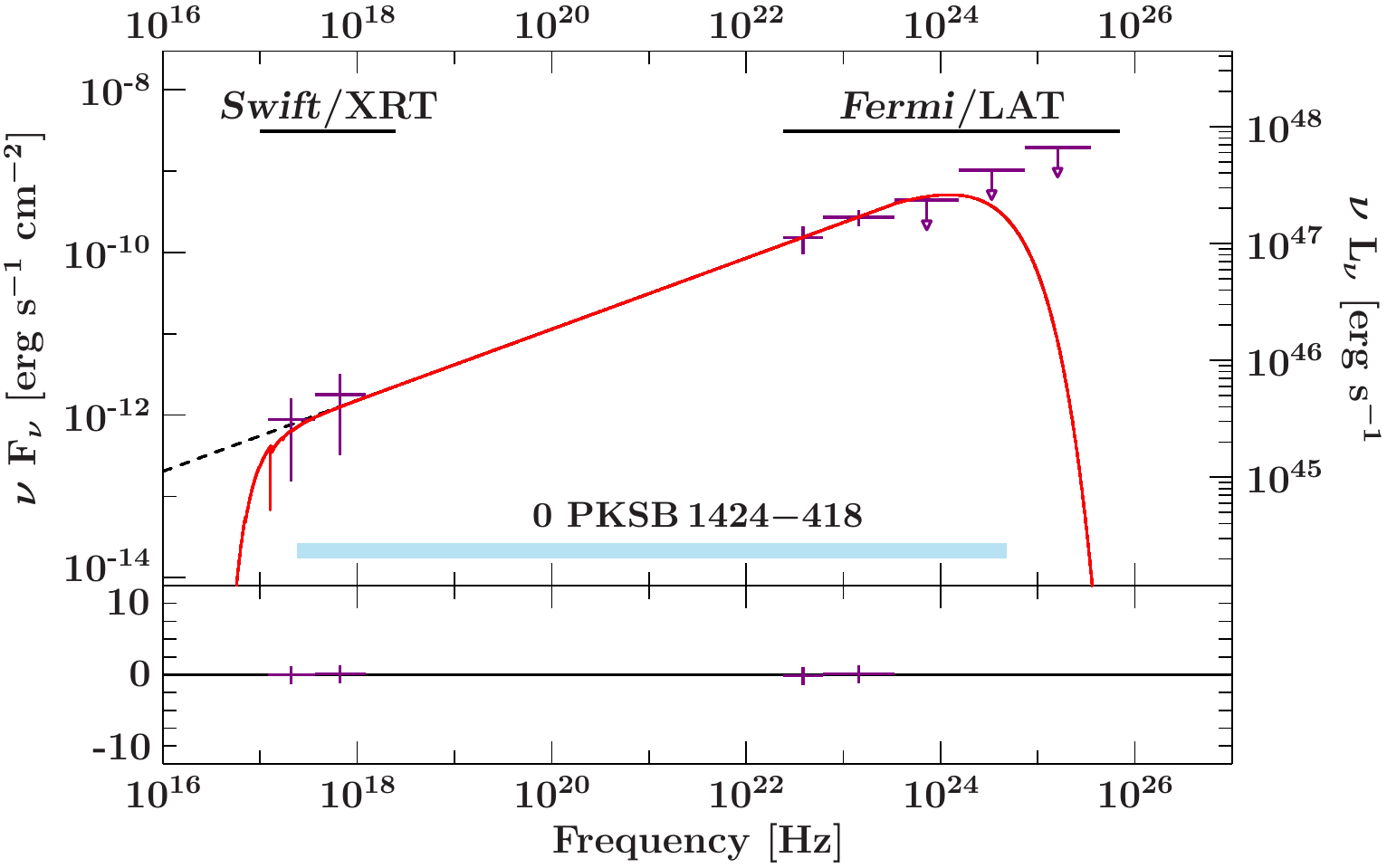}
\figsetgrpnote{Quasi-simultaneous SEDs of flares listed in Table\/\ref{Tab_neutrino_expectation}
. The red line indicates a log-parabola parametrization, including intrinsic absorption in X-rays and an exponential cut-off. The dashed black line shows the same parametrization without the absorption. The shadowed      area represents the integration range from $10\,\text{keV}$ to $20\,\text{GeV}$.}
\figsetgrpend

\figsetgrpstart
\figsetgrpnum{3.42}
\figsetgrptitle{Flare 42 from 3C\,279}
\figsetplot{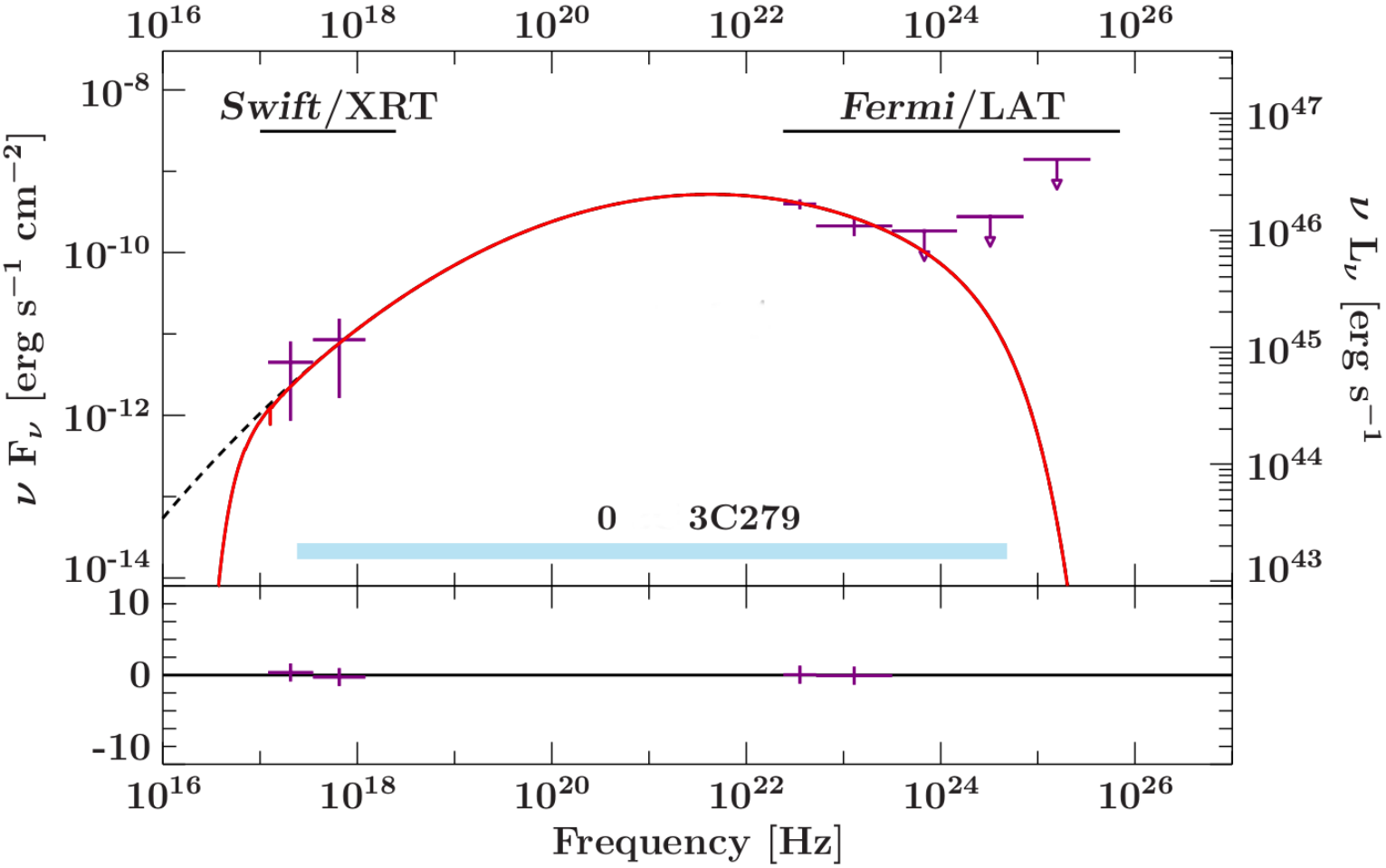}
\figsetgrpnote{Quasi-simultaneous SEDs of flares listed in Table\/\ref{Tab_neutrino_expectation}
. The red line indicates a log-parabola parametrization, including intrinsic absorption in X-rays and an exponential cut-off. The dashed black line shows the same parametrization without the absorption. The shadowed      area represents the integration range from $10\,\text{keV}$ to $20\,\text{GeV}$.}
\figsetgrpend

\figsetgrpstart
\figsetgrpnum{3.43}
\figsetgrptitle{Flare 43 from PKS\,1424$-$418}
\figsetplot{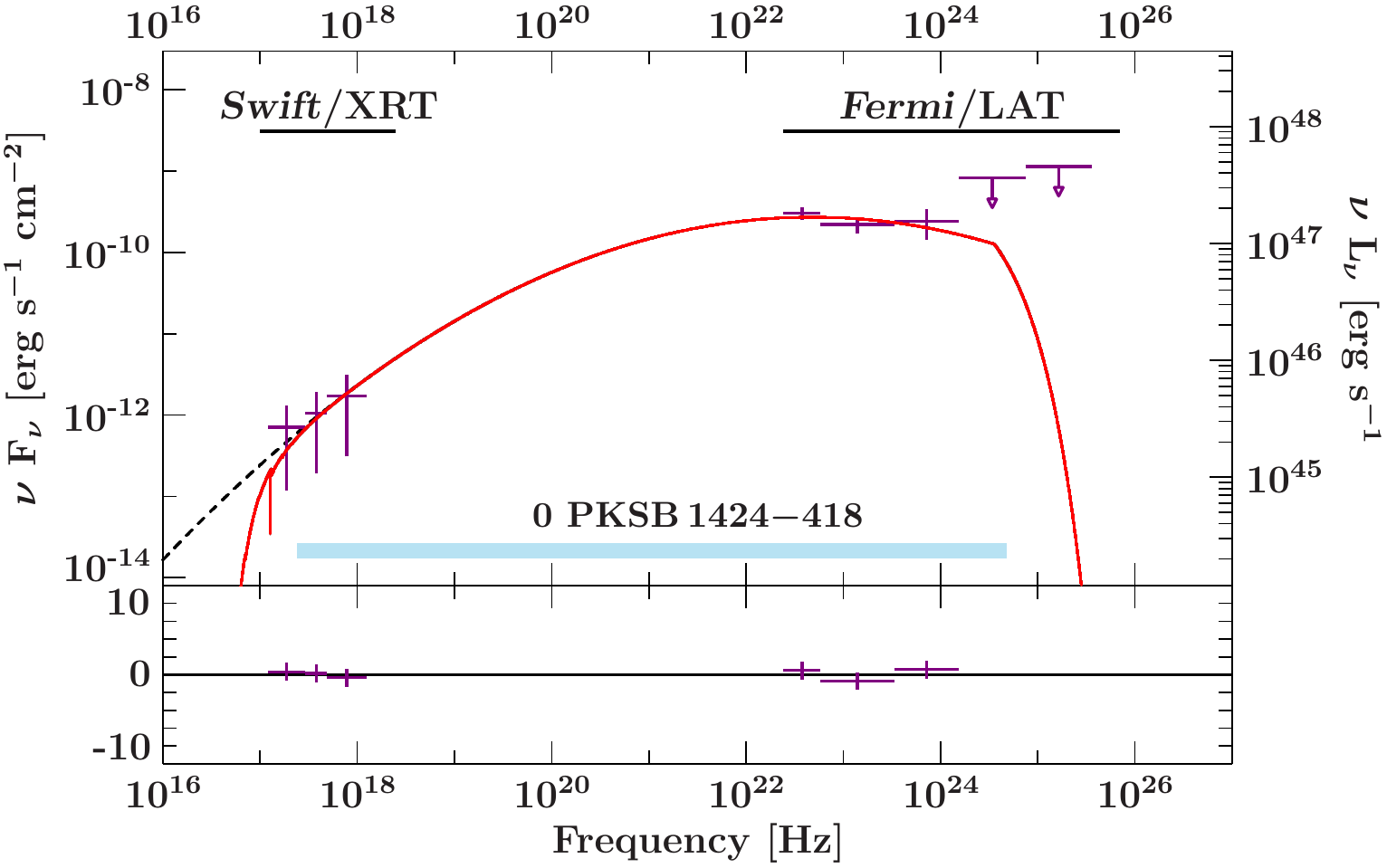}
\figsetgrpnote{Quasi-simultaneous SEDs of flares listed in Table\/\ref{Tab_neutrino_expectation}
. The red line indicates a log-parabola parametrization, including intrinsic absorption in X-rays and an exponential cut-off. The dashed black line shows the same parametrization without the absorption. The shadowed      area represents the integration range from $10\,\text{keV}$ to $20\,\text{GeV}$.}
\figsetgrpend

\figsetgrpstart
\figsetgrpnum{3.44}
\figsetgrptitle{Flare 44 from PKS\,1510$-$089}
\figsetplot{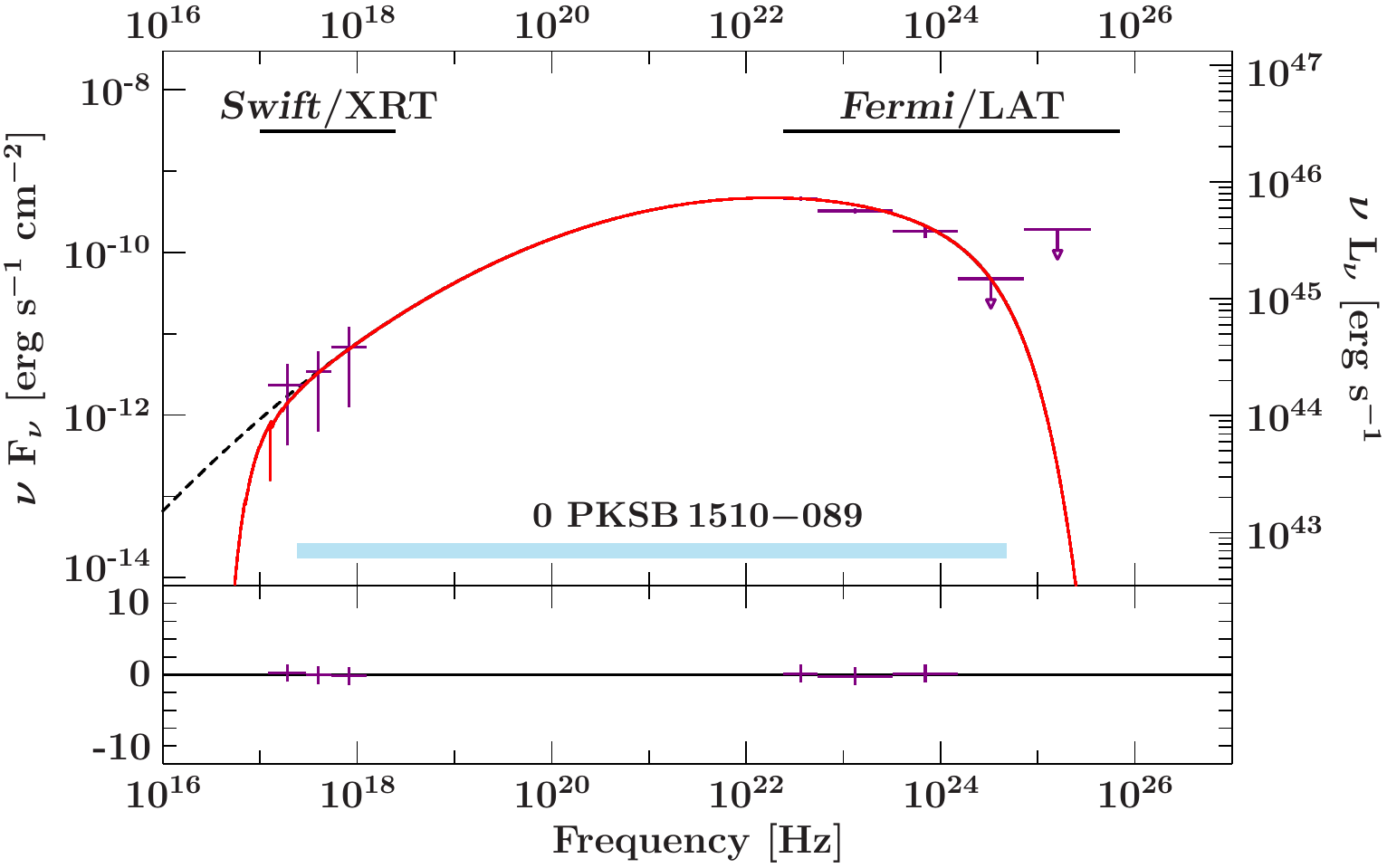}
\figsetgrpnote{Quasi-simultaneous SEDs of flares listed in Table\/\ref{Tab_neutrino_expectation}
. The red line indicates a log-parabola parametrization, including intrinsic absorption in X-rays and an exponential cut-off. The dashed black line shows the same parametrization without the absorption. The shadowed      area represents the integration range from $10\,\text{keV}$ to $20\,\text{GeV}$.}
\figsetgrpend

\figsetgrpstart
\figsetgrpnum{3.45}
\figsetgrptitle{Flare 45 from PKS\,1510$-$089}
\figsetplot{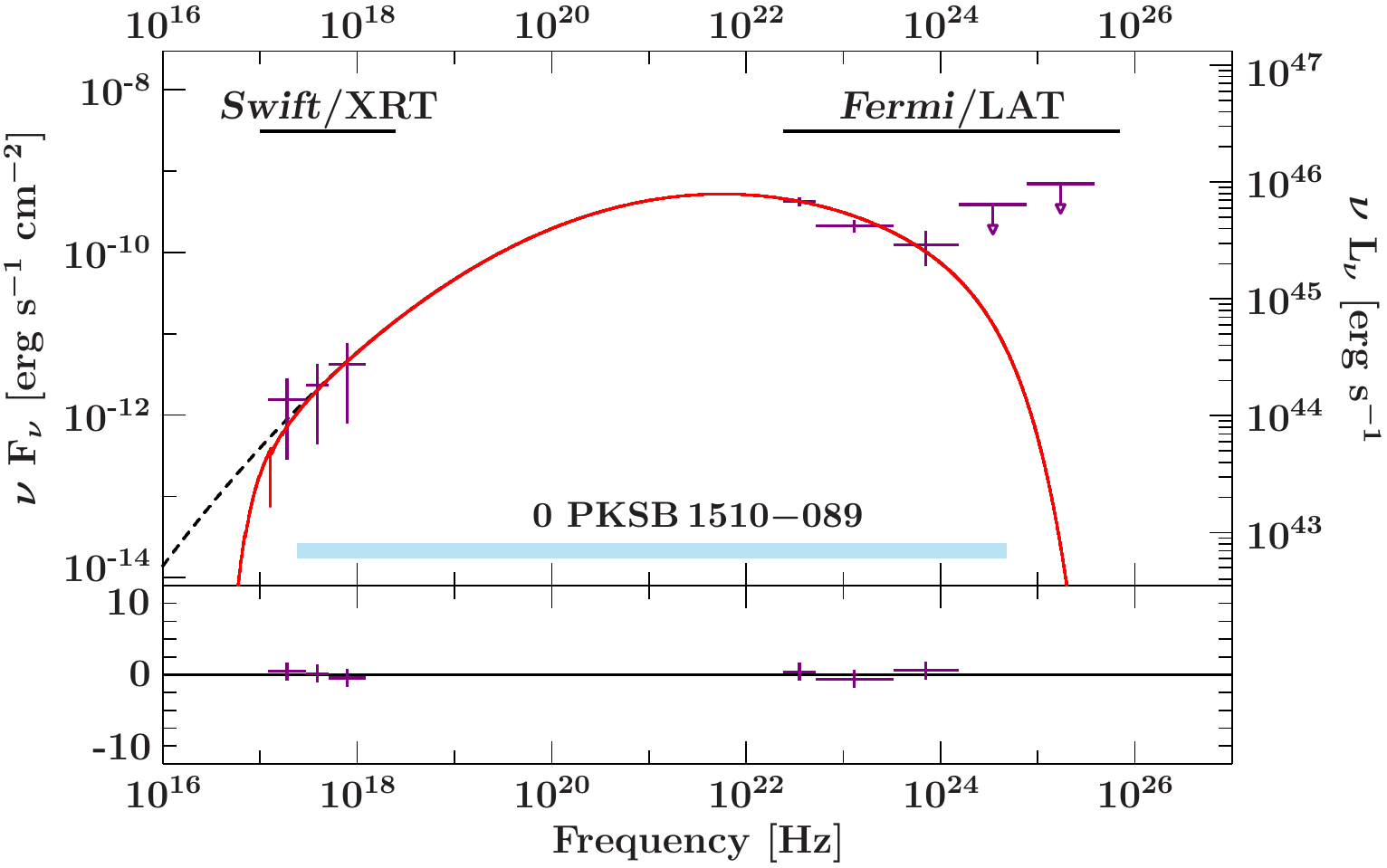}
\figsetgrpnote{Quasi-simultaneous SEDs of flares listed in Table\/\ref{Tab_neutrino_expectation}
. The red line indicates a log-parabola parametrization, including intrinsic absorption in X-rays and an exponential cut-off. The dashed black line shows the same parametrization without the absorption. The shadowed      area represents the integration range from $10\,\text{keV}$ to $20\,\text{GeV}$.}
\figsetgrpend

\figsetgrpstart
\figsetgrpnum{3.46}
\figsetgrptitle{Flare 46 from PKS\,1510$-$089}
\figsetplot{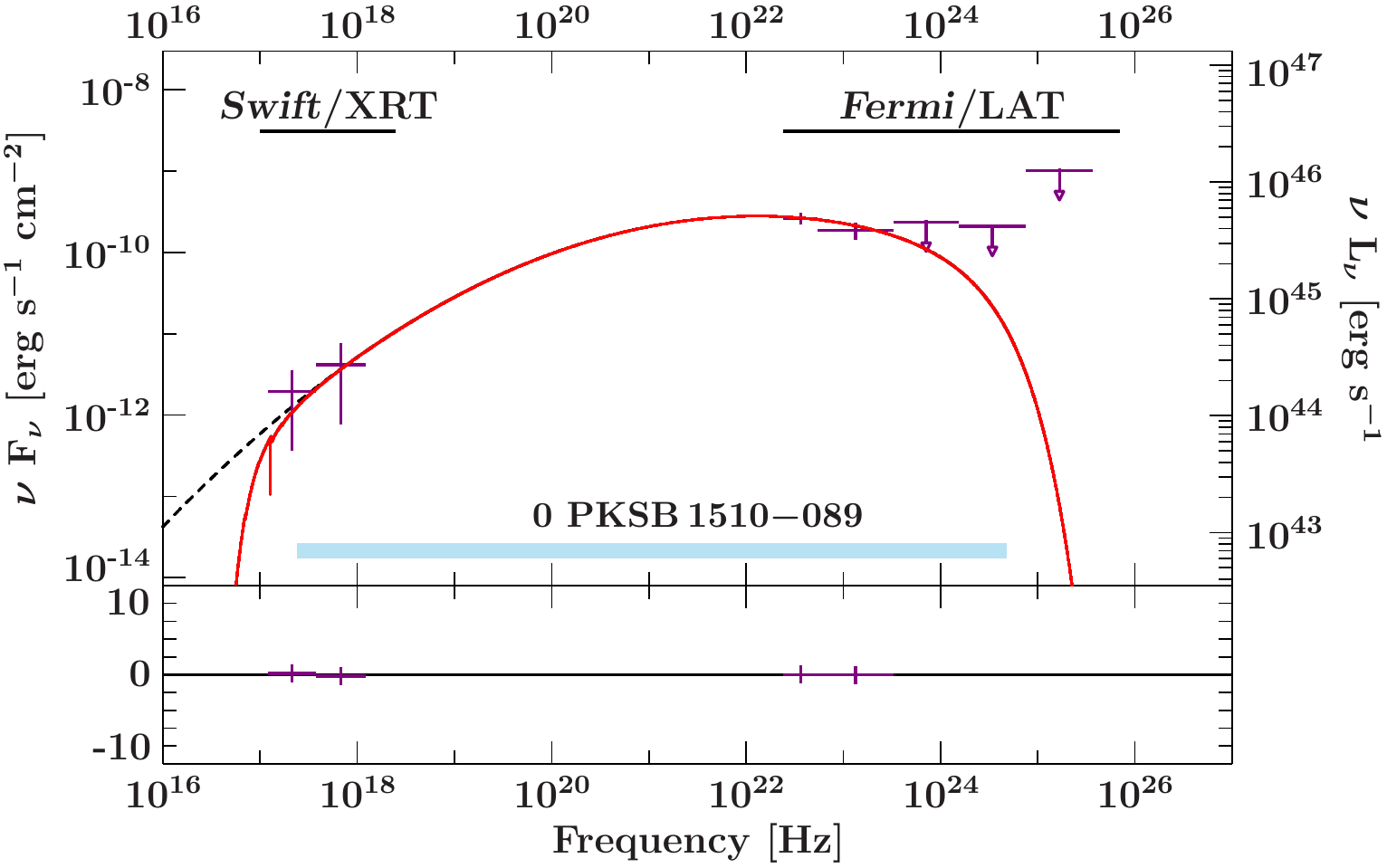}
\figsetgrpnote{Quasi-simultaneous SEDs of flares listed in Table\/\ref{Tab_neutrino_expectation}
. The red line indicates a log-parabola parametrization, including intrinsic absorption in X-rays and an exponential cut-off. The dashed black line shows the same parametrization without the absorption. The shadowed      area represents the integration range from $10\,\text{keV}$ to $20\,\text{GeV}$.}
\figsetgrpend

\figsetgrpstart
\figsetgrpnum{3.47}
\figsetgrptitle{Flare 47 from PKS\,1329$-$049}
\figsetplot{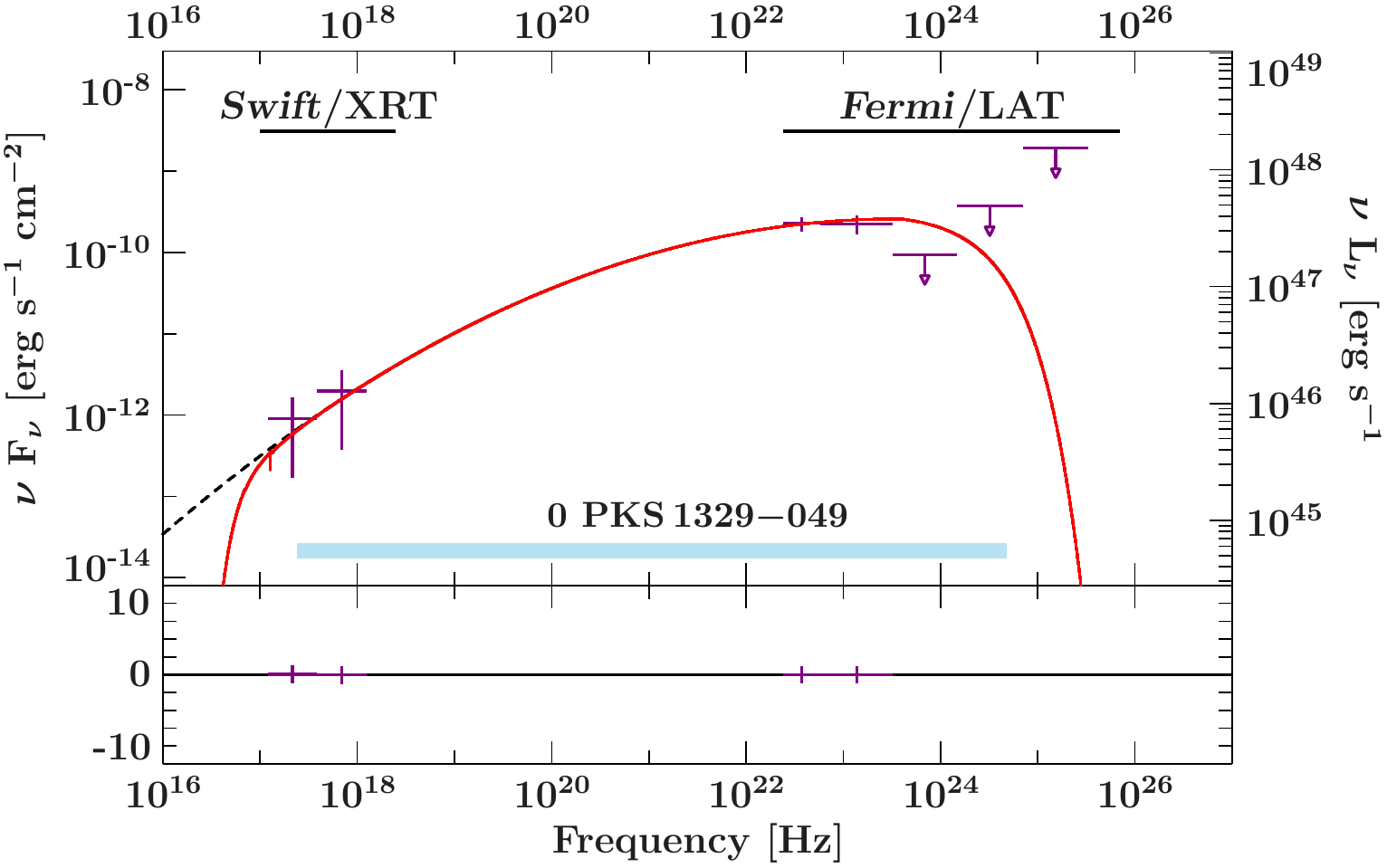}
\figsetgrpnote{Quasi-simultaneous SEDs of flares listed in Table\/\ref{Tab_neutrino_expectation}
. The red line indicates a log-parabola parametrization, including intrinsic absorption in X-rays and an exponential cut-off. The dashed black line shows the same parametrization without the absorption. The shadowed      area represents the integration range from $10\,\text{keV}$ to $20\,\text{GeV}$.}
\figsetgrpend

\figsetgrpstart
\figsetgrpnum{3.48}
\figsetgrptitle{Flare 48 from PKS\,1329$-$049}
\figsetplot{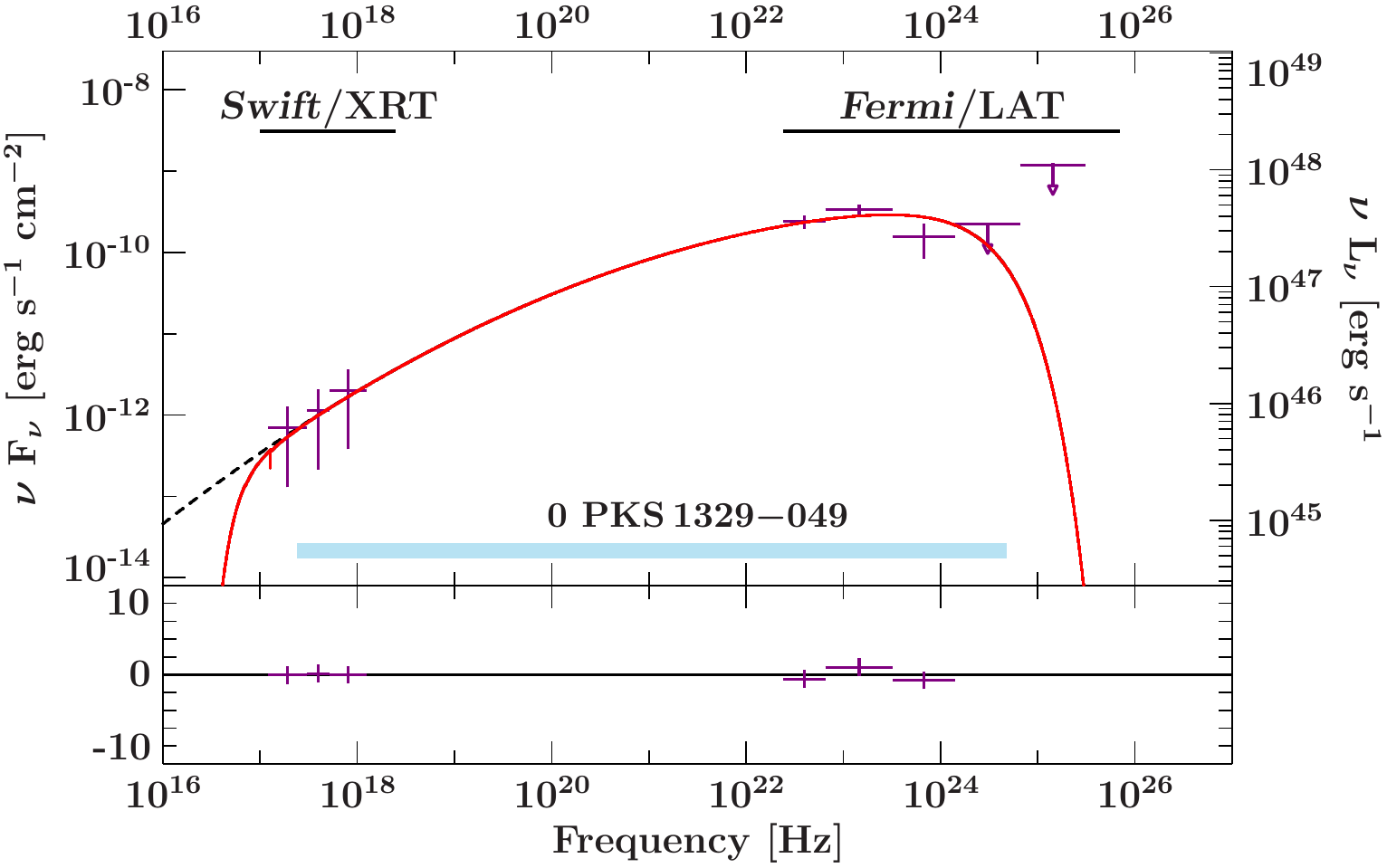}
\figsetgrpnote{Quasi-simultaneous SEDs of flares listed in Table\/\ref{Tab_neutrino_expectation}
. The red line indicates a log-parabola parametrization, including intrinsic absorption in X-rays and an exponential cut-off. The dashed black line shows the same parametrization without the absorption. The shadowed      area represents the integration range from $10\,\text{keV}$ to $20\,\text{GeV}$.}
\figsetgrpend

\figsetgrpstart
\figsetgrpnum{3.49}
\figsetgrptitle{Flare 49 from PKS\,1424$-$418}
\figsetplot{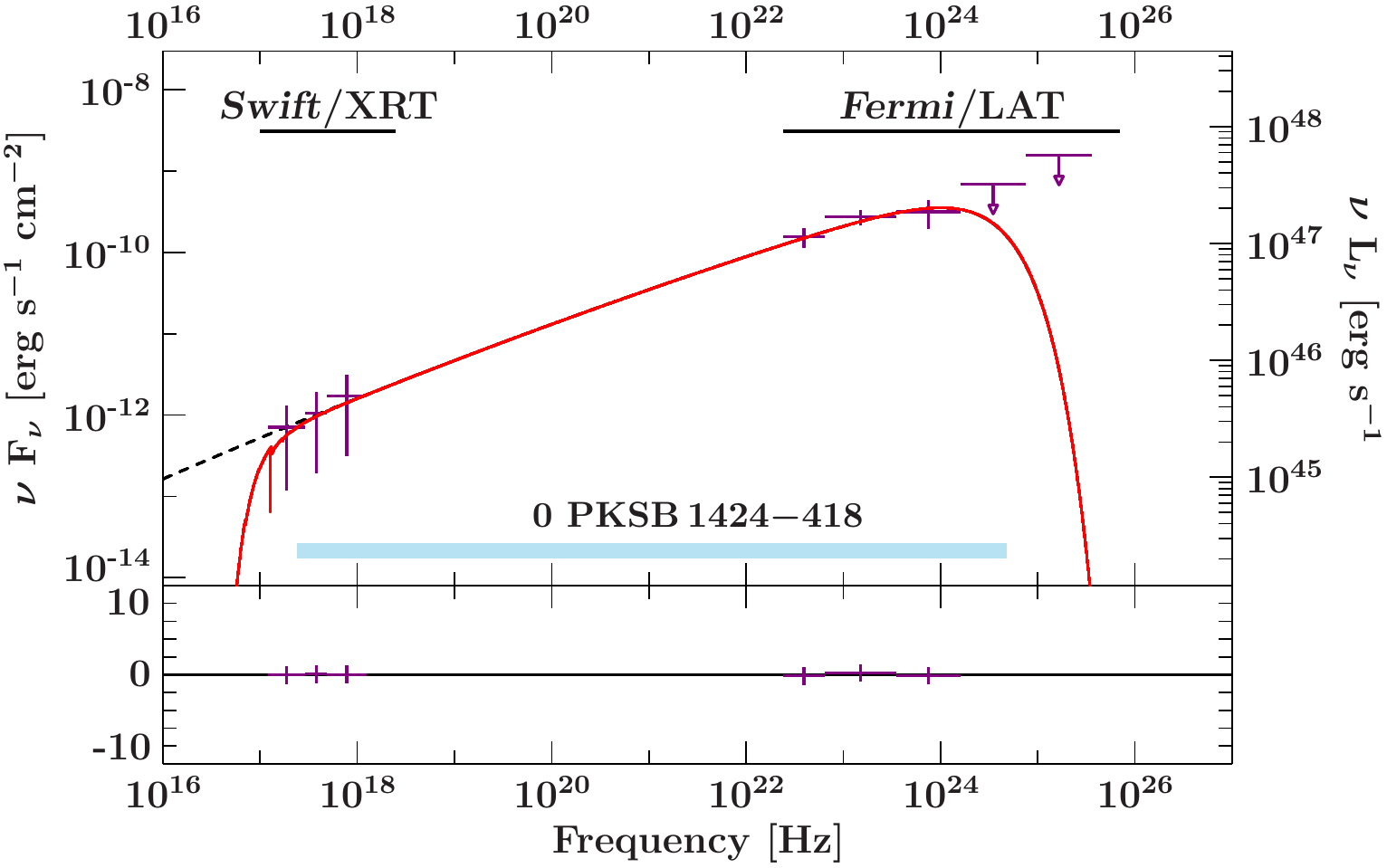}
\figsetgrpnote{Quasi-simultaneous SEDs of flares listed in Table\/\ref{Tab_neutrino_expectation}
. The red line indicates a log-parabola parametrization, including intrinsic absorption in X-rays and an exponential cut-off. The dashed black line shows the same parametrization without the absorption. The shadowed      area represents the integration range from $10\,\text{keV}$ to $20\,\text{GeV}$.}
\figsetgrpend

\figsetgrpstart
\figsetgrpnum{3.50}
\figsetgrptitle{Flare 50 from PKS\,1510$-$089}
\figsetplot{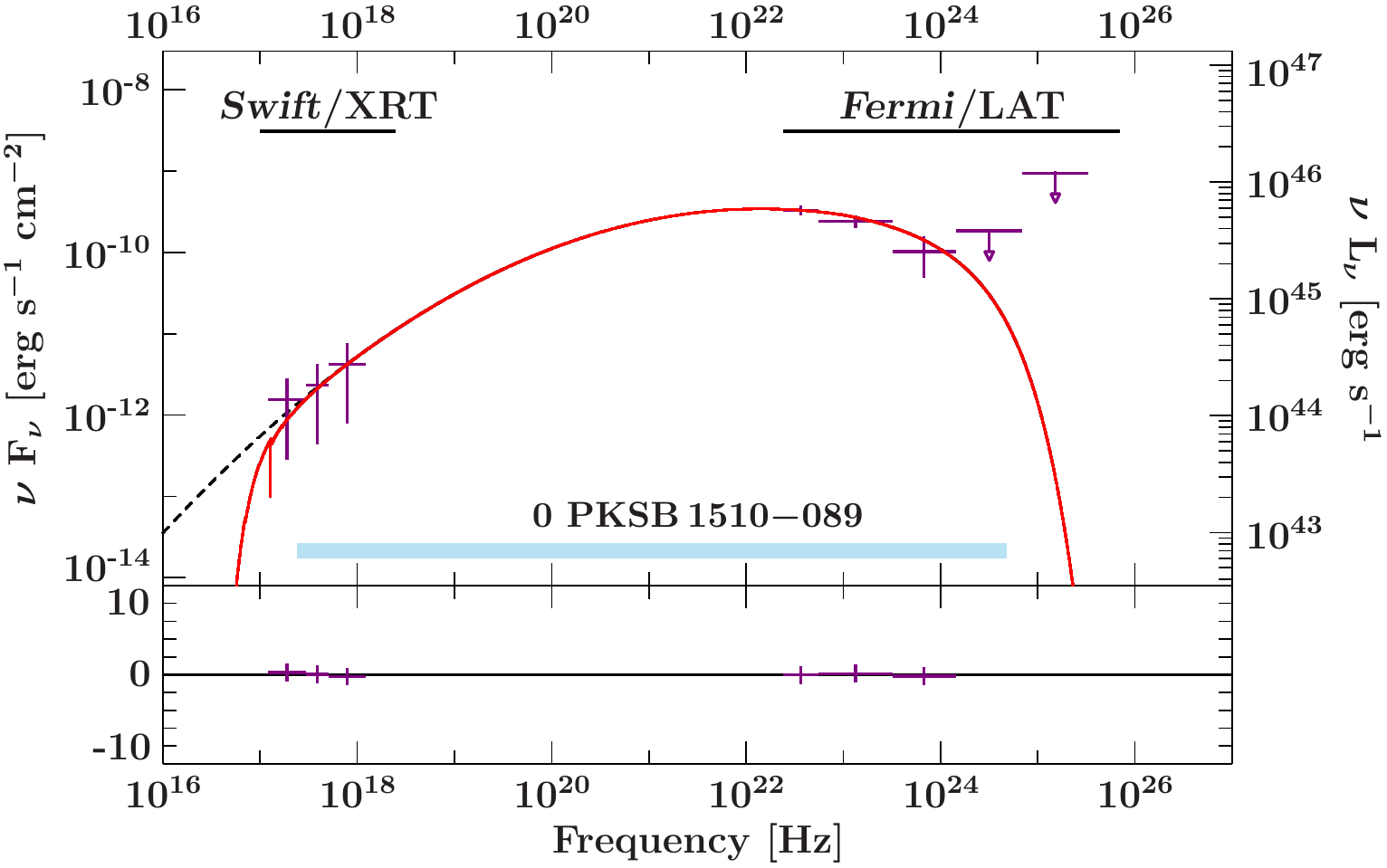}
\figsetgrpnote{Quasi-simultaneous SEDs of flares listed in Table\/\ref{Tab_neutrino_expectation}
. The red line indicates a log-parabola parametrization, including intrinsic absorption in X-rays and an exponential cut-off. The dashed black line shows the same parametrization without the absorption. The shadowed      area represents the integration range from $10\,\text{keV}$ to $20\,\text{GeV}$.}
\figsetgrpend

\section{Results}\label{Section Results}
\hspace{-13pt}
The presented flare identification is optimized to search for periods of successive light curve intervals above a source-specific threshold. Therefore, flare durations of only a few days dominate. The main motivation to choose such a flare identification is to calculate the neutrino prediction from the brightest short-term flares observed by \textit{Fermi}/LAT. Flares that show the highest $\gamma$-ray flux time-averaged over the duration of the flare 
\begin{align}
 \langle\Phi_{\gamma} \rangle = {\dfrac{1}{\Delta t}} \int\limits_{0}^{\Delta t} \Phi_{\gamma} (t) dt 
 \label{averaged flux}
\end{align}
have been selected. Table\,\ref{Tab_neutrino_expectation} lists the results of the neutrino expectation for a subsample of 50 flares identified following Equ.\,\ref{averaged flux}.
Selected flares are highlighted in blue, while all other flares identified by Equ.\,\ref{flare_selection_1} are marked in gray. 
In order to compare the neutrino expectation of flares with various durations, it is 
necessary to normalize the neutrino predictions of individual flares according to their duration. The neutrino rate per day is given by:
\begin{align}
\langle \dot{N}_{\nu}^{\mathrm{flare}} \rangle = \frac{\mathrm{N}_{\nu}^{\mathrm{pred}}}{\Delta t}.
 \label{flare_significance}
\end{align}
While longer flares lead to higher absolute neutrino numbers (see Eq.\,\ref{N_max_formula}), the chance for a correlated detection of a high-energy neutrino and a $\gamma$-ray flare is expected to be higher for shorter flare durations.
The neutrino rate therefore allows us to compare the neutrino prediction for flares with various time spans.
While the majority of flares seem to have similar neutrino rates, 3C\,279 underwent a massive flare in 2015 \citep{Paliya_2015} that clearly dominates the sample by producing the highest time-averaged integrated $\gamma$-ray flux as well as the highest neutrino rate.
The two blazars 3C\,279 and PKS\,1510$-$089 alone account for a majority of 42 flares, while 
the 50 highest-ranked flares  are produced by a group of only 
seven different sources: 3C\,279 (10 flares), PKS\,1510$-$089 (32 flares), 
PKS\,0402$-$362 (1 flare), CTA\,102 (1 flare), 3C\,454.3 (\mbox{1 flare)}, PKS\,1424$-$418 (3 flares), 
and PKS\,1329$-$049 (2 flares).
Flare 38, caused by 3C\,454.3, provides the highest absolute neutrino prediction due to its brightness and duration of multiple months. A duration of more than 200 days is reached, as this flare accounts for one of the brightest and longest blazar outbursts ever observed by \textit{Fermi}/LAT. To reach such a long flare duration, not a single light curve interval is allowed to violate the selection criterion introduced in Sect.\,\ref{Flare_identification}.
Comparing this flare to the $``$BigBird$"$ event studied by \citet{Kadler_2016}, the chance for a correlated high-energy neutrino detection is also expected to be small, due to the long duration of this flare.
Comparing the neutrino rates, we see that in order to maximize the chance for an astrophysical neutrino detection, flares need to have a high integrated $\gamma$-ray energy flux on a relatively short time scale.

\begin{table*}[ptbh]
 \caption{Neutrino expectation for the 50 best-ranked flares, selected according to their (time-averaged) integrated $\gamma$-ray energy flux $\langle\Phi_{\gamma}\rangle$ (Equ.\,\ref{averaged flux}).
 $\text{N}_{\nu}^{\text{max}}$ (Equ.\,\ref{N_max_formula}) corresponds to the maximum number of neutrino events a flare would be able to create under the most optimistic conditions.
 $N_{\nu}^{\text{pred}}$ (Equ.\,\ref{N_pred_formula}) corresponds to the predicted neutrino number for each flare, scaling the maximum expectation down by a constant factor accounting for additional physical assumptions. 
 Finally, $\langle \dot{N}_{\nu}^{\mathrm{flare}} \rangle $ (Equ.\,\ref{flare_significance})
 is the neutrino rate per flare, which describes the duration-normalized neutrino prediction of each flare.
  $\text{t}_{\text{min}}$ and $\text{t}_{\text{max}}$ indicate the start and stop time of a flare in Modified Julian Day (MJD). 
  Flares for which quasi-simultaneous data are available are marked by $\star$.
  SEDs for all flares are available as supplementary material online.}
 \label{Tab_neutrino_expectation}
 \centering
 \begin{small}
\begin{tabular}{l c c c c c c c c}
\hline\hline 
\textbf{Source}	&
 \begin{tabular}{c}
    \textbf{\tiny{Flare}}\\
    \textbf{\tiny{Number}}
   \end{tabular} & 
    \begin{tabular}{c}
    \textbf{$\langle\Phi_{\gamma} \rangle$}\\
    \textbf{\tiny{$[\text{erg cm}^{-2}s^{-1}day^{-1}]$}}
  \end{tabular}
     &   \begin{tabular}{c}
    \textbf{\tiny{$\text{t}_{\text{min}}$}}\\
    \textbf{\tiny{[MJD]}}
  \end{tabular}  
  &   \begin{tabular}{c}
    \textbf{\tiny{$\text{t}_{\text{max}}$}}\\
    \textbf{\tiny{[MJD]}}
   \end{tabular}
  &\textbf{$\text{N}_{\nu}^{\text{max}}$}
  &	  \begin{tabular}{c}
    \textbf{\tiny{$\text{N}_{\nu}^{\text{pred}}$}}\
    \textbf{\tiny{$\times 10^{-2}$}}
   \end{tabular}
  	&
   \begin{tabular}{c}
    \textbf{\tiny{Duration}}\\
    \textbf{\tiny{[Days]}}
   \end{tabular}
&	
    \textbf{\tiny{$\langle \dot{N}_{\nu}^{\mathrm{flare}} \rangle \times 10^{-3}$ }}
\\\hline
3C\,279$^\star$	&	1&	260238&		57186 &		57192 	&	0.797	&1.99&	6&	3.32\\
PKS\,1510$-$089$^\star$&	2&	192902	&	55849 &		55854 	&	0.306&	0.764&	5&	1.53\\
PKS\,1510$-$089$^\star$&	3&	151569	&	55866 &		55877 	&	0.586&	1.46&	11&	1.33\\
PKS\,1510$-$089$^\star$&	4&	151262	&	55856 &		55857 	&	0.0405&	0.101&	1&	1.01\\
3C\,279$^\star$	&	5&	138636	&	56717 &		56718 	&	0.0272&	0.0681&	1&	0.681\\
3C\,279	&	6&	128078	&	56749 	&	56754 	&	0.214&	0.535&	5&	1.07\\
PKS\,1510$-$089&	7&	125857	&	57241 &		57251 	&	0.393&	0.982&	10&	0.982\\
3C\,279$^\star$	&	8&	119379	&	56866 &		56868 	&	0.0993	&0.248&	2&	1.24\\
PKS\,1510$-$089&	9&	119033	&	56553 &		56557 	&	0.159&	0.398&	4&	0.995\\
PKS\,1510$-$089&	10&	116956	&	55766 &		55768 	&	0.0605&	0.151&	2&	0.757\\
PKS\,1510$-$089$^\star$&	11&	114301	&	57098 &		57099 	&	0.0422&	0.105&	1&	1.05\\
PKS\,1510$-$089$^\star$&	12&	111097	&	56563 &		56565 	&	0.0652&	0.163&	2&	0.815\\
PKS\,1510$-$089$^\star$&	13&	110472	&	55843 &		55844 	&	0.0305&	0.0763&	1&	0.763\\
PKS\,1510$-$089$^\star$&	14&	109350	&	57148 &		57149 	&	0.0310&	0.0775&	1&	0.775\\
PKS\,1510$-$089$^\star$&	15&	104003	&	56542 &		56543 	&	0.0283&	0.0707&	1&	0.707\\
PKS\,1510$-$089&	16&	102145	&	55742 &		55746 	&	0.0889&	0.222&	4&	0.555\\
PKS\,1510$-$089$^\star$&	17&	93862	&	56610 &		56611 	&	0.0223&	0.0560&	1&	0.560\\
PKS\,1510$-$089$^\star$&	18&	93711	&	55738 &		55740 	&	0.0480&	0.120&	2&	0.600\\
PKS\,1510$-$089&	19&	91983	&	55954 &		55994 	&	1.61&	4.03&	40&	1.01\\
PKS\,1510$-$089&	20&	91625	&	55997 &		56006 	&	0.403&	1.01&	9&	1.12\\
PKS\,1510$-$089&	21&	91131	&	57113 &		57129 	&	0.655&	1.64&	16&	1.02\\
PKS\,1510$-$089&	22&	89860	&	57152 	&	57172 	&	0.693&	1.73&	20&	0.866\\
PKS\,1510$-$089$^\star$&	23&	88048	&	57609 	&	57610 	&	0.0224&	0.0559&	1&	0.559\\
PKS\,1510$-$089$^\star$&	24&	87449	&	57201 	&	57202 	&	0.0307&	0.0767&	1&	0.767\\
PKS\,1510$-$089&	25&	86571	&	55790 	&	55791 	&	0.0228&	0.0570&	1&	0.570\\
PKS\,1510$-$089$^\star$&	26&	85220	&	57101 	&	57102 	&	0.0189&	0.0473&	1&	0.473\\
3C\,279$^\star$	&	27&	84223	&	57153 	&	57155 	&	0.0653	&0.163&	2&	0.816\\
PKS\,0402$-$362&	28&	84194	&	55824 	&	55831 	&	0.204&	0.510&	7&	0.729\\
PKS\,1510$-$089$^\star$&	29&	83764	&	57131 	&	57132 	&	0.0360&	0.0900&	1&	0.900\\
PKS\,1510$-$089&	30&	81956	&	57198 	&	57199 	&	0.0148&	0.0370&	1&	0.370\\
CTA\,102$^\star$	&	31&	80537	&	57558 	&	57559 	&	0.0108&	0.0270&	1&	0.270\\
3C\,279$^\star$	&	32&	79636	&	57212 	&	57215 	&	0.0789&	0.197&	3&	0.657\\
PKS\,1510$-$089$^\star$&	33&	77287	&	56832 	&	56833 	&	0.0223&	0.0557&	1&	0.557\\
3C\,279	&	34&	75503	&	56646 	&	56655 	&	0.236&	0.590&	9&	0.656\\
3C\,279$^\star$	&	35&	75455	&	56757 	&	56758 	&	0.0129&	0.0322&	1&	0.322\\
3C\,279	&	36&	74919	&	56742 	&	56747 	&	0.147&	0.368&	5&	0.736\\
PKS\,1510$-$089$^\star$&	37&	74815	&	57055 	&	57056 	&	0.0269&	0.0674&	1&	0.674\\
3C\,454.3&	38&	74620	&	55408 	&	55648 	&	11.71&	29.28&	240&	1.22\\
PKS\,1510$-$089$^\star$&	39&	74337	&	56571 	&	56572 	&	0.0193&	0.0482&	1&	0.482\\
PKS\,1510$-$089$^\star$&	40&	73629	&	57210 	&	57213 	&	0.0892	&0.223&	3&	0.744\\
PKS\,1424$-$41$^\star$&	41&	73574	&	56435 	&	56436 	&	0.0154	&0.0387&	1&0.387\\
3C\,279$^\star$	&	42&	72654	&	55466 	&	55468 	&	0.0573&	0.143&	2&	0.716\\
PKS\,1424$-$41$^\star$&	43&	72058	&	56394 	&	56396 	&	0.0317&	0.0794&	2&	0.397\\
PKS\,1510$-$089&	44&	69019	&	57215 	&	57223 	&	0.217&	0.542&	8&	0.678\\
PKS\,1510$-$089$^\star$&	45&	68479	&	56536 	&	56539 	&	0.0823&	0.206&	3&	0.686\\
PKS\,1510$-$089$^\star$&	46&	67473	&	56954 	&	56955 	&	0.0162&	0.0404&	1&	0.404\\
PKS\,1329$-$049$^\star$&	47&	67355	&	55467 	&	55468 	&	0.0137&	0.0341&	1&	0.341\\
PKS\,1329$-$049$^\star$&	48&	66954	&	55445 	&	55447 	&	0.0289&	0.0722&	2&	0.361\\
PKS\,1424$-$41$^\star$&	49&	66732	&	57385 	&	57386 	&	0.0123&	0.0308&	1&	0.308\\
PKS\,1510$-$089$^\star$&	50&	66603	&	56567 	&	56569 	&	0.0393&	0.0983&	2&	0.491\\
\hline 
\end{tabular}
\end{small}
 \end{table*}

\section{Discussion}\label{Discussion}
\hspace{-12pt}
In this section, we compare our results to previous estimates done by \citet{Halzen_2016} and \citet{IceCube_2018a}.
Additionally, we discuss limitations of the adopted method and derive the total detection potential of the  sample.

\subsection{Comparison to previous work}
\hspace{-12pt}
Even if the neutrino prediction from most blazar flares is small, we showed that individual extremely bright flares can provide a non-negligible neutrino detection rate, if they generate enough integrated keV--GeV flux on a relatively short time range. \citet{Halzen_2016} studied the neutrino prediction from bright blazar flares and suggested flare number one in Table\,\ref{Tab_neutrino_expectation} to be the most promising candidate for a correlated neutrino-blazar detection.  While we can confirm this finding, our 
calorimetric argumentation indicates the expected number of neutrino events from this specific flare to be significantly smaller. While \citet{Halzen_2016} did consider different neutrino spectral energy ranges and interaction processes, they overestimate the expected neutrino flux by not scaling the expected neutrino number down to a more realistic expectation. \\
In 2017 the \citet{IceCube_2018a} identified the extremely high-energy neutrino event IC\,170922A
to be temporally and spatially consistent with enhanced $\gamma$-ray activity of the blazar TXS\,0506+056, resulting in the most significant neutrino blazar association so far.
This association is based on a long-term outburst activity of TXS\,0506+056, lasting several months, our analysis indicates that a short flare association would be highly unlikely.
Since the $\gamma$-ray flare presented in \citet{IceCube_2018a} is one to two orders of magnitude fainter than the flares studied in Table\,\ref{Tab_neutrino_expectation}, the only way for this flare to build up the necessary fluence for a comparable neutrino prediction 
is via an increase in flare duration. 
\citet{Foteini_2019} have arrived at similar 
conclusions regarding the expected number of neutrino detections for a 
sample of flares from BL Lac objects with current neutrino telescopes.

\subsection{Total detection potential}
\hspace{-13pt}
To study the total detection potential of all 50 flares in this sample, we stacked all flares listed in Table\,\ref{Tab_neutrino_expectation}. While  11 flares show a neutrino rate $\left(\langle \dot{N}_{\nu}^{\mathrm{flare}} \rangle\right)$ of more than 0.001 event per day, only 2 flares  produce a maximum neutrino prediction $\left(\text{N}_{\nu}^{\text{max}}  \right)$ of $>\,1$ event. The maximum neutrino prediction from most of the studied flares is $\ll\,1$ event, making it unlikely to detect high-energy neutrinos from these flares. By calculating the Poisson probability to detect one single coincident high-energy neutrino event with IceCube, we find that all studied 50 flares in total yield only a probability of 50\,\%,
which is consistent with the non-detection of any such events with IceCube.
Extending this sample to e.g.\,100 flares  would therefore only marginally increase the total probability, as no substantial increase is obtained by adding more, fainter short-term flares.

\subsection{Limitations of the Adopted Method} \label{Limitations}
\hspace{-12pt}
Here, we discuss limitations of our method and other caveats regarding the
neutrino expectation.

\subsubsection{Calorimetric Neutrino Output}
\hspace{-14pt}
The neutrino calculation presented in this work is based on a purely hadronic origin of the observed high-energy radiation. Furthermore, only photo-pionproduction is taken into account, 
neglecting other processes such as leptonic interactions in the jet, which also contribute to the observed high-energy emission.
Bethe-Heitler interactions \citep{Petropoulou_2015} could also contribute to the observed X-ray to $\gamma$-ray fluence.
FSRQs  are favored against BL\,Lac objects due to their strong optical and UV components, which serve as targets for $p\gamma$ interactions \citep{Mannheim_1993, Mannheim_1995}. Some authors \citep{Righi_2017}, however, consider BL\,Lac type objects as promising
neutrino candidates, where optical seed photons are assumed to originate from a spine-sheath structured jet \citep{Tavecchio_2015, Ansoldi_2018}.
The  empirical scaling factor is derived following \citet{Kadler_2016}.
\citet{Krauss_2018} showed that the actual neutrino prediction is expected to be even smaller, considering a power-law neutrino spectrum and a variable hadronic contribution to the observed emission. As the maximum neutrino prediction from bright short-term blazar flares  is small (typically $<\,1$ event), introducing a more realistic estimate for the number of predicted neutrino events does not change the main findings of this paper. 

\subsubsection{Neutrino Spectrum}
\hspace{-14pt}
Eq.~\eqref{N_max_formula} assumes all neutrinos to be produced with a given energy, following a $\delta$-distribution.
The neutrino expectation listed in Table\,\ref{Tab_neutrino_expectation} therefore only takes neutrinos generated exclusively with an energy of 2\,PeV into account, neglecting contributions from lower energies as well as energy uncertainties.
Eq.~\eqref{N_pred_formula} includes a factor to scale the maximum neutrino expectation down, assuming the neutrino energy distribution to be smeared out between about 0.03\,PeV to 10\,PeV.
 This reduces the amount of predicted neutrinos, while ignoring contributions from other energies. By changing the energy range of this smearing out, we can vary the scaled neutrino prediction by a factor of about $20\,\%$.
By taking a more realistic power-law neutrino spectrum into account, \citet{Krauss_2018} showed that the maximum neutrino prediction reduces significantly compared to a delta-peaked neutrino spectrum. As the probability of a neutrino event to be of astrophysical origin gets close to unity only for the highest energies, we did not integrate over a wider neutrino energy range.  
Given the fact that short-term blazar flares in general fail in providing enough fluence for possible neutrino detections by orders of magnitude, changes in the assumed neutrino spectrum do only have a negligible effect on the findings of this paper.

\subsubsection{Multiwavelength Coverage}
\hspace{-14pt}
The multiwavelength coverage is not ideal. In fact, several of the flares listed in Table\,\ref{Tab_neutrino_expectation} are only observed by either a single \textit{Swift}/XRT pointing or no X-ray observation at all during the duration of the flare.
Several flares show upper limits in the \textit{Fermi}/LAT spectrum, being significantly (TS\,$>$\,25) detected only in two to three energy bands.
Achieving (quasi-)simultaneous data within different energy bands for most of the flares is highly challenging, as data taking cycles of different instruments are only barely synchronized and pointed instruments such as \textit{Swift}/XRT
have a given reaction time to observe new targets.
This sparse multiwavelength coverage may lead to an inaccurate calculation of the total high-energy fluence and therefore adds an additional uncertainty on the neutrino expectation.
We investigated this uncertainty by also using the 6.5 years averaged \textit{Swift}/XRT spectrum for flares for which quasi-simultaneous data are available. The variations in the maximum neutrino prediction were  only on the order of $\Delta \text{N}_{\nu}^{\text{max}} \sim 0.002$.
However, for the majority of flares from 3C\,279 and PKS\,1510$-$089 quasi-simultaneous multiwavelength data are available, which in the end results in a small systematic uncertainty of only a few percent.

\subsubsection{Fluence Integration Range}
\hspace{-14pt}
The maximum number of neutrino events expected from a specific flare depends on several factors, such as the neutrino spectrum and flare duration. However, the primary ingredient in Eq.~\eqref{N_max_formula} is the total integrated high-energy emission (assuming purely hadronic processes).
The exact energy interval of 10\,keV to 20\,GeV does not affect the neutrino expectation significantly, as long as most of the high-energy hump is covered. This clearly indicates that the maximum neutrino expectation depends on the total high-energy fluence instead of the emission in some energy bands like e.g.\,the \textit{Fermi}/LAT flux \citep{Krauss_2018}. 


\subsubsection{Flare Identification}
\hspace{-15pt}
The flare identification described in Sec.\,\ref{Flare_identification} is optimized for bright flares with durations on the order of days to weeks. Intra Day Variability (IDV) cannot be resolved due to the choice of daily time binning of the light curves \citep{Foschini_2011, Ackermann_2016}.
While for each source an individual ground level flux is calculated, no variation in time is included in the ground state calculation. This mostly affects sources with a steady increase in flux over several years (e.g. PKS\,1424$-$418), while 
in the current analysis this assumption results in a larger ground level flux uncertainty.
As the flux uncertainty of each light curve bin is taken into account in the selection of flaring periods, time ranges with higher test statistic and thus smaller error are preferably selected. This selection introduces a bias towards the brightest blazar flares observed by \textit{Fermi}/LAT.
However, when it comes to the study of bright blazar flares, \citet{Meyer_2019} finds similar results by using a much higher test statistic threshold for the selection of flaring periods.\\
Short-term flares are preferably selected, as the flare duration is sharply defined by the number of successive light curve bins that fulfill Eq.~\eqref{flare_selection_1}. 
Longer periods of enhanced activity with varying flux levels or even non-detections are split into multiple (short-term) flares, which in the total ranking are all treated independently.
A first approach of introducing an equivalent selection criterion for longer outburst periods on the order of weeks to years could be achieved by allowing some flux values within a longer outburst period to violate  Eq.~\eqref{flare_selection_1}. 
On the one hand this modification would reduce the number of short-term flares into which a long-term outburst period is divided, while on the other hand it would introduce a systematic offset on the duration of short-term flaring states, which also affect the neutrino expectation. However, as only three sources from our studied sample of bright \textit{Fermi}/LAT blazars show extended long-term emission, this modification only marginally affects the  flare selection.\\
By assuming Gaussian fluctuations, we can estimate the number of false positive identified daily bins in each light curve. At a threshold of TS\,$\geq 9$, we estimate the fraction of false positive bins to be $0.5\%$.
Thus, considering the full 6.5 years time range, about 12 daily bins are expected to be false positive detections. By taking also the fraction of maximum $9\%$ of light curve bins which contribute the the top 50 flares into account, only a single daily flare per light curve is expected as a flare positive detection.

\section{Conclusion}\label{Conclusion}
\hspace{-14pt}
We have studied the calorimetric neutrino expectation of individual blazar flares on different time scales. Following the conclusions from \citet{Krauss_2014, Krauss_2018} and \citet{Kadler_2016}, blazar flares, which provide the highest keV-to-GeV fluence, would be the best candidates for a high-energy neutrino association\footnote{A balance between keV-to-GeV fluence and integration time is required. This statement mainly applies for periods of enhanced activity measured by \textit{Fermi}/LAT.}. The fluence of individual flares is derived by integrating over the peak of the high-energy hump of the blazar SED, taking mostly quasi-simultaneous multiwavelength data during flaring periods into account. The estimates presented in this work are based on an equal energetic balance of neutrinos and $\gamma$-rays, which clearly overestimates the number of predicted neutrino events. Even by using over-optimistic assumptions, we showed that the absolute neutrino expectation for most short-term blazar flares is too small to be considered for a dedicated correlation analysis. By studying an all-sky sample of 150 bright \textit{Fermi}/LAT-detected blazars, we found that potential  high-energy neutrino detectable blazar flares are associated with a small group of only seven blazars.
Thus we conclude that even in the (unrealistic) maximum case, the majority of short-term blazar flares are not able to generate a substantial detectable neutrino flux, at least with the current generation of neutrino telescopes. Future neutrino point source searches should therefore concentrate on periods of maximum fluence, rather than on short-term temporal and spatial associations.\\
In addition to searching for a correlation of high-energy neutrinos and $\gamma$-ray flares with various durations, neutrino point source studies should in addition consider the hadronic emission to peak more towards the X-ray or optical regime.
The All-sky Medium Energy Gamma-ray Observatory (AMEGO) will survey the entire sky in an energy range of $200\,$keV to $10\,$GeV and provide crucial information for the understanding of the emission mechanisms in blazar jets \citep{Perkins_2017, Roopesh_2019}.
Because the flux of high-energy neutrino events observed in IceCube is small, the possibility of associated neutrino detections for the brightest short-term blazar flares cannot be neglected completely. 

\begin{acknowledgements}
\hspace{-14pt}
The work of M. Kreter and M. B\"ottcher is supported by the South African 
Research Chairs Initiative (grant no. 64789) of the Department of Science and 
Innovation and the National Research Foundation\footnote{Any opinion, finding 
and conclusion or recommendation expressed in this material is that of the authors 
and the NRF does not accept any liability in this regard.} of South Africa.
F. Krau{\ss} was supported as an Eberly Research Fellow by the Eberly College of Science at the Pennsylvania State University.
 We thank J.E.~Davis for the development of the \texttt{slxfig}
  module that has been used to prepare the figures in this work.
  This research has made use of a collection of
  ISIS scripts provided by the Dr. Karl Remeis-Observatory, Bamberg,
  Germany at \url{http://www.sternwarte.uni-erlangen.de/isis/}.
  \\The \textit{Fermi} LAT Collaboration acknowledges generous ongoing support
from a number of agencies and institutes that have supported both the
development and the operation of the LAT as well as scientific data analysis.
These include the National Aeronautics and Space Administration and the
Department of Energy in the United States, the Commissariat \`a l'Energie Atomique
and the Centre National de la Recherche Scientifique / Institut National de Physique
Nucl\'eaire et de Physique des Particules in France, the Agenzia Spaziale Italiana
and the Istituto Nazionale di Fisica Nucleare in Italy, the Ministry of Education,
Culture, Sports, Science and Technology (MEXT), High Energy Accelerator Research
Organization (KEK) and Japan Aerospace Exploration Agency (JAXA) in Japan, and
the K.~A.~Wallenberg Foundation, the Swedish Research Council and the
Swedish National Space Board in Sweden.
Additional support for science analysis during the operations phase is gratefully
acknowledged from the Istituto Nazionale di Astrofisica in Italy and the Centre
National d'\'Etudes Spatiales in France. This work performed in part under DOE
Contract DE-AC02-76SF00515.
  We acknowledge the use of public data from the {\it Swift} data archive. 
\end{acknowledgements}

\bibliography{References.bib}

\begin{thebibliography}{}
\expandafter\ifx\csname natexlab\endcsname\relax\def\natexlab#1{#1}\fi

\bibitem[{{Aartsen} {et~al.}(2015){Aartsen}, {Ackermann}, {Adams}, {Aguilar},
  {Ahlers}, {Ahrens}, {Altmann}, {Anderson}, {Archinger}, {Arguelles}, {Arlen},
  {Auffenberg}, {Bai}, {Barwick}, {Baum}, {Bay}, {Baker}, {Beatty}, {Becker
  Tjus}, {Becker}, {BenZvi}, {Berghaus}, {Berley}, {Bernardini}, {Bernhard},
  {Besson}, {Binder}, {Bindig}, {Bissok}, {Blaufuss}, {Blumenthal}, {Boersma},
  {Bohm}, {Bos}, {Bose}, {B{\"o}ser}, {Botner}, {Brayeur}, {Bretz}, {Brown},
  {Buzinsky}, {Casey}, {Casier}, {Cheung}, {Chirkin}, {Christov}, {Christy},
  {Clark}, {Classen}, {Clevermann}, {Coenders}, {Cowen}, {Cruz Silva},
  {Daughhetee}, {Davis}, {Day}, {de Andr{\'e}}, {De Clercq}, {Dembinski}, {De
  Ridder}, {Desiati}, {de Vries}, {de Wasseige}, {de With}, {DeYoung},
  {D{\'\i}az-V{\'e}lez}, {Dumm}, {Dunkman}, {Eagan}, {Eberhardt}, {Ehrhardt},
  {Eichmann}, {Eisch}, {Euler}, {Evenson}, {Fadiran}, {Fazely}, {Fedynitch},
  {Feintzeig}, {Felde}, {Filimonov}, {Finley}, {Fischer-Wasels}, {Flis},
  {Frantzen}, {Fuchs}, {Gaisser}, {Gaior}, {Gallagher}, {Gerhardt}, {Gier},
  {Gladstone}, {Gl{\"u}senkamp}, {Goldschmidt}, {Golup}, {Gonzalez}, {Goodman},
  {G{\'o}ra}, {Grant}, {Gretskov}, {Groh}, {Gro{\ss}}, {Ha}, {Haack}, {Haj
  Ismail}, {Hallen}, {Hallgren}, {Halzen}, {Hanson}, {Hebecker}, {Heereman},
  {Heinen}, {Helbing}, {Hellauer}, {Hellwig}, {Hickford}, {Hignight}, {Hill},
  {Hoffman}, {Hoffmann}, {Homeier}, {Hoshina}, {Huang}, {Huelsnitz}, {Hulth},
  {Hultqvist}, {In}, {Ishihara}, {Jacobi}, {Jacobsen}, {Japaridze}, {Jero},
  {Jurkovic}, {Kaminsky}, {Kappes}, {Karg}, {Karle}, {Kauer}, {Keivani},
  {Kelley}, {Kheirandish}, {Kiryluk}, {Kl{\"a}s}, {Klein}, {K{\"o}hne},
  {Kohnen}, {Kolanoski}, {Koob}, {K{\"o}pke}, {Kopper}, {Kopper}, {Koskinen},
  {Kowalski}, {Krings}, {Kroll}, {Kroll}, {Kunnen}, {Kurahashi}, {Kuwabara},
  {Labare}, {Lanfranchi}, {Larsen}, {Larson}, {Lesiak-Bzdak}, {Leuermann},
  {L{\"u}nemann}, {Madsen}, {Maggi}, {Mahn}, {Maruyama}, {Mase}, {Matis},
  {Maunu}, {McNally}, {Meagher}, {Medici}, {Meli}, {Meures}, {Miarecki},
  {Middell}, {Middlemas}, {Milke}, {Miller}, {Mohrmann}, {Montaruli}, {Morse},
  {Nahnhauer}, {Naumann}, {Niederhausen}, {Nowicki}, {Nygren}, {Obertacke},
  {Olivas}, {Omairat}, {O'Murchadha}, {Palczewski}, {Paul}, {Pepper},
  {P{\'e}rez de los Heros}, {Pfendner}, {Pieloth}, {Pinat}, {Posselt}, {Price},
  {Przybylski}, {P{\"u}tz}, {Quinnan}, {R{\"a}del}, {Rameez}, {Rawlins},
  {Redl}, {Rees}, {Reimann}, {Relich}, {Resconi}, {Rhode}, {Richman}, {Riedel},
  {Robertson}, {Rodrigues}, {Rongen}, {Rott}, {Ruhe}, {Ruzybayev}, {Ryckbosch},
  {Saba}, {Sand er}, {Sandroos}, {Santander}, {Sarkar}, {Schatto}, {Scheriau},
  {Schmidt}, {Schmitz}, {Schoenen}, {Sch{\"o}neberg}, {Sch{\"o}nwald},
  {Schukraft}, {Schulte}, {Schulz}, {Seckel}, {Sestayo}, {Seunarine},
  {Shanidze}, {Smith}, {Soldin}, {Spiczak}, {Spiering}, {Stamatikos}, {Stanev},
  {Stanisha}, {Stasik}, {Stezelberger}, {Stokstad}, {St{\"o}{\ss}l},
  {Strahler}, {Str{\"o}m}, {Strotjohann}, {Sullivan}, {Sutherland}, {Taavola},
  {Taboada}, {Tamburro}, {Ter-Antonyan}, {Terliuk}, {Te{\v{s}}i{\'c}}, {Tilav},
  {Toale}, {Tobin}, {Tosi}, {Tselengidou}, {Unger}, {Usner}, {Vallecorsa}, {van
  Eijndhoven}, {Vand enbroucke}, {van Santen}, {Vanheule}, {Vehring}, {Voge},
  {Vraeghe}, {Walck}, {Wallraff}, {Weaver}, {Wellons}, {Wendt}, {Westerhoff},
  {Whelan}, {Whitehorn}, {Wichary}, {Wiebe}, {Wiebusch}, {Williams}, {Wissing},
  {Wolf}, {Wood}, {Woschnagg}, {Xu}, {Xu}, {Xu}, {Yanez}, {Yodh}, {Yoshida},
  {Zarzhitsky}, {Ziemann}, {Zoll}, \& {IceCube Collaboration}}]{Aartsen_2015b}
{Aartsen}, M.~G., {Ackermann}, M., {Adams}, J., {et~al.} 2015, \apj, 807, 46

\bibitem[{{Aartsen} {et~al.}(2016){Aartsen}, {Abraham}, {Ackermann}, {Adams},
  {Aguilar}, {Ahlers}, {Ahrens}, {Altmann}, {Anderson}, {Ansseau}, \&
  et~al.}]{Aartsen_2016b}
{Aartsen}, M.~G., {Abraham}, K., {Ackermann}, M., {et~al.} 2016, \apj, 824, 115

\bibitem[{Aartsen {et~al.}(2016)Aartsen, {Abraham}, {Ackermann}, {Adams},
  {Aguilar}, {Ahlers}, {Ahrens}, {Altmann}, {Anderson}, {Ansseau}, \&
  et~al.}]{Aartsen_2016}
Aartsen, M.~G., {Abraham}, K., {Ackermann}, M., {et~al.} 2016, \prl, 117,
  241101

\bibitem[{{Acero} {et~al.}(2015){Acero}, {Ackermann}, {Ajello}, {Albert},
  {Atwood}, {Axelsson}, {Baldini}, {Ballet}, {Barbiellini}, {Bastieri},
  {Belfiore}, {Bellazzini}, {Bissaldi}, {Blandford}, {Bloom}, {Bogart},
  {Bonino}, {Bottacini}, {Bregeon}, {Britto}, {Bruel}, {Buehler}, {Burnett},
  {Buson}, {Caliandro}, {Cameron}, {Caputo}, {Caragiulo}, {Caraveo},
  {Casandjian}, {Cavazzuti}, {Charles}, {Chaves}, {Chekhtman}, {Cheung},
  {Chiang}, {Chiaro}, {Ciprini}, {Claus}, {Cohen-Tanugi}, {Cominsky}, {Conrad},
  {Cutini}, {D'Ammando}, {de Angelis}, {DeKlotz}, {de Palma}, {Desiante},
  {Digel}, {Di Venere}, {Drell}, {Dubois}, {Dumora}, {Favuzzi}, {Fegan},
  {Ferrara}, {Finke}, {Franckowiak}, {Fukazawa}, {Funk}, {Fusco}, {Gargano},
  {Gasparrini}, {Giebels}, {Giglietto}, {Giommi}, {Giordano}, {Giroletti},
  {Glanzman}, {Godfrey}, {Grenier}, {Grondin}, {Grove}, {Guillemot}, {Guiriec},
  {Hadasch}, {Harding}, {Hays}, {Hewitt}, {Hill}, {Horan}, {Iafrate}, {Jogler},
  {J{\'o}hannesson}, {Johnson}, {Johnson}, {Johnson}, {Johnson}, {Kamae},
  {Kataoka}, {Katsuta}, {Kuss}, {La Mura}, {Landriu}, {Larsson}, {Latronico},
  {Lemoine-Goumard}, {Li}, {Li}, {Longo}, {Loparco}, {Lott}, {Lovellette},
  {Lubrano}, {Madejski}, {Massaro}, {Mayer}, {Mazziotta}, {McEnery},
  {Michelson}, {Mirabal}, {Mizuno}, {Moiseev}, {Mongelli}, {Monzani},
  {Morselli}, {Moskalenko}, {Murgia}, {Nuss}, {Ohno}, {Ohsugi}, {Omodei},
  {Orienti}, {Orlando}, {Ormes}, {Paneque}, {Panetta}, {Perkins},
  {Pesce-Rollins}, {Piron}, {Pivato}, {Porter}, {Racusin}, {Rando}, {Razzano},
  {Razzaque}, {Reimer}, {Reimer}, {Reposeur}, {Rochester}, {Romani},
  {Salvetti}, {S{\'a}nchez-Conde}, {Saz Parkinson}, {Schulz}, {Siskind},
  {Smith}, {Spada}, {Spandre}, {Spinelli}, {Stephens}, {Strong}, {Suson},
  {Takahashi}, {Takahashi}, {Tanaka}, {Thayer}, {Thayer}, {Thompson},
  {Tibaldo}, {Tibolla}, {Torres}, {Torresi}, {Tosti}, {Troja}, {Van Klaveren},
  {Vianello}, {Winer}, {Wood}, {Wood}, {Zimmer}, \& {Fermi-LAT
  Collaboration}}]{Acero_2015}
{Acero}, F., {Ackermann}, M., {Ajello}, M., {et~al.} 2015, \apjs, 218, 23

\bibitem[{Acero {et~al.}(2016)Acero, {Ackermann}, {Ajello}, {Albert},
  {Baldini}, {Ballet}, {Barbiellini}, {Bastieri}, {Bellazzini}, {Bissaldi},
  {Bloom}, {Bonino}, {Bottacini}, {Brandt}, {Bregeon}, {Bruel}, {Buehler},
  {Buson}, {Caliandro}, {Cameron}, {Caragiulo}, {Caraveo}, {Casandjian},
  {Cavazzuti}, {Cecchi}, {Charles}, {Chekhtman}, {Chiang}, {Chiaro}, {Ciprini},
  {Claus}, {Cohen-Tanugi}, {Conrad}, {Cuoco}, {Cutini}, {D'Ammando}, {de
  Angelis}, {de Palma}, {Desiante}, {Digel}, {Di Venere}, {Drell}, {Favuzzi},
  {Fegan}, {Ferrara}, {Focke}, {Franckowiak}, {Funk}, {Fusco}, {Gargano},
  {Gasparrini}, {Giglietto}, {Giordano}, {Giroletti}, {Glanzman}, {Godfrey},
  {Grenier}, {Guiriec}, {Hadasch}, {Harding}, {Hayashi}, {Hays}, {Hewitt},
  {Hill}, {Horan}, {Hou}, {Jogler}, {J{\'o}hannesson}, {Kamae}, {Kuss},
  {Landriu}, {Larsson}, {Latronico}, {Li}, {Li}, {Longo}, {Loparco},
  {Lovellette}, {Lubrano}, {Maldera}, {Malyshev}, {Manfreda}, {Martin},
  {Mayer}, {Mazziotta}, {McEnery}, {Michelson}, {Mirabal}, {Mizuno}, {Monzani},
  {Morselli}, {Nuss}, {Ohsugi}, {Omodei}, {Orienti}, {Orlando}, {Ormes},
  {Paneque}, {Pesce-Rollins}, {Piron}, {Pivato}, {Rain{\`o}}, {Rando},
  {Razzano}, {Razzaque}, {Reimer}, {Reimer}, {Remy}, {Renault},
  {S{\'a}nchez-Conde}, {Schaal}, {Schulz}, {Sgr{\`o}}, {Siskind}, {Spada},
  {Spandre}, {Spinelli}, {Strong}, {Suson}, {Tajima}, {Takahashi}, {Thayer},
  {Thompson}, {Tibaldo}, {Tinivella}, {Torres}, {Tosti}, {Troja}, {Vianello},
  {Werner}, {Wood}, {Wood}, {Zaharijas}, \& {Zimmer}}]{Acero_2016}
Acero, F., {Ackermann}, M., {Ajello}, M., {et~al.} 2016, \apjs, 223, 26

\bibitem[{{Achterberg} {et~al.}(2008){Achterberg}, {Ackermann}, {Adams},
  {Ahrens}, {Andeen}, {Auffenberg}, {Bahcall}, {Bai}, {Baret}, {Barwick}, \&
  et~al.}]{Achterberg_2008}
{Achterberg}, A., {Ackermann}, M., {Adams}, J., {et~al.} 2008, \apj, 674, 357

\bibitem[{{Ackermann} {et~al.}(2011){Ackermann}, {Ajello}, {Allafort},
  {Antolini}, {Atwood}, {Axelsson}, {Baldini}, {Ballet}, {Barbiellini},
  {Bastieri}, {Bechtol}, {Bellazzini}, {Berenji}, {Blandford}, {Bloom},
  {Bonamente}, {Borgland}, {Bottacini}, {Bouvier}, {Bregeon}, {Brigida},
  {Bruel}, {Buehler}, {Burnett}, {Buson}, {Caliandro}, {Cameron}, {Caraveo},
  {Casand jian}, {Cavazzuti}, {Cecchi}, {Charles}, {Cheung}, {Chiang},
  {Ciprini}, {Claus}, {Cohen-Tanugi}, {Conrad}, {Costamante}, {Cutini}, {de
  Angelis}, {de Palma}, {Dermer}, {Digel}, {Silva}, {Drell}, {Dubois},
  {Escande}, {Favuzzi}, {Fegan}, {Ferrara}, {Finke}, {Focke}, {Fortin},
  {Frailis}, {Fukazawa}, {Funk}, {Fusco}, {Gargano}, {Gasparrini}, {Gehrels},
  {Germani}, {Giebels}, {Giglietto}, {Giommi}, {Giordano}, {Giroletti},
  {Glanzman}, {Godfrey}, {Grenier}, {Grove}, {Guiriec}, {Gustafsson},
  {Hadasch}, {Hayashida}, {Hays}, {Healey}, {Horan}, {Hou}, {Hughes},
  {Iafrate}, {J{\'o}hannesson}, {Johnson}, {Johnson}, {Kamae}, {Katagiri},
  {Kataoka}, {Kn{\"o}dlseder}, {Kuss}, {Lande}, {Larsson}, {Latronico},
  {Longo}, {Loparco}, {Lott}, {Lovellette}, {Lubrano}, {Madejski}, {Mazziotta},
  {McConville}, {McEnery}, {Michelson}, {Mitthumsiri}, {Mizuno}, {Moiseev},
  {Monte}, {Monzani}, {Moretti}, {Morselli}, {Moskalenko}, {Murgia},
  {Nakamori}, {Naumann-Godo}, {Nolan}, {Norris}, {Nuss}, {Ohno}, {Ohsugi},
  {Okumura}, {Omodei}, {Orienti}, {Orlando}, {Ormes}, {Ozaki}, {Paneque},
  {Parent}, {Pesce-Rollins}, {Pierbattista}, {Piranomonte}, {Piron}, {Pivato},
  {Porter}, {Rain{\`o}}, {Rando}, {Razzano}, {Razzaque}, {Reimer}, {Reimer},
  {Ritz}, {Rochester}, {Romani}, {Roth}, {Sanchez}, {Sbarra}, {Scargle},
  {Schalk}, {Sgr{\`o}}, {Shaw}, {Siskind}, {Spandre}, {Spinelli}, {Strong},
  {Suson}, {Tajima}, {Takahashi}, {Takahashi}, {Tanaka}, {Thayer}, {Thayer},
  {Thompson}, {Tibaldo}, {Tinivella}, {Torres}, {Tosti}, {Troja}, {Uchiyama},
  {Vandenbroucke}, {Vasileiou}, {Vianello}, {Vitale}, {Waite}, {Wallace},
  {Wang}, {Winer}, {Wood}, {Wood}, \& {Zimmer}}]{Ackermann_2011}
{Ackermann}, M., {Ajello}, M., {Allafort}, A., {et~al.} 2011, \apj, 743, 171

\bibitem[{{Ackermann} {et~al.}(2015){Ackermann}, {Ajello}, {Atwood}, {Baldini},
  {Ballet}, {Barbiellini}, {Bastieri}, {Becerra Gonzalez}, {Bellazzini},
  {Bissaldi}, {Bland ford}, {Bloom}, {Bonino}, {Bottacini}, {Brandt},
  {Bregeon}, {Britto}, {Bruel}, {Buehler}, {Buson}, {Caliandro}, {Cameron},
  {Caragiulo}, {Caraveo}, {Carpenter}, {Casandjian}, {Cavazzuti}, {Cecchi},
  {Charles}, {Chekhtman}, {Cheung}, {Chiang}, {Chiaro}, {Ciprini}, {Claus},
  {Cohen-Tanugi}, {Cominsky}, {Conrad}, {Cutini}, {D'Abrusco}, {D'Ammando}, {de
  Angelis}, {Desiante}, {Digel}, {Di Venere}, {Drell}, {Favuzzi}, {Fegan},
  {Ferrara}, {Finke}, {Focke}, {Franckowiak}, {Fuhrmann}, {Fukazawa},
  {Furniss}, {Fusco}, {Gargano}, {Gasparrini}, {Giglietto}, {Giommi},
  {Giordano}, {Giroletti}, {Glanzman}, {Godfrey}, {Grenier}, {Grove},
  {Guiriec}, {Hewitt}, {Hill}, {Horan}, {Itoh}, {J{\'o}hannesson}, {Johnson},
  {Johnson}, {Kataoka}, {Kawano}, {Krauss}, {Kuss}, {La Mura}, {Larsson},
  {Latronico}, {Leto}, {Li}, {Li}, {Longo}, {Loparco}, {Lott}, {Lovellette},
  {Lubrano}, {Madejski}, {Mayer}, {Mazziotta}, {McEnery}, {Michelson},
  {Mizuno}, {Moiseev}, {Monzani}, {Morselli}, {Moskalenko}, {Murgia}, {Nuss},
  {Ohno}, {Ohsugi}, {Ojha}, {Omodei}, {Orienti}, {Orland o}, {Paggi},
  {Paneque}, {Perkins}, {Pesce-Rollins}, {Piron}, {Pivato}, {Porter},
  {Rain{\`o}}, {Rando}, {Razzano}, {Razzaque}, {Reimer}, {Reimer}, {Romani},
  {Salvetti}, {Schaal}, {Schinzel}, {Schulz}, {Sgr{\`o}}, {Siskind},
  {Sokolovsky}, {Spada}, {Spandre}, {Spinelli}, {Stawarz}, {Suson},
  {Takahashi}, {Takahashi}, {Tanaka}, {Thayer}, {Thayer}, {Tibaldo}, {Torres},
  {Torresi}, {Tosti}, {Troja}, {Uchiyama}, {Vianello}, {Winer}, {Wood}, \&
  {Zimmer}}]{Ackermann_2015}
{Ackermann}, M., {Ajello}, M., {Atwood}, W.~B., {et~al.} 2015, \apj, 810, 14

\bibitem[{{Ackermann} {et~al.}(2016){Ackermann}, {Anantua}, {Asano}, {Baldini},
  {Barbiellini}, {Bastieri}, {Becerra Gonzalez}, {Bellazzini}, {Bissaldi},
  {Blandford}, {Bloom}, {Bonino}, {Bottacini}, {Bruel}, {Buehler}, {Caliandro},
  {Cameron}, {Caragiulo}, {Caraveo}, {Cavazzuti}, {Cecchi}, {Cheung}, {Chiang},
  {Chiaro}, {Ciprini}, {Cohen-Tanugi}, {Costanza}, {Cutini}, {D'Ammando}, {de
  Palma}, {Desiante}, {Digel}, {Di Lalla}, {Di Mauro}, {Di Venere}, {Drell},
  {Favuzzi}, {Fegan}, {Ferrara}, {Fukazawa}, {Funk}, {Fusco}, {Gargano},
  {Gasparrini}, {Giglietto}, {Giordano}, {Giroletti}, {Grenier}, {Guillemot},
  {Guiriec}, {Hayashida}, {Hays}, {Horan}, {J{\'o}hannesson}, {Kensei},
  {Kocevski}, {Kuss}, {La Mura}, {Larsson}, {Latronico}, {Li}, {Longo},
  {Loparco}, {Lott}, {Lovellette}, {Lubrano}, {Madejski}, {Magill}, {Maldera},
  {Manfreda}, {Mayer}, {Mazziotta}, {Michelson}, {Mirabal}, {Mizuno},
  {Monzani}, {Morselli}, {Moskalenko}, {Nalewajko}, {Negro}, {Nuss}, {Ohsugi},
  {Orlando}, {Paneque}, {Perkins}, {Pesce-Rollins}, {Piron}, {Pivato},
  {Porter}, {Principe}, {Rando}, {Razzano}, {Razzaque}, {Reimer}, {Scargle},
  {Sgr{\`o}}, {Sikora}, {Simone}, {Siskind}, {Spada}, {Spinelli}, {Stawarz},
  {Thayer}, {Thompson}, {Torres}, {Troja}, {Uchiyama}, {Yuan}, \&
  {Zimmer}}]{Ackermann_2016}
{Ackermann}, M., {Anantua}, R., {Asano}, K., {et~al.} 2016, \apjl, 824, L20

\bibitem[{{Alvarez-Mu{\~n}iz} \& {M{\'e}sz{\'a}ros}(2004)}]{Alvarez_2004}
{Alvarez-Mu{\~n}iz}, J., \& {M{\'e}sz{\'a}ros}, P. 2004, \prd, 70, 123001

\bibitem[{{Ansoldi} {et~al.}(2018){Ansoldi}, {Antonelli}, {Arcaro}, {Baack},
  {Babi{\'c}}, {Banerjee}, {Bangale}, {Barres de Almeida}, {Barrio}, {Becerra
  Gonz{\'a}lez}, {Bednarek}, {Bernardini}, {Berse}, {Berti}, {Besenrieder},
  {Bhattacharyya}, {Bigongiari}, {Biland}, {Blanch}, {Bonnoli}, {Carosi},
  {Ceribella}, {Chatterjee}, {Colak}, {Colin}, {Colombo}, {Contreras},
  {Cortina}, {Covino}, {Cumani}, {D'Elia}, {Da Vela}, {Dazzi}, {De Angelis},
  {De Lotto}, {Delfino}, {Delgado}, {Di Pierro}, {Dom{\'\i}nguez}, {Dominis
  Prester}, {Dorner}, {Doro}, {Einecke}, {Elsaesser}, {Fallah Ramazani},
  {Fattorini}, {Fern{\'a}ndez-Barral}, {Ferrara}, {Fidalgo}, {Foffano},
  {Fonseca}, {Font}, {Fruck}, {Gallozzi}, {Garc{\'\i}a L{\'o}pez},
  {Garczarczyk}, {Gaug}, {Giammaria}, {Godinovi{\'c}}, {Guberman}, {Hadasch},
  {Hahn}, {Hassan}, {Hayashida}, {Herrera}, {Hoang}, {Hrupec}, {Inoue},
  {Ishio}, {Iwamura}, {Konno}, {Kubo}, {Kushida}, {Lamastra}, {Lelas}, {Leone},
  {Lindfors}, {Lombardi}, {Longo}, {L{\'o}pez}, {Maggio}, {Majumdar},
  {Makariev}, {Maneva}, {Manganaro}, {Mannheim}, {Maraschi}, {Mariotti},
  {Mart{\'\i}nez}, {Masuda}, {Mazin}, {Mielke}, {Minev}, {Miranda}, {Mirzoyan},
  {Moralejo}, {Moreno}, {Moretti}, {Neustroev}, {Niedzwiecki}, {Nievas
  Rosillo}, {Nigro}, {Nilsson}, {Ninci}, {Nishijima}, {Noda}, {Nogu{\'e}s},
  {Paiano}, {Palacio}, {Paneque}, {Paoletti}, {Paredes}, {Pedaletti},
  {Pe{\~n}il}, {Peresano}, {Persic}, {Pfrang}, {Prada Moroni}, {Prandini},
  {Puljak}, {Garcia}, {Rhode}, {Rib{\'o}}, {Rico}, {Righi}, {Rugliancich},
  {Saha}, {Saito}, {Satalecka}, {Schweizer}, {Sitarek}, {{\v{S}}nidari{\'c}},
  {Sobczynska}, {Stamerra}, {Strzys}, {Suri{\'c}}, {Tavecchio}, {Temnikov},
  {Terzi{\'c}}, {Teshima}, {Torres-Alb{\'a}}, {Tsujimoto}, {Vanzo}, {Vazquez
  Acosta}, {Vovk}, {Ward}, {Will}, {Zari{\'c}}, \& {Cerruti}}]{Ansoldi_2018}
{Ansoldi}, S., {Antonelli}, L.~A., {Arcaro}, C., {et~al.} 2018, \apjl, 863, L10

\bibitem[{{Becker Tjus} {et~al.}(2014){Becker Tjus}, {Eichmann}, {Halzen},
  {Kheirandish}, \& {Saba}}]{Becker_2014}
{Becker Tjus}, J., {Eichmann}, B., {Halzen}, F., {Kheirandish}, A., \& {Saba},
  S.~M. 2014, \prd, 89, 123005

\bibitem[{{Blanco} \& {Hooper}(2017)}]{Blanco_2017}
{Blanco}, C., \& {Hooper}, D. 2017, Journal of Cosmology and Astroparticle
  Physics, 2017, 017

\bibitem[{{B{\"o}ttcher} {et~al.}(2013){B{\"o}ttcher}, {Reimer}, {Sweeney}, \&
  {Prakash}}]{Bottcher_2013}
{B{\"o}ttcher}, M., {Reimer}, A., {Sweeney}, K., \& {Prakash}, A. 2013, \apj,
  768, 54

\bibitem[{{Feldman} \& {Cousins}(1998)}]{Feldman_1998}
{Feldman}, G.~J., \& {Cousins}, R.~D. 1998, \prd, 57, 3873

\bibitem[{{Foschini} {et~al.}(2011){Foschini}, {Ghisellini}, {Tavecchio},
  {Bonnoli}, \& {Stamerra}}]{Foschini_2011}
{Foschini}, L., {Ghisellini}, G., {Tavecchio}, F., {Bonnoli}, G., \&
  {Stamerra}, A. 2011, \aap, 530, A77

\bibitem[{{Gao} {et~al.}(2019){Gao}, {Fedynitch}, {Winter}, \&
  {Pohl}}]{Gao_2019}
{Gao}, S., {Fedynitch}, A., {Winter}, W., \& {Pohl}, M. 2019, Nature Astronomy,
  3, 88

\bibitem[{{Halzen} \& {Kheirandish}(2016)}]{Halzen_2016}
{Halzen}, F., \& {Kheirandish}, A. 2016, \apj, 831, 12

\bibitem[{{Hooper}(2016)}]{Hooper_2016}
{Hooper}, D. 2016, Journal of Cosmology and Astroparticle Physics, 2016, 002

\bibitem[{{Hooper} {et~al.}(2019){Hooper}, {Linden}, \&
  {Vieregg}}]{Hooper_2019}
{Hooper}, D., {Linden}, T., \& {Vieregg}, A. 2019, Journal of Cosmology and
  Astroparticle Physics, 2019, 012

\bibitem[{{Houck} \& {Denicola}(2000)}]{Houck2000}
{Houck}, J.~C., \& {Denicola}, L.~A. 2000, {ISIS: An Interactive Spectral
  Interpretation System for High Resolution X-Ray Spectroscopy}

\bibitem[{{IceCube ~Collaboation}(2018)}]{IceCube_2018b}
{IceCube ~Collaboation}. 2018, Science, 361, 147

\bibitem[{{IceCube Collaboration}(2013)}]{IceCube_2013}
{IceCube Collaboration}. 2013, Science, 342, 1242856

\bibitem[{{IceCube Collaboration} {et~al.}(2018){IceCube Collaboration},
  {Aartsen}, {Ackermann}, {Adams}, {Aguilar}, {Ahlers}, {Ahrens}, {Al Samarai},
  {Altmann}, {Andeen}, \& et~al.}]{IceCube_2018a}
{IceCube Collaboration}, {Aartsen}, M.~G., {Ackermann}, M., {et~al.} 2018,
  Science, 361, eaat1378

\bibitem[{{Kadler} {et~al.}(2016){Kadler}, {Krau{\ss}}, {Mannheim}, {Ojha},
  {M{\"u}ller}, {Schulz}, {Anton}, {Baumgartner}, {Beuchert}, {Buson},
  {Carpenter}, {Eberl}, {Edwards}, {Eisenacher Glawion}, {Els{\"a}sser},
  {Gehrels}, {Gr{\"a}fe}, {Gulyaev}, {Hase}, {Horiuchi}, {James}, {Kappes},
  {Kappes}, {Katz}, {Kreikenbohm}, {Kreter}, {Kreykenbohm}, {Langejahn},
  {Leiter}, {Litzinger}, {Longo}, {Lovell}, {McEnery}, {Natusch}, {Phillips},
  {Pl{\"o}tz}, {Quick}, {Ros}, {Stecker}, {Steinbring}, {Stevens}, {Thompson},
  {Tr{\"u}stedt}, {Tzioumis}, {Weston}, {Wilms}, \& {Zensus}}]{Kadler_2016}
{Kadler}, M., {Krau{\ss}}, F., {Mannheim}, K., {et~al.} 2016, Nature Physics,
  12, 807

\bibitem[{{Kimura} {et~al.}(2015){Kimura}, {Murase}, \& {Toma}}]{Kimura_2015}
{Kimura}, S.~S., {Murase}, K., \& {Toma}, K. 2015, \apj, 806, 159

\bibitem[{{Krauss}(2016)}]{Krauss2016}
{Krauss}, F. 2016, PhD thesis, ECAP/FAU

\bibitem[{{Krau{\ss}} {et~al.}(2014){Krau{\ss}}, {Kadler}, {Mannheim},
  {Schulz}, {Tr{\"u}stedt}, {Wilms}, {Ojha}, {Ros}, {Anton}, {Baumgartner},
  {Beuchert}, {Blanchard}, {B{\"u}rkel}, {Carpenter}, {Eberl}, {Edwards},
  {Eisenacher}, {Els{\"a}sser}, {Fehn}, {Fritsch}, {Gehrels}, {Gr{\"a}fe},
  {Gro{\ss}berger}, {Hase}, {Horiuchi}, {James}, {Kappes}, {Katz},
  {Kreikenbohm}, {Kreykenbohm}, {Langejahn}, {Leiter}, {Litzinger}, {Lovell},
  {M{\"u}ller}, {Phillips}, {Pl{\"o}tz}, {Quick}, {Steinbring}, {Stevens},
  {Thompson}, \& {Tzioumis}}]{Krauss_2014}
{Krau{\ss}}, F., {Kadler}, M., {Mannheim}, K., {et~al.} 2014, \aap, 566, L7

\bibitem[{{Krau{\ss}} {et~al.}(2018){Krau{\ss}}, {Deoskar}, {Baxter}, {Kadler},
  {Kreter}, {Langejahn}, {Mannheim}, {Polko}, {Wang}, \& {Wilms}}]{Krauss_2018}
{Krau{\ss}}, F., {Deoskar}, K., {Baxter}, C., {et~al.} 2018, \aap, 620, A174

\bibitem[{{Loeb} \& {Waxman}(2006)}]{Loeb_2006}
{Loeb}, A., \& {Waxman}, E. 2006, J. Cosmology Astropart. Phys, 5, 003

\bibitem[{{Mannheim}(1993)}]{Mannheim_1993}
{Mannheim}, K. 1993, \aap, 269, 67

\bibitem[{Mannheim(1995)}]{Mannheim_1995}
Mannheim, K. 1995, Astroparticle Physics, 3, 295

\bibitem[{{Mannheim} \& {Biermann}(1989)}]{Mannheim_1989}
{Mannheim}, K., \& {Biermann}, P.~L. 1989, \aap, 221, 211

\bibitem[{Mannheim \& {Biermann}(1992)}]{Mannheim_1992}
Mannheim, K., \& {Biermann}, P.~L. 1992, \aap, 253, L21

\bibitem[{{Massaro} {et~al.}(2004){Massaro}, {Perri}, {Giommi}, \&
  {Nesci}}]{Massaro2004}
{Massaro}, E., {Perri}, M., {Giommi}, P., \& {Nesci}, R. 2004, \aap, 413, 489

\bibitem[{{Mattox} {et~al.}(1996){Mattox}, {Bertsch}, {Chiang}, {Dingus},
  {Digel}, {Esposito}, {Fierro}, {Hartman}, {Hunter}, {Kanbach}, {Kniffen},
  {Lin}, {Macomb}, {Mayer-Hasselwander}, {Michelson}, {von Montigny},
  {Mukherjee}, {Nolan}, {Ramanamurthy}, {Schneid}, {Sreekumar}, {Thompson}, \&
  {Willis}}]{Mattox_1996}
{Mattox}, J.~R., {Bertsch}, D.~L., {Chiang}, J., {et~al.} 1996, \apj, 461, 396

\bibitem[{{Meyer} {et~al.}(2019){Meyer}, {Scargle}, \&
  {Blandford}}]{Meyer_2019}
{Meyer}, M., {Scargle}, J.~D., \& {Blandford}, R.~D. 2019, \apj, 877, 39

\bibitem[{{M{\"u}cke} {et~al.}(2000){M{\"u}cke}, {Engel}, {Rachen},
  {Protheroe}, \& {Stanev}}]{Mucke_2000}
{M{\"u}cke}, A., {Engel}, R., {Rachen}, J.~P., {Protheroe}, R.~J., \& {Stanev},
  T. 2000, Computer Physics Communications, 124, 290

\bibitem[{{Murase} \& {Waxman}(2016)}]{Murase_2016}
{Murase}, K., \& {Waxman}, E. 2016, \prd, 94, 103006

\bibitem[{{Oikonomou} {et~al.}(2019{\natexlab{a}}){Oikonomou}, {Murase},
  {Padovani}, {Resconi}, \& {M{\'e}sz{\'a}ros}}]{Foteini_2019}
{Oikonomou}, F., {Murase}, K., {Padovani}, P., {Resconi}, E., \&
  {M{\'e}sz{\'a}ros}, P. 2019{\natexlab{a}}, \mnras, 489, 4347

\bibitem[{{Oikonomou} {et~al.}(2019{\natexlab{b}}){Oikonomou}, {Murase}, \&
  {Petropoulou}}]{Oikonomou_2019}
{Oikonomou}, F., {Murase}, K., \& {Petropoulou}, M. 2019{\natexlab{b}}, in
  European Physical Journal Web of Conferences, Vol. 210, European Physical
  Journal Web of Conferences, 03006

\bibitem[{{Ojha} {et~al.}(2019){Ojha}, {Zhang}, {Kadler}, {Neilson}, {Kreter},
  {McEnery}, {Buson}, {Caputo}, {Coppi}, {D'Ammando}, {De Angelis}, {Fang},
  {Giannios}, {Guiriec}, {Guo}, {Kopp}, {Krauss}, {Li}, {Meyer}, {Moiseev},
  {Petropoulou}, {Prescod-Weinstein}, {Rani}, {Shrader}, {Venters}, \&
  {Wadiasingh}}]{Roopesh_2019}
{Ojha}, R., {Zhang}, H., {Kadler}, M., {et~al.} 2019, \baas, 51, 431

\bibitem[{{Padovani} {et~al.}(2016){Padovani}, {Resconi}, {Giommi}, {Arsioli},
  \& {Chang}}]{Padovani_2016}
{Padovani}, P., {Resconi}, E., {Giommi}, P., {Arsioli}, B., \& {Chang}, Y.~L.
  2016, \mnras, 457, 3582

\bibitem[{{Paliya}(2015)}]{Paliya_2015}
{Paliya}, V.~S. 2015, \apjl, 808, L48

\bibitem[{{Perkins} {et~al.}(2018){Perkins}, {Ajello}, {Hartmann},
  {Marcotulli}, {Meyer}, {Paliya}, \& {Venters}}]{Perkins_2017}
{Perkins}, J.~S., {Ajello}, M., {Hartmann}, D., {et~al.} 2018, in PoS, Vol.
  ICRC2017, {AMEGO: Active Galactic Nuclei}, 598

\bibitem[{{Petropoulou} {et~al.}(2016){Petropoulou}, {Coenders}, \&
  {Dimitrakoudis}}]{Petropoulou_2016}
{Petropoulou}, M., {Coenders}, S., \& {Dimitrakoudis}, S. 2016, Astroparticle
  Physics, 80, 115

\bibitem[{{Petropoulou} \& {Mastichiadis}(2015)}]{Petropoulou_2015}
{Petropoulou}, M., \& {Mastichiadis}, A. 2015, \mnras, 447, 36

\bibitem[{{Petropoulou} {et~al.}(2019){Petropoulou}, {Murase}, {Brindley Fox},
  \& {Kawai}}]{Petropoulou_2019}
{Petropoulou}, M., {Murase}, K., {Brindley Fox}, D., \& {Kawai}, N. 2019, in
  AAS/High Energy Astrophysics Division, AAS/High Energy Astrophysics Division,
  106.35

\bibitem[{{Razzaque} {et~al.}(2003){Razzaque}, {M{\'e}sz{\'a}ros}, \&
  {Waxman}}]{Razzaque_2003}
{Razzaque}, S., {M{\'e}sz{\'a}ros}, P., \& {Waxman}, E. 2003, \prd, 68, 083001

\bibitem[{{Reimer} {et~al.}(2019){Reimer}, {B{\"o}ttcher}, \&
  {Buson}}]{Reimer_2019}
{Reimer}, A., {B{\"o}ttcher}, M., \& {Buson}, S. 2019, \apj, 881, 46

\bibitem[{{Righi} {et~al.}(2017){Righi}, {Tavecchio}, \& {Guetta}}]{Righi_2017}
{Righi}, C., {Tavecchio}, F., \& {Guetta}, D. 2017, \aap, 598, A36

\bibitem[{{Stecker} \& {Salamon}(1996)}]{Stecker_1996}
{Stecker}, F.~W., \& {Salamon}, M.~H. 1996, \ssr, 75, 341

\bibitem[{{Stone} {et~al.}(2008){Stone}, {Gardiner}, {Teuben}, {Hawley}, \&
  {Simon}}]{Stone_2008}
{Stone}, J.~M., {Gardiner}, T.~A., {Teuben}, P., {Hawley}, J.~F., \& {Simon},
  J.~B. 2008, \apjs, 178, 137

\bibitem[{{Tavecchio} \& {Ghisellini}(2015)}]{Tavecchio_2015}
{Tavecchio}, F., \& {Ghisellini}, G. 2015, \mnras, 451, 1502

\bibitem[{{The ANTARES Collaboration}(2015)}]{ANTARES_2015}
{The ANTARES Collaboration}. 2015, J. Cosmology Astropart. Phys, 2015, 014

\bibitem[{{Turley} {et~al.}(2016){Turley}, {Fox}, {Murase}, {Falcone},
  {Barnaba}, {Coutu}, {Cowen}, {Filippatos}, {Hanna}, {Keivani}, {Messick},
  {M{\'e}sz{\'a}ros}, {Mostaf{\'a}}, {Oikonomou}, {Shoemaker}, {Toomey},
  {Te{\v{s}}i{\'c}}, \& ophysical Multimessenger
  Observatory~Network}]{Turley_2016}
{Turley}, C.~F., {Fox}, D.~B., {Murase}, K., {et~al.} 2016, \apj, 833, 117

\bibitem[{{Verner} {et~al.}(1996){Verner}, {Ferland}, {Korista}, \&
  {Yakovlev}}]{Verner1996}
{Verner}, D.~A., {Ferland}, G.~J., {Korista}, K.~T., \& {Yakovlev}, D.~G. 1996,
  \apj, 465, 487

\bibitem[{{Waxman} \& {Bahcall}(1997)}]{Waxman_1997}
{Waxman}, E., \& {Bahcall}, J. 1997, Physical Review Letters, 78, 2292

\bibitem[{{Waxman} \& {Bahcall}(2000)}]{Waxman_2000}
{Waxman}, E., \& {Bahcall}, J.~N. 2000, \apj, 541, 707

\bibitem[{Wilks(1938)}]{Wilks_1938}
Wilks, S.~S. 1938, Annals Math. Statist., 9, 60

\bibitem[{{Wilms} {et~al.}(2000){Wilms}, {Allen}, \& {McCray}}]{Wilms2000}
{Wilms}, J., {Allen}, A., \& {McCray}, R. 2000, \apj, 542, 914

\end{thebibliography}

\end{document}